\documentclass{emulateapj}
\usepackage{apjfonts}

\newcommand{\oii}{[O{\sc ii}] }

\slugcomment{ApJ, in press}

\shorttitle{Star forming galaxies}
\shortauthors{Poggianti et al.}

\begin{document}


\title{The evolution of the star formation activity in galaxies and
 its dependence on environment
\footnote{Based on observations obtained
 at the ESO Very Large Telescope (VLT) as part of the Large programme
 166.A-0162 (the ESO Distant Cluster Survey). Based on observations
 made with the NASA/ESA Hubble Space Telescope, obtained at the Space
 Telescope Science Institute, which is operated by the Association of
 Universities for Research in Astronomy, Inc., under NASA contract NAS
 5-26555. These observations are associated with proposal \#9476.}}


\author{Bianca M. Poggianti$^1$}
\email{poggianti@pd.astro.it}

\author{Anja von der Linden$^2$, Gabriella De Lucia$^2$, Vandana Desai$^3$, Luc Simard$^4$, Claire Halliday$^5$, Alfonso Arag\'on-Salamanca$^6$, Richard Bower$^7$, Jesus Varela$^1$, Philip Best$^8$, Douglas I. Clowe$^9$, Julianne Dalcanton$^3$, Pascale Jablonka$^{10}$, Bo Milvang-Jensen$^{11}$, Roser Pello$^{12}$, Gregory Rudnick$^9$, Roberto Saglia$^{11}$, Simon D.M. White$^2$, Dennis Zaritsky$^9$}


\affil{$^1$INAF-Astronomical Observatory of Padova, Italy, $^2$Max-Planck-Institut fur Astrophysik, Garching, Germany, $^3$Astronomy Department, University of Washington, Box 351580, Seattle, WA 98195, $^4$Herzberg Institute of Astrophysics, National Research Council of Canada, Victoria, BC V9E 2E7, Canada, $^5$Institut fuer Astrophysik, Friedrich-Hund-Platz 1, 37077 Goettingen, Germany, $^6$School of Physics and Astrophysics, University of Nottingham, University Park, Nottingham NG7 2RD, United Kingdom, $^7$Department of Physics, University of Durham, South Road, DH1 3LE Durham, UK, $^8$Institute for Astronomy, Royal Observatory Edinburgh, Blackford hill, Edinburgh EH9 3HJ, UK, $^9$Steward Observatory, University of Arizona, 933 North Cherry Avenue, Tucson, AZ 85721, $^{10}$Observatoire de Gen\`eve, Laboratoire d'Astrophysique
 Ecole Polytechnique Federale de Lausanne (EPFL),
           CH-1290 Sauverny, Switzerland. On leave from GEPI, CNRS-UMR8111, Observatoire de Paris, section de Meudon, 5 Place Jules Janssen, F-92195 Meudon Cedex, France, $^{11}$Max-Planck Institut fur extraterrestrische Physik, Giessenbachstrasse, D-85748, Garching, Germany, $^{12}$Laboratoire d'Astrophysique, UMR 5572, Observatoire Midi-Pyrenees, 14 Avenue E. Belin, 31400 Toulouse, France}

\begin{abstract}
We study how the proportion of star-forming galaxies evolves between
$z=0.8$ and $z=0$ as a function of galaxy environment, using the \oii line in
emission as a signature of ongoing star formation.
Our high-z dataset comprises 16 clusters, 10 groups and another 250
galaxies in poorer groups and the field at $z=0.4-0.8$ 
from the ESO Distant Cluster Survey, plus another 9
massive clusters at similar redshifts.  As a local comparison, we use 
samples of galaxy systems
selected from the Sloan Digital Sky Survey at $0.04< z < 0.08$.  
At high-z most systems follow a broad anticorrelation
between the fraction of star-forming galaxies and the system velocity
dispersion.  At face value, this suggests that at
$z=0.4-0.8$ the mass of the system largely determines the proportion of
galaxies with ongoing star formation.  At these redshifts the
strength of star formation (as measured by the \oii equivalent width)
in star-forming galaxies is also found to vary systematically with
environment.  Sloan clusters have much lower fractions of star-forming
galaxies than clusters at $z=0.4-0.8$ and, in contrast with the distant
clusters, show a plateau for velocity
dispersions $\ge 550 \rm \, km \, s^{-1}$, where the fraction of galaxies
with \oii emission
does not vary systematically with velocity dispersion. 
We quantify the evolution of the proportion of star-forming galaxies 
as a function of the system velocity dispersion and find it is
strongest in intermediate-mass systems
($\sigma \sim 500-600 \, \rm km \, s^{-1}$ at z=0).  
To understand the origin of the
observed trends, we use the Press-Schechter formalism and the
Millennium Simulation and show that galaxy star formation histories
may be closely related to the growth history of clusters and groups.
We consider a scenario in which the population of passive galaxies
(those devoid of ongoing star formation at the time they are observed)
consists of two components: ``primordial'' passive galaxies
whose stars all formed at $z>2.5$ and ``quenched'' galaxies whose star
formation has been truncated due to the dense environment at later
times. 
We propose a scheme that is able
to account for the observed relations between the star-forming fraction
and $\sigma$ in clusters at high- and low-z.
If this scenario is roughly correct, 
the link between galaxy properties and environment is
extremely simple to predict purely from a knowledge of the growth
of dark matter structures.  
\end{abstract}





\section{Introduction}

The universe as a whole was more actively forming stars in the past
than today (Lilly et al. 1996, Madau, Pozzetti \& Dickinson 1998, Hopkins 2004,
Schiminovich et al. 2005).  Studies of galaxies in clusters, groups
and in the general field indicate an increased star formation activity
at higher redshifts, in all environments. However, a complete mapping
of the average star formation activity with redshift as a function of
environment has still not been achieved.

A large number of studies, during the last thirty years, have showed that 
distant clusters generally contain many star-forming galaxies.  In fact, the
first evidence for galaxy evolution in clusters, and for galaxy
evolution in general, has been the detection of evolution in the
star formation activity of cluster galaxies, as revealed by photometry
and spectroscopy.

Historically, the higher incidence of star--forming galaxies in
distant clusters compared to nearby clusters was first discovered by
photometric studies of the proportion of blue galaxies -- the
so--called Butcher--Oemler effect (Butcher \& Oemler 1978, 1984, Smail
et al. 1998, Margoniner \& de Carvalho 2000, Ellingson et al. 2001,
Kodama \& Bower 2001, Margoniner et al. 2001).

In agreement with the photometric results, spectroscopic studies of
distant clusters have found significant populations of emission-line
galaxies (Dressler \& Gunn 1982, 1983, Couch \& Sharples 1997,
Dressler \& Gunn 1992, Couch et al. 1994, Dressler et al. 1999, Fisher
et al. 1998, Postman et al. 1998, 2001, Balogh et al. 1997, 1998,
Poggianti et al. 1999, Tran et al. 2005, Demarco et al. 2005, Moran et 
al. 2005 to name a
few).  In contrast, nearby rich clusters (such as Coma) generally are
``known'' to have relatively few emission line galaxies.  Increased
star formation activity in distant clusters is also indicated by the
emission properties of composite cluster-integrated spectra (Dressler
et al. 2004).  In parallel to
the cluster studies, the fraction of star-forming galaxies has been
found to be higher at $z=0.3-0.5$ than at $z=0$ also in groups
(Allington-Smith et al. 1993, Wilman et al. 2005b).

While these observations have qualitatively shown that star--forming
galaxies were more common in the past than
today, {\it quantifying} this evolution has proved to be very hard.
At any given redshift, the properties of cluster galaxies display a
large cluster to cluster variance. Disentangling
cosmic evolution from cluster--to--cluster variations in a 
quantitative fashion has not been
possible to date due to the relatively small samples of clusters
studied in detail at different redshifts.  This difficulty in
measuring how the fraction of star--forming galaxies evolves with
redshift as a function of the cluster properties has affected all
types of studies, photometric and spectroscopic, both those based on the \oii
line from spectroscopic multislit surveys and $\rm H\alpha$
cluster-wide studies (Couch et al. 2001, Finn et
al. 2004, 2005, Kodama et al. 2004, Umeda et al. 2004). This might
be the reason why a quantitative detection of a clear evolution with
redshift in the fraction of star-forming galaxies has been elusive so
far (Nakata et al. 2005).

Knowing how galaxy properties depend on cluster and group properties
at different redshifts is therefore a necessary condition to
assess the amount of evolution with redshift,
even before attempting
to shed some light on how this evolution depends on environment.  
General trends were soon discovered by the early studies of nearby clusters, 
such as the fact that richer, more centrally
concentrated, relaxed clusters tend to have proportionally fewer
star-forming galaxies than less rich, irregular, unrelaxed clusters.
However, an exact portrait of how the star formation
activity in galaxies depends on the cluster characteristics is still
lacking. For example, apparently contrasting results have been
found in the literature regarding the presence (Martinez et al. 2002,
Biviano et al. 1997, Zabludoff \& Mulchaey 1998, Margoniner et
al. 2001, Goto et al. 2003) or absence (Smail et al. 1998, Andreon
\& Ettori 1999, Ellingson et al. 2001, Fairley et al. 2002, De Propris
et al. 2004, Goto 2005, Wilman et al. 2005a) of a relation between
galaxy properties and global cluster/group properties such as velocity
dispersion, X-ray luminosity and richness.

In this paper we analyze how the fraction of actively star-forming
galaxies varies with environment and redshift, comparing samples of
clusters and groups at $z=0.4$ to $0.8$ with samples in the local
universe. This study is based on the ESO Distant Cluster Survey, a
photometric and spectroscopic survey of distant clusters described in
\S2. Deriving the proportion of actively star-forming galaxies as
those with \oii emission in EDisCS and other high-z samples (\S3) and
comparing it with low redshift samples from the Sloan Digital Sky
Survey (\S4), we present how the fraction of star-forming galaxies
evolves between $z=0.4-0.8$ and $z=0$ as a function of the
cluster/group velocity dispersion (\S5). In \S5.3 we discuss the
incidence of \oii emitters in the poorest groups and the field, and in
\S5.4 we show how the distributions of the equivalent widths of \oii
vary with environment. Galaxy systems that strongly deviate from the
trends followed by most groups/clusters are discussed in \S5.5.  Star
formation activity and galaxy Hubble types are compared in \S5.6.
Finally, we propose a possible scenario accounting for the observed
trends and discuss its major implications in \S6.

Throughout the paper, 
line equivalent widths and cluster velocity dispersions 
are given in the rest frame. We use $H_{0}=70 \, \rm km \, s^{-1} \, Mpc^{-1}$,
$h=H_0/100$, ${\Omega}_m = 0.3$ and ${\Omega}_{\lambda}=0.7$.

\section{The EDisCS dataset}

Our study is based on data obtained by the ESO Distant Cluster Survey
(hereafter, EDisCS), a photometric and spectroscopic survey of
galaxies in 20 fields containing galaxy clusters at $z=0.4-1$. 
The goal of this project is to
study cluster and cluster galaxy evolution, characterizing the
structure,
stellar populations, internal kinematics, luminosities and masses of
galaxies in high redshift clusters. 

Candidate clusters were selected from the Las Campanas Distant Cluster
Survey (LCDCS) of Gonzalez et al. (2001).
Candidates
were identified by the LCDCS as a surface brightness excess using
a very wide filter ($\sim$ 4500-7500 \AA). The
EDisCS sample of 20 clusters was built from the 30 highest surface
brightness candidates in the LCDCS, confirming the presence of an
apparent cluster and of a possible red sequence with VLT 20min
exposures in two filters (White et al. 2005). 

Deep optical photometry with FORS2/VLT, near-IR
photometry with SOFI/NTT and multislit spectroscopy with FORS2/VLT
have been obtained for the 20 fields. ACS/HST mosaic imaging
of 10 of the highest redshift clusters has also been acquired (Desai et al. 
2006). 

An overview of the goals and strategy of the survey is given in White
et al. (2005) where the optical ground--based photometry is presented
in detail.  This consists of V, R and I imaging for the 10 highest
redshift cluster candidates, aimed to provide a sample
at $z \sim 0.8$ (hereafter the high-z sample) and B, V and
I imaging for 10 intermediate--redshift candidates, aimed to provide
a sample at $z \sim 0.5$ (hereafter the
mid-z sample).\footnote{In practice, 
the redshift distributions of the high-z and the mid-z
samples partly overlap, as can be seen in Table~1.}
A weak-shear analysis of gravitational lensing
by our clusters based on these data is presented in Clowe et al. (2005).

\begin{table*}
\begin{center}
{\scriptsize
\caption{EDisCS clusters.\label{tbl1}}
\begin{tabular}{llclcclcccrc}
\tableline\tableline
&&&&&&&&&&& \\
Cluster & Cluster & $z$ & $\sigma$ $\pm{\delta}_{\sigma}$ & $N_{mem}$ & $N_{\oii}$ & Imaging & $R_{200}$ & $f_{\oii}$ &  $f_{\oii}^{uncorr}$ &  $f_{\oii}^{lens}$ & $f_{\oii}^{\rm 1Mpc}$ \\
&&&&&&& (Mpc) &&&& \\
\tableline
 Cl\,1232.5-1250     &  Cl\,1232     & 0.5414  & 1080 $_{-89}^{+119}$ &  54  & 51 & BVIJK+ & 1.99 & 0.32$\pm$0.08 &  0.31 &  0.32 & 0.34 \\ 
 Cl\,1216.8-1201     &  Cl\,1216     & 0.7943  & 1018 $_{-77}^{+73}$  &  67  & 57 & VRIJK+ & 1.61 & 0.53$\pm$0.14 &  0.46 &  0.53 & 0.44 \\ 
 Cl\,1138.2-1133     &  Cl\,1138     & 0.4798  &  737 $_{-56}^{+77}$  &  48  & 24 & VRIJK+ & 1.41 & 0.59$\pm$0.16 &  0.62 &  0.63$^\star$ & 0.63 \\  
 Cl\,1411.1-1148     &  Cl\,1411     & 0.5201  &  709 $_{-105}^{+180}$&  26  & 18 & BVIK   & 1.32 & 0.24$\pm$0.11 &  0.22 &  0.24 & 0.24 \\ 
 Cl\,1301.7-1139     &  Cl\,1301     & 0.4828  &  681 $_{-86}^{+86}$  &  37  & 28 & BVIK   & 1.30 & 0.62$\pm$0.15 &  0.61 &  0.61$^\star$ & 0.57  \\ 
 Cl\,1354.2-1230     &  Cl\,1354     & 0.7627  &  668 $_{-80}^{+161}$ &  21  & 14 & VRIJK+ & 1.08 & 0.80$\pm$0.22 &  0.71 &  0.82$^\star$ & 0.75 \\  
 Cl\,1353.0-1137     &  Cl\,1353     & 0.5883  &  663 $_{-91}^{+179}$ &  22  & 16 & BVIK   & 1.19 & 0.45$\pm$0.17 &  0.44 &  0.34 & 0.34 \\ 
 Cl\,1054.4-1146     &  Cl\,1054-11  & 0.6972  &  589 $_{-70}^{+78}$  &  49  & 28 & VRIJK+ & 0.99 & 0.70$\pm$0.16 &  0.68 &  0.73 & 0.70 \\ 
 Cl\,1227.9-1138     &  Cl\,1227     & 0.6355  &  572 $_{-54}^{+96}$  &  22  & 12 & VRIJK+ & 1.00 & 0.69$\pm$0.24 &  0.67 &  0.67 & 0.69 \\ 
 Cl\,1202.7-1224     &  Cl\,1202     & 0.4244  &  540 $_{-83}^{+139}$ &  21  & 14 & BVIK   & 1.07 & 0.31$\pm$0.14 &  0.29 &  0.31 & 0.31 \\  
 Cl\,1059.2-1253     &  Cl\,1059     & 0.4561  &  517 $_{-40}^{+71}$  &  41  & 28 & BVIK   & 1.00 & 0.56$\pm$0.14 &  0.57 &  0.56$^\star$ & 0.56 \\ 
 Cl\,1054.7-1245     &  Cl\,1054-12  & 0.7498  &  504 $_{-65}^{+113}$ &  36  & 20 & VRIJK+ & 0.82 & 0.52$\pm$0.15 &  0.45 &  0.63 & 0.55 \\  
 Cl\,1018.8-1211     &  Cl\,1018     & 0.4732  &  474 $_{-57}^{+75}$  &  33  & 20 & BVIK   & 0.91 & 0.56$\pm$0.17 &  0.55 &  0.48$^\star$ & 0.46 \\ 
 Cl\,1040.7-1155     &  Cl\,1040     & 0.7043  &  418 $_{-46}^{+55}$  &  30  & 13 & VRIJK+ & 0.70 & 0.71$\pm$0.23 &  0.69 &  0.71 & 0.71 \\ 
 Cl\,1037.9-1243     &  Cl\,1037     & 0.5789  &  315 $_{-37}^{+76}$  &  19  &  8 & VRIJK+ & 0.57 & 0.90$\pm$0.33 &  0.88 &  0.92$^\star$ & 0.92 \\ 
 Cl\,1103.7-1245b    &  Cl\,1103     & 0.7029  &  242 $_{-104}^{+126}$&  11  &  3 & VRIJK+ & 0.41 & 1.00$\pm$0.58 &  1.00 &  --- & 1.00 \\ 
 Cl\,1420.3-1236     &  Cl\,1420     & 0.4959  &  225 $_{-62}^{+77}$  &  27  &  9 & BVIK   & 0.43 & 0.00$\pm$0.11 &  0.00 &  0.38$^\star$ & 0.34 \\ 
 Cl\,1119.3-1129     &  Cl\,1119     & 0.5500  &  165 $_{-19}^{+34}$  &  21  & 10 & BVI    & 0.30 & 0.26$\pm$0.20 &  0.40 &  0.21 & 0.21 \\ 
\tableline
\end{tabular}
}
\tablecomments{Col. (1): Cluster name.  Col. (2): Short cluster name. 
Col. (3) Cluster redshift. Col. (4) Cluster velocity dispersion. 
Redshifts and velocity dispersions are taken from Halliday et al. (2004) and
Milvang-Jensen et al. (2006). 
Col. (5) Number of spectroscopically confirmed members. 
Col. (6) Number of members used for computing the \oii fraction of Col.(9). 
Col. (7) Available imaging. A + sign indicates those clusters
with HST imaging. Col. (8) $R_{200}$ in Mpc. Col. (9) \oii fraction within
$R_{200}$ corrected
for completeness. Col. (10) \oii fraction within
$R_{200}$ uncorrected for completeness.
Col. (11) \oii fraction computed within a radius $R_{200}$ derived from the lensing
estimate of $\sigma$ (Clowe et al. 2005). An asterisk indicates systems
with additional mass structures along the line of sight, whose
lensing $\sigma$ is probably overestimated.
Col. (12) \oii fraction computed within a radius = 1 Mpc.
}
\end{center}
\end{table*}

Typically 4hrs-- (high-z sample) and 2hrs--exposure (mid-z sample)
spectra of $>100$ galaxies per cluster field were obtained.
Spectroscopic targets were selected from I-band catalogues.  At
the redshifts of our clusters this corresponds to $\sim 5000 \pm500$
\AA $\,$ rest frame. 
Conservative rejection criteria based on photometric redshifts were
used in the selection of spectroscopic targets
to reject a significant fraction of non--members while
retaininig a spectroscopic sample of cluster galaxies  
equivalent to a purely I-band selected one. {\it A posteriori}, we 
verified that these criteria have excluded
at most 1\% of the cluster galaxies (Halliday et al. 2004
and Milvang-Jensen et al. 2006).
The spectroscopic selection, observations and spectroscopic
catalogs are presented in detail in Halliday et al. (2004) and
Milvang-Jensen et al. (2006).

As explained in White et al. (2005), deep spectroscopy was not obtained
for two of the EDisCS fields (Cl\,1122 and Cl\,1238), hence they
have not been included in the present study.  In the following we
consider the other 18 EDisCS fields with high quality
spectroscopy. For each field, 
Table~1 lists the cluster name, redshift, velocity
dispersion and number of spectroscopically confirmed members
of the structure that was targeted for spectroscopy and that forms
the basis of our study.

\section{Deriving the [OII] fractions in clusters at high redshift}

In this paper we wish to investigate the incidence of actively star--forming
galaxies as a function of cluster velocity dispersion, 
and how this evolves with redshift.
We do this by analyzing the proportion of galaxies with a significant
\oii emission line at 3727 \AA,  
a reliable signal of ongoing star formation.
Dust and metallicity
variations affect significantly the {\it strength} of the [O{\sc
ii}] line, and a quantitative estimate of the star formation rate 
from the line flux depends
on slit coverage of the galaxy area and spectral extraction
method.
However, when the limit for line detection is sufficiently low, 
the simple {\it presence or absence} 
of this line in emission provides a clean estimate of the
incidence of star-forming galaxies in different environments and at
different redshifts.\footnote{If an AGN is present, this can
contribute to the emission line flux. However, in the great majority
of cases an AGN with an emission line spectrum is associated with some
level of star formation activity (e.g. Heckman et al. 1995, Cid
Fernandes et al. 2004 and references therein), therefore the
contamination of the population of \oii emitters from passive
galaxies is bound to be negligible.}

EDisCS spectra have
a dispersion of 1.32\AA/pixel and 1.66\AA/pixel depending on the
observing run, with a FWHM resolution of $\sim 6$\AA, corresponding
to rest frame 3.3 \AA $\,$ at z=0.8 and 4.3 \AA $\,$ at z=0.4.
The equivalent widths of \oii were measured on the spectra
with a line-fitting technique that follows the one used by the MORPHS
collaboration as in Dressler et al. (1999).
With this method each 1D spectrum is
inspected interactively. Each 2D spectrum was also inspected to confirm
the presence of an eventual line in 1D: this is especially useful to
assess the reality of weak \oii lines.

We classify as star--forming galaxies those with an equivalent width
(EW) of \oii $< - 3$ \AA $\,$ rest frame, adopting the convention that
EWs are negative when in emission. This is a reasonable limit for
galaxies with weak but still detectable current star formation
activity, for example nearby Sa galaxies.  Measuring the line strength of 
147 repeated spectra of galaxies (those observed more than once in
different masks/runs), we find that all of them fall into either the
star-forming or non--star-forming class in all repeated cases, proving that
the discrimination between galaxies with EW(\oii) $>$ and $<-3$ \AA
$\,$ is robust from our spectra.

For each EDisCS cluster, we have computed the fraction of
star--forming cluster members as the fraction of spectroscopically
confirmed members\footnote{After having excluded the repeated spectra
of galaxies that were observed more than once.} 
with a rest frame EW([O{\sc ii}])$\le -3$ \AA. Errorbars on the
fractions have been computed using Poissonian statistics. 
We consider only galaxies located within the projected
radius delimiting a sphere with interior mean density 200 times
the critical density ($R_{200}$) and with an absolute V magnitude
brighter than $M_{Vlim}$. $M_{Vlim}$ was varied with redshift
between -20.5 at $z=0.8$ and
-20.1 at $z=0.4$ to account for passive evolution.
Our spectroscopy would allow an analysis for galaxies
up to 0.5 mag fainter than these limits,
but for this study $M_{Vlim}$ was chosen to carry out a comparison with
the Sloan dataset (see below). 
Rest frame absolute magnitudes were estimated for each EDisCS galaxy as
in Rudnick et al. (2003) and are given in Rudnick et al. (2006). 
$R_{200}$ was computed from the cluster velocity dispersion $\sigma$
as in Finn et al. (2005):

\begin{equation}
R_{200} = 1.73 \, \frac{\sigma}{1000 \, \rm km \, s^{-1}} \,
{\frac{1}{\sqrt{{\Omega}_{\Lambda} + {\Omega}_{0}(1+z)^3}}} \, h^{-1} \, \rm Mpc
\end{equation}

 \begin{figure*}[t]
 \vspace{-7cm}
 \centerline{\hspace{6cm}\includegraphics[width= 1.5\columnwidth]{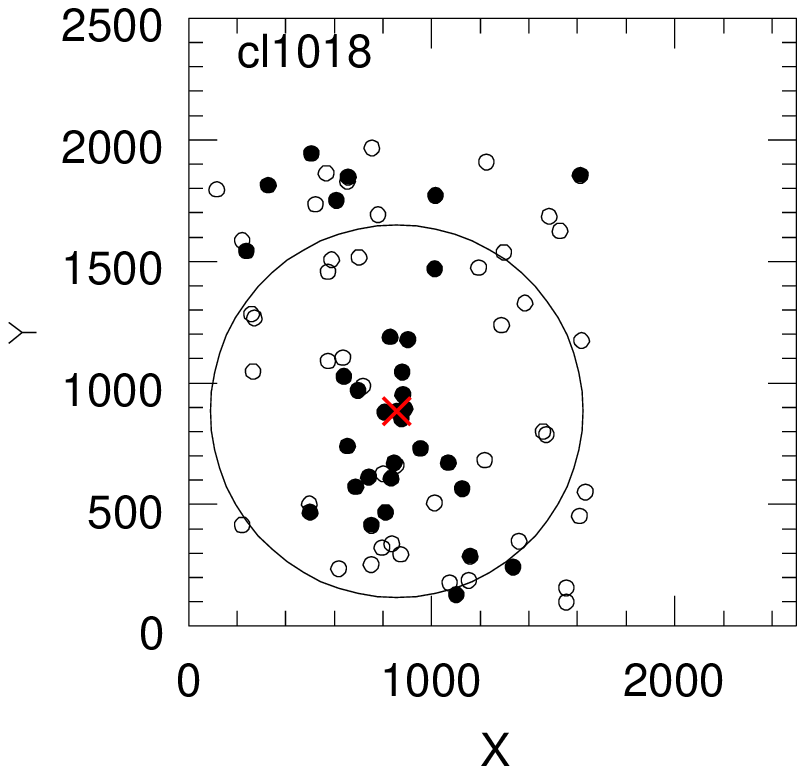}\hfill\hspace{-8cm}\includegraphics[width= 1.5\columnwidth]{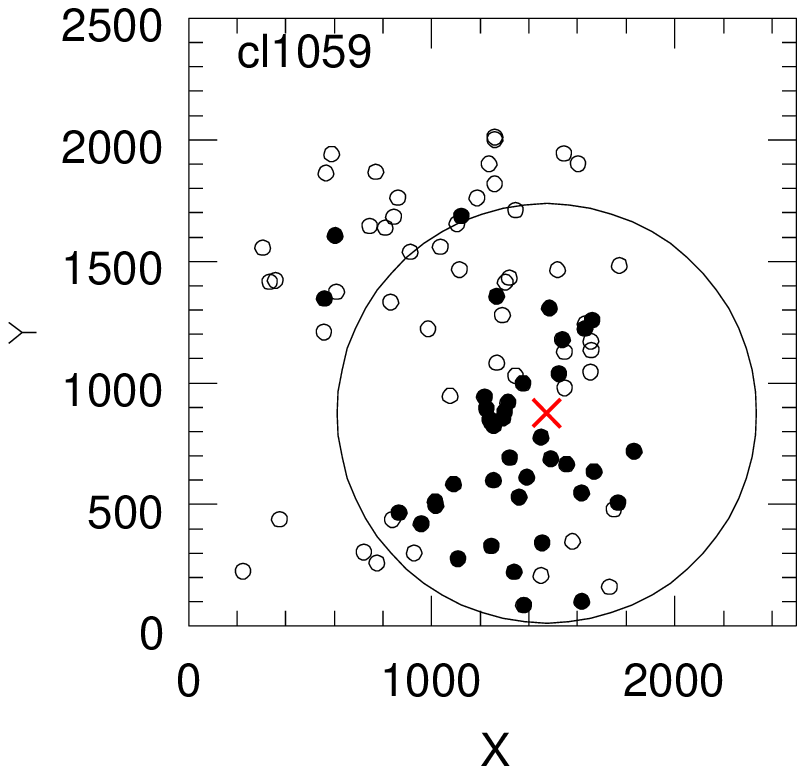}\hfill\hspace{-8cm}\includegraphics[width= 1.5\columnwidth]{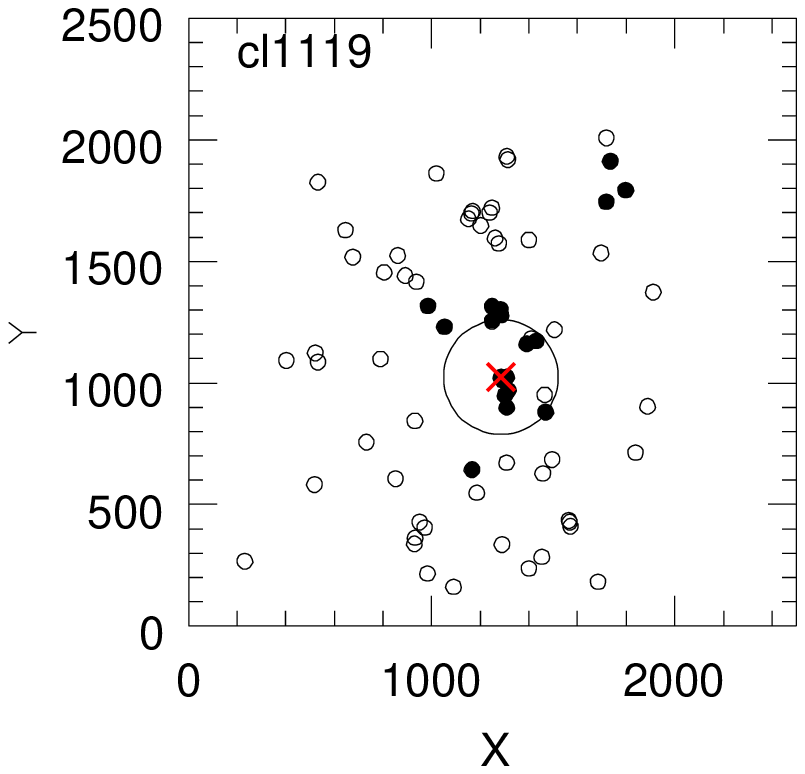}}
 \vspace{-8cm}
 \centerline{\hspace{6cm}\includegraphics[width= 1.5\columnwidth]{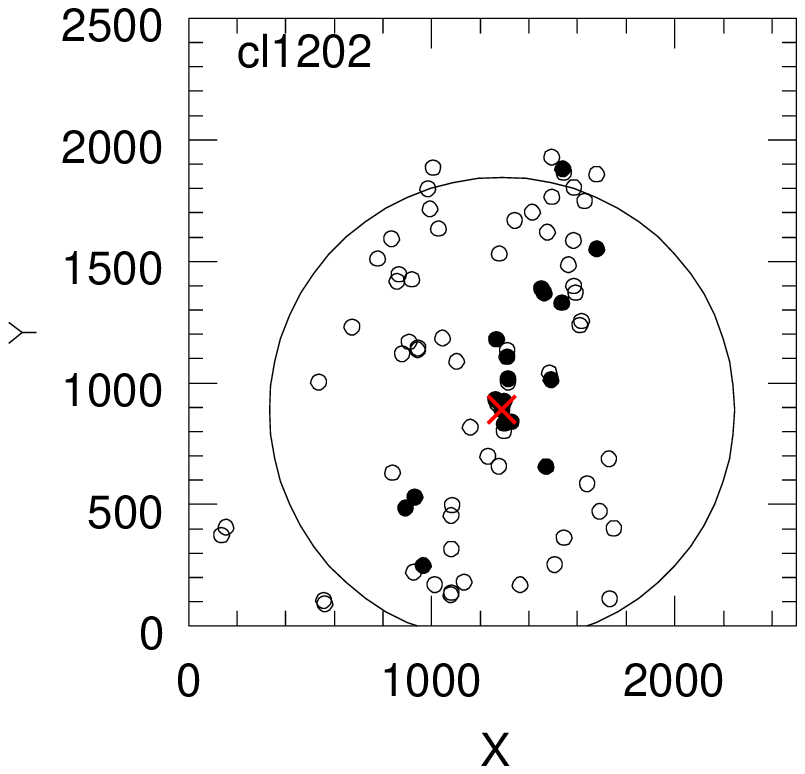}\hfill\hspace{-8cm}\includegraphics[width= 1.5\columnwidth]{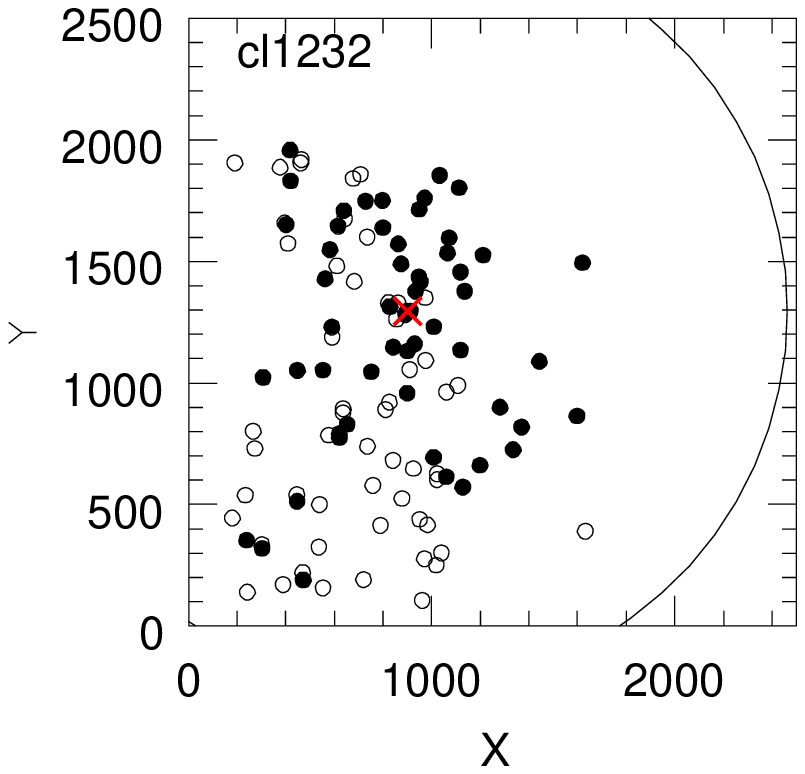}\hfill\hspace{-8cm}\includegraphics[width= 1.5\columnwidth]{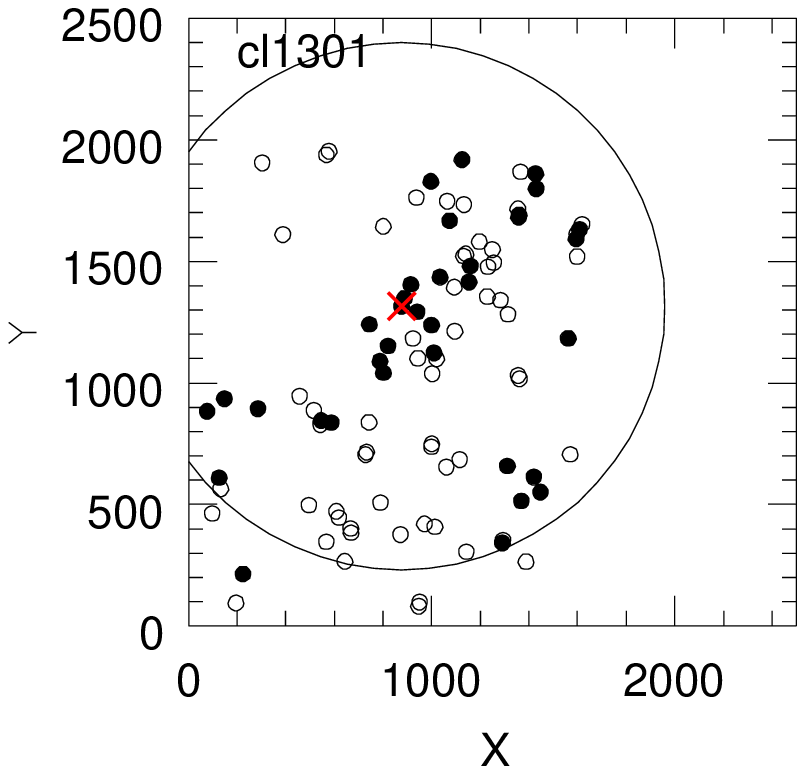}}
 \vspace{-8cm}
 \centerline{\hspace{6cm}\includegraphics[width= 1.5\columnwidth]{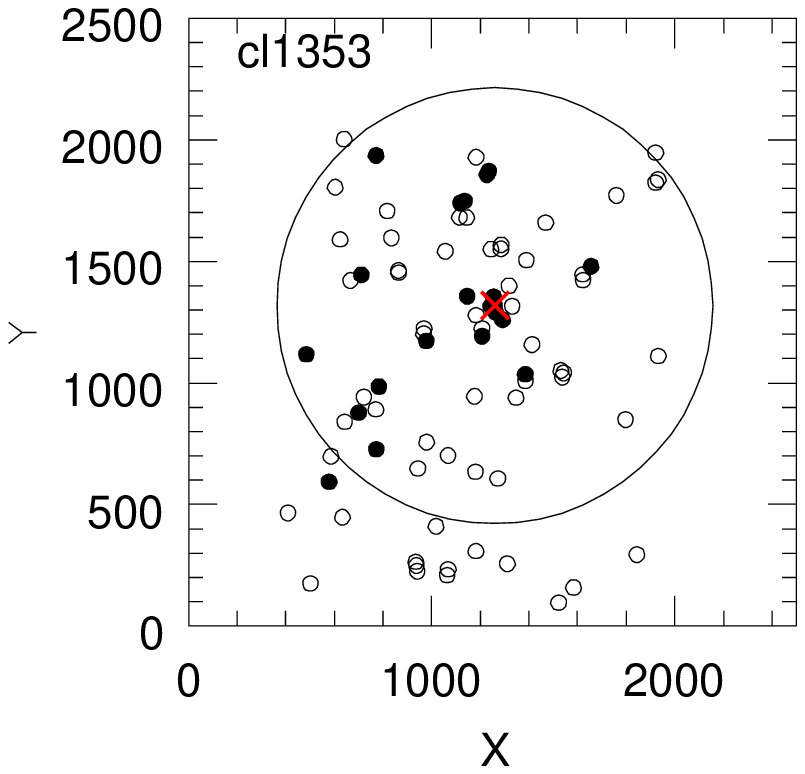}\hfill\hspace{-8cm}\includegraphics[width= 1.5\columnwidth]{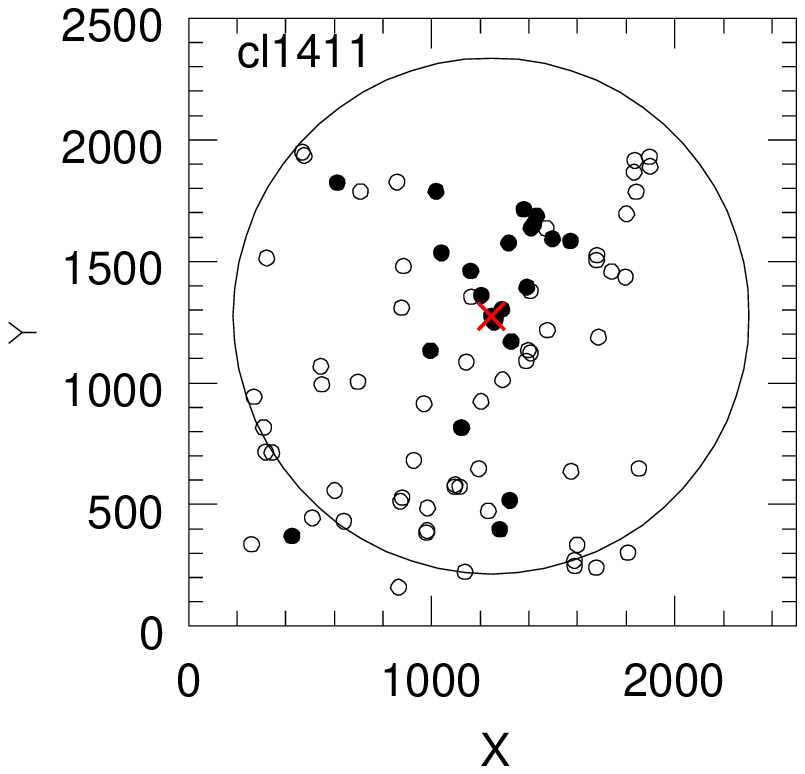}\hfill\hspace{-8cm}\includegraphics[width= 1.5\columnwidth]{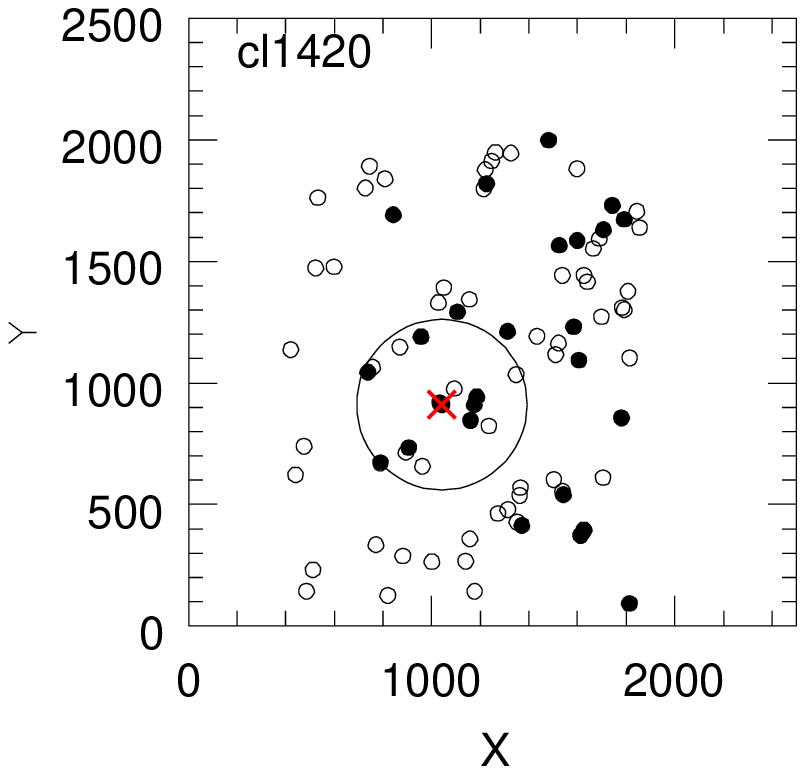}}
 \vspace{-1cm}
 \caption{XY pixel positions of objects with spectra
 in the EDisCs mid-z fields. Filled dots represent spectroscopically
 confirmed cluster members. The circle with radius $R_{200}$, centered
 on the BCG,  is shown. The axis units are pixels = 0.2\arcsec. \label{xy1}}
 \end{figure*}

 \begin{figure*}[t]
 \vspace{-7cm}
 \centerline{\hspace{6cm}\includegraphics[width= 1.5\columnwidth]{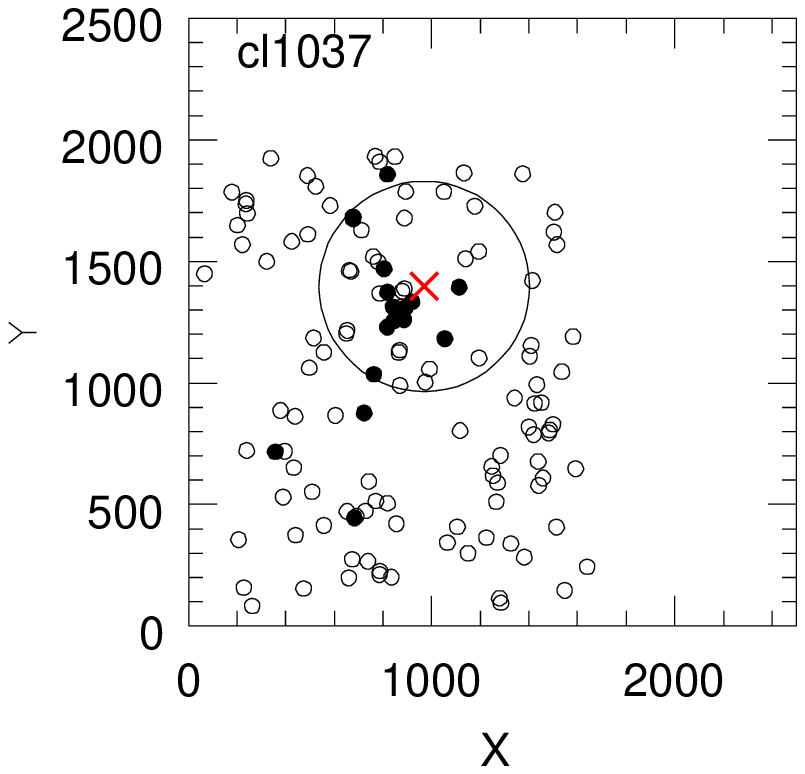}\hfill\hspace{-8cm}\includegraphics[width= 1.5\columnwidth]{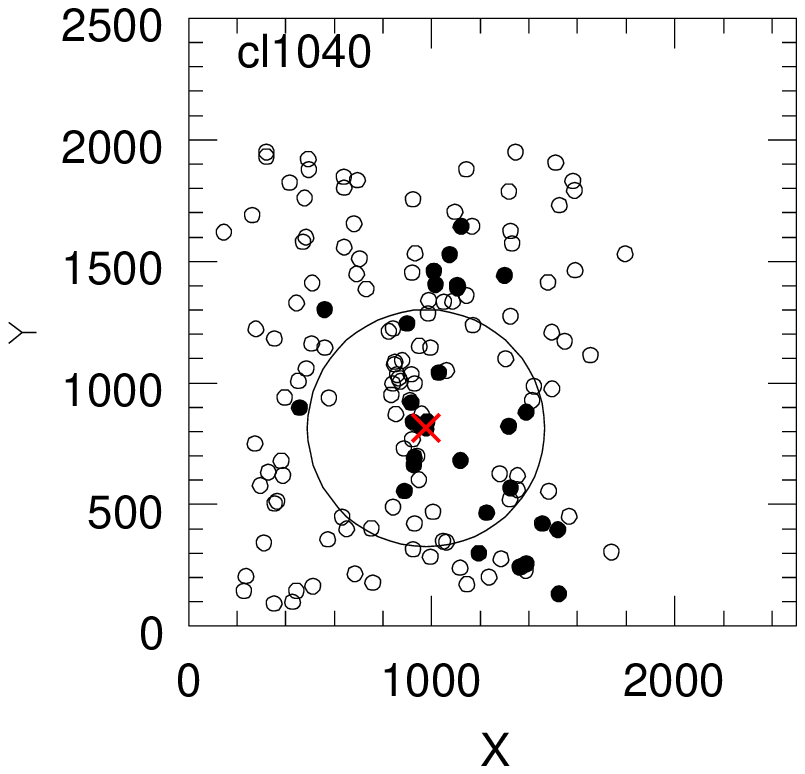}\hfill\hspace{-8cm}\includegraphics[width= 1.5\columnwidth]{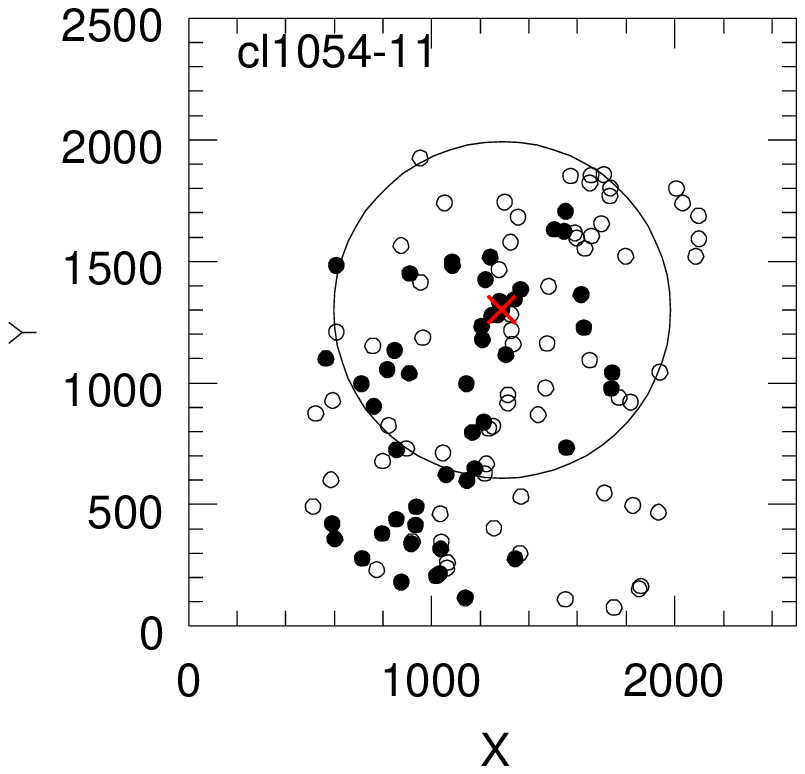}}
 \vspace{-8cm}
 \centerline{\hspace{6cm}\includegraphics[width= 1.5\columnwidth]{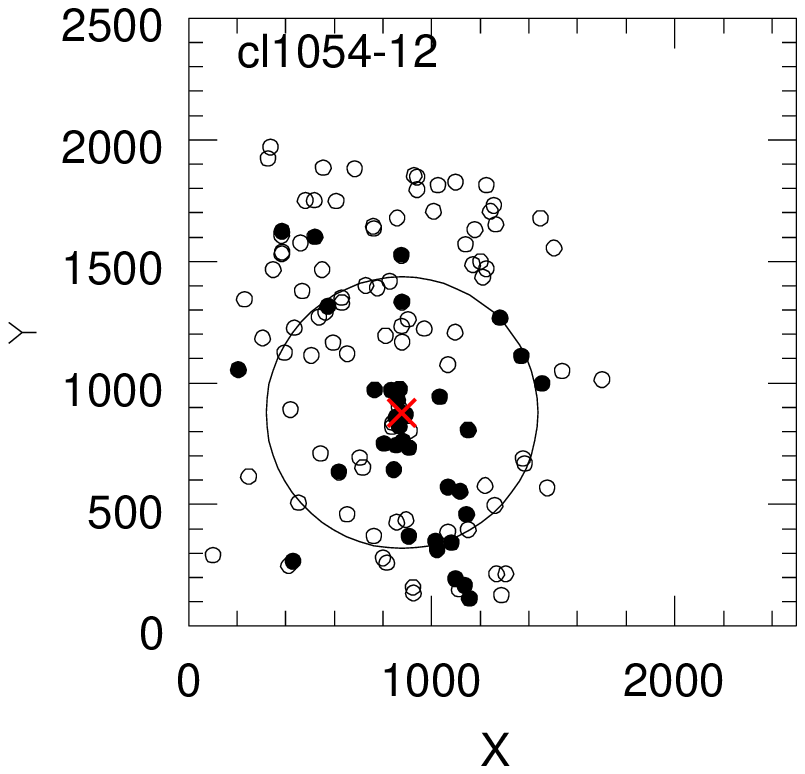}\hfill\hspace{-8cm}\includegraphics[width= 1.5\columnwidth]{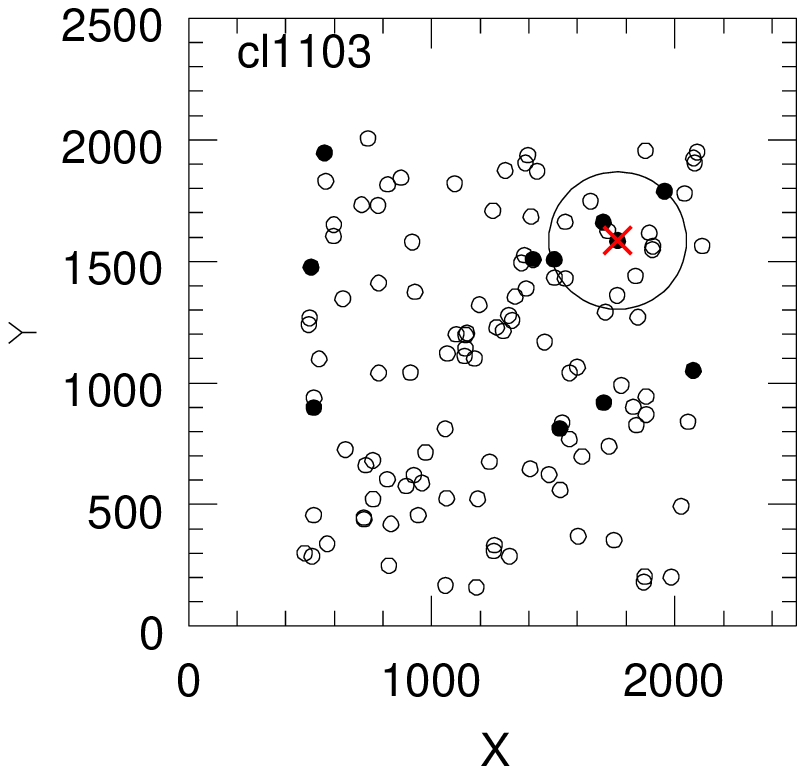}\hfill\hspace{-8cm}\includegraphics[width= 1.5\columnwidth]{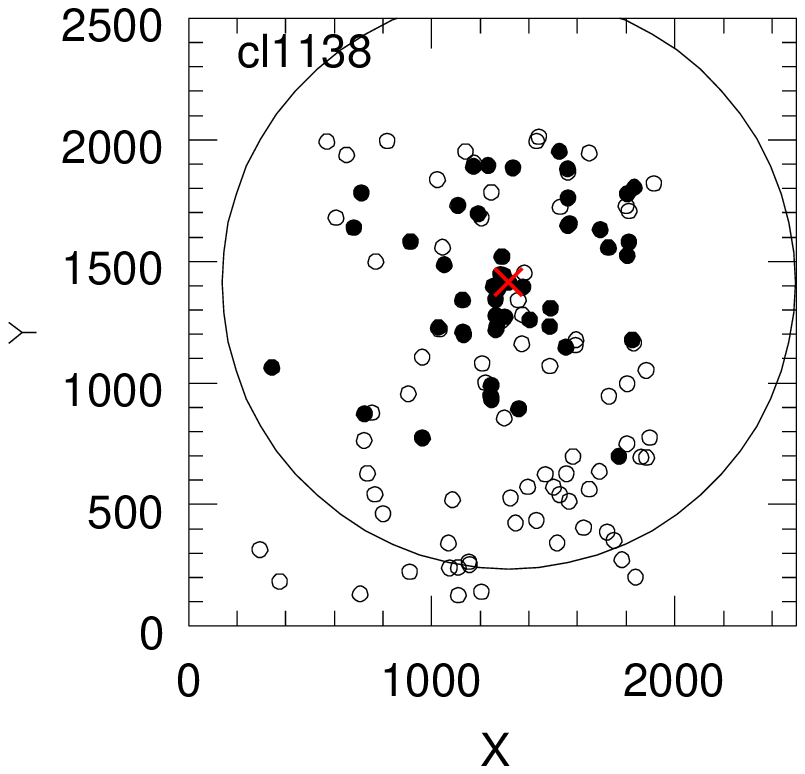}}
 \vspace{-8cm}
 \centerline{\hspace{6cm}\includegraphics[width= 1.5\columnwidth]{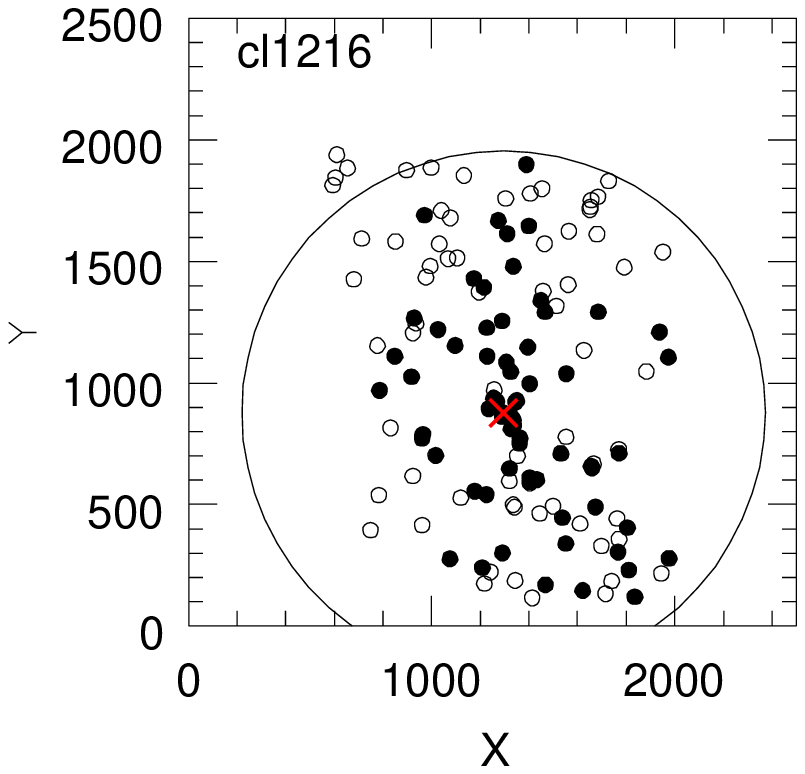}\hfill\hspace{-8cm}\includegraphics[width= 1.5\columnwidth]{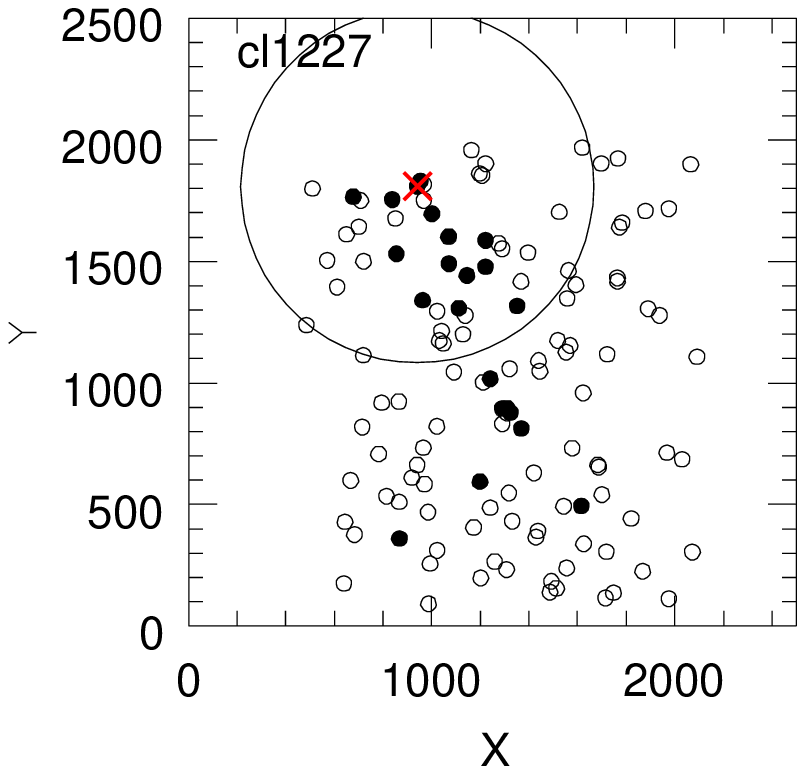}\hfill\hspace{-8cm}\includegraphics[width= 1.5\columnwidth]{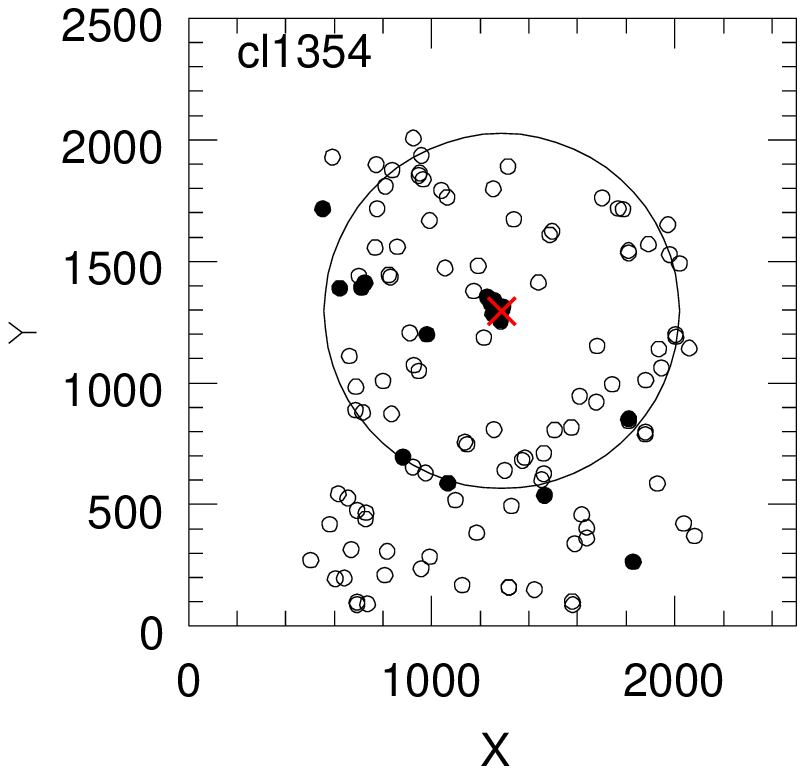}}
 \vspace{-1cm}
 \caption{Same as Fig.~\ref{xy1} but for the high-z EDisCS fields.\label{xy2}}
 \end{figure*}

The cluster center was assumed to coincide with
the Brightest Cluster Galaxy (BCG), that was identified interactively on the
EDisCS VLT images. A list of the BCGs can be found in White et al. (2005).
For most of the clusters, our spectroscopy samples at least out to 
the cluster $R_{200}$ or beyond (Figs.~\ref{xy1} and ~\ref{xy2}). 
A few clusters have incomplete radial
sampling due to their large
projected radii (most notably, Cl\,1232, Cl\,1411 and Cl\,1138), 
or to the BCG location close to the FORS2 field edge
(Cl\,1227): when relevant, these cases will be commented separately. 

The fractions of [O{\sc ii}] emitters were computed weighting each
galaxy for incompleteness of the spectroscopic catalog, taking into
account the completeness as a function of galaxy magnitude and
position, as described in Appendix A.  Ignoring these weights,
however, does not affect significantly our results, as discussed in
\S5.  In Appendix B, we show that the color distributions of the final
spectroscopic sample and its parent photometric sample are
indistinguishable according to a KS test, and therefore no color bias
is present in the spectroscopic sample we are using.

The exposure times of the EDisCS spectroscopy 
(4hrs and 2hrs of VLT for the high- and mid-z samples,
respectively) 
were chosen to allow not only a redshift
determination, but also a detailed spectroscopic analysis
of emission and absorption features. As a consequence,
the fraction of spectroscopic targets for which no redshift
could be derived is negligible:
only 3\% of the spectra 
brighter than the magnitude limit used here did not yield
a redshift.\footnote{As discussed in Halliday et al. (2004), 
the spectra show 
that most of these are bright lower redshift galaxies 
(non-members of our clusters) observed in a red rest-frame 
spectral region that is featureless and thus makes it hard to derive
a secure redshift.} Hence, no correction is required
to account for the success rate (percentage of
spectra providing a redshift) as a function of magnitude or color.

The \oii fractions derived as described above are given in 
column 9 of Table~1.
For comparison, the table also lists the \oii fractions computed
without applying completeness corrections (column 10),
or with different radial criteria:
within $R_{200}$ as derived from the $\sigma$ based on the
weak lensing analysis of Clowe et al. (2005) (column 11)
and within a fixed metric radius equal to 1 Mpc (column 12).
For all structures except one (Cl\,1420), 
the different estimates of the \oii fraction
are compatible within the errors.

\subsection{Other high-z cluster samples}

The EDisCS dataset is homogeneous for cluster and galaxy selection and
data quality, thus an internal comparison among clusters is
straightforward.  A comparison with other spectroscopic surveys of
distant clusters requires much more caution, as a
number of conditions need to be met: such a survey should cover out to
$R_{200}$ and be representative of a magnitude-limited sample of
galaxies selected in the rest-frame at 4500-5500 \AA.  A
reliable determination of the cluster velocity dispersion should be
available, as well as accurate EW([O{\sc ii}]) measurements highly
complete down to 3 \AA $\,$ for galaxies down to the absolute magnitude
limit $M_{Vlim}$. These are demanding requirements that are largely
fulfilled by surveys of just a few distant clusters in the literature.

The list of additional distant clusters we include in
our analysis is given in Table~2. Seven of these clusters are taken
from the MORPHS survey (Dressler et al. 1999, Poggianti et al. 1999, hereafter
D99 and P99) and are at redshifts covering the low redshift end
of the EDisCS redshift range ($z=0.38-0.55$). Two other clusters
are MS1054-03 at z=0.83 taken from van Dokkum et al. (2000) (hereafter vD00),
and Cl1324+3011 at z=0.76 from Postman et al. (2001) (hereafter POL01).

\begin{table*}
\begin{center}
{\scriptsize
\caption{Other distant clusters.\label{tbl2}}
\begin{tabular}{lclcccccrr}
\tableline\tableline
Cluster & $z$ & $\sigma$ $\pm{\delta}_{\sigma}$ & $N_{\oii}$ & Imaging & $R_{200}$(Mpc) & FOV & $f_{\oii}$ & Ref1\tablenotemark{a} & Ref2\tablenotemark{b} \\
\tableline
 Cl\,1447       & 0.3762  &  838 $_{-163}^{+163}$ &  21  & $r/i$, $M_V=-19.8$  & 1.70  & $1.2 \times 1.2 \, R_{200}$ & 0.56$\pm$0.16 &  D99,P99 & GM01 \\
 Cl\,0024       & 0.3928  &  911 $_{-107}^{+81}$  & 107  & $r/i$, $M_V=-19.8$  & 1.83  & $1.1 \times 1.1 \, R_{200}$ & 0.36$\pm$0.06 & D99,P99 & GM01 \\
 Cl\,0939       & 0.4060  & 1067 $_{-96}^{+89}$   &  71  & $r/i$, $M_V=-19.8$  & 2.13  & $1.0 \times 1.0 \, R_{200}$ & 0.26$\pm$0.06 & D99,P99 & GM01 \\
 Cl\,0303       & 0.4184  &  876 $_{-140}^{+144}$ &  51  & $r/i$, $M_V=-19.8$  & 1.73  & $1.3 \times 1.3 \, R_{200}$ & 0.55$\pm$0.10 & D99,P99 & GM01 \\
 3C295          & 0.4593  & 1642 $_{-187}^{+224}$ &  25  & $r/i$, $M_V=-19.8$  & 3.18  & $0.7 \times 0.7 \, R_{200}$ & 0.18$\pm$0.08 & D99,P99 & GM01 \\
 Cl\,1601       & 0.5388  &  646 $_{-87}^{+84}$   &  58  & $r/i$, $M_V=-19.8$  & 1.19  & $2.0 \times 2.0 \, R_{200}$ & 0.15$\pm$0.05 & D99,P99 & GM01 \\
 Cl\,0016       & 0.5459  &  984 $_{-95}^{+130}$  &  29  & $r/i$, $M_V=-19.8$  & 1.81  & $1.4 \times 1.4 \, R_{200}$ & 0.14$\pm$0.07 & D99,P99 & GM01 \\
\tableline\tableline
 MS1054-03      & 0.8315  & 1150 $_{-97}^{+97}$   &  71  & $I_{814}$, $M_V=-20.5$   & 1.78  & $0.5 \times 0.8 \, R_{200}$ & 0.31$\pm$0.06 & vD00    & vD00 \\
\tableline\tableline
 Cl\,1324+3011  & 0.7565  & 1016 $_{-93}^{+126}$  &  27  & R, $M_V=-20.4$    & 1.71  & $0.3 \times 0.9 \, R_{200}$ & 0.47$\pm$0.14 & PLO01   & LOP02 \\
\tableline
\end{tabular}
}
\tablenotetext{a}{Source for the \oii measurements and completeness functions. D99=Dressler et al. 1999; P99=Poggianti et al. 1999; vD00=van Dokkum et al. 2000; PLO01=Postman, Lubin \& Oke 2001.}
\tablenotetext{b}{Source for the cluster velocity dispersion. GM01=Girardi \& Mezzetti 2001; LOP02=Lubin, Oke \& Postman 2002.}
\tablecomments{Col. (1) Cluster name. Col. (2) Cluster redshift.
Col. (3) Cluster velocity dispersion.
Col. (4) Number of galaxies
members of the cluster used for the calculation of the \oii fraction.
Col. (5) Photometric band used for selection of spectroscopic
targets and magnitude limit we adopted to be compatible with
EDisCS. Col. (6) $R_{200}$ in Mpc. Col. (7) Field-of-view of the
spectroscopic coverage. Col. (8) \oii fraction. Col. (9)-(10)
References. 
An additional cluster presented in POL01 has later been
shown to be composed of 4 distinct clusters for which a spectroscopic
catalog should become available in the future (Gal \& Lubin 2004).
A cluster from Postman et al. (1998) was not included because its
completeness function was not available. 
The cluster and the group at $z =0.59$ in the
MS2053 field of Tran et al. (2005) have $\sigma=865$ and
$f_{\oii}=0.34$, and $\sigma=282$ and $f_{\oii}=0.67$, respectively,
where $f_{\oii}$ is given by the authors for galaxies with
EW(\oii)$<-5$ \AA,
down to approximately the same galaxy magnitude limit and radius
adopted here.
Since their EW limit is higher than the 3 \AA $\,$ limit we have
used, we decided not to include these structures in
Fig.~\ref{main}, but we note that these points would roughly follow
the trend of the other distant clusters in the plot.
Any other cluster from the
literature at $z \ge 0.4$ could not be included because missing one or
more necessary pieces of information (\oii catalogs, completeness
information, etc.).}
\end{center}
\end{table*}

Measurements of the EW(\oii) were taken from these authors, assuming
their spectroscopic catalogs are highly complete for EW(\oii)$< -3$ \AA.
This is the case for the MORPHS sample, as discussed in D99, and appears
to be a reasonable assumption also for vD00 and POL01, given the EW
distributions and errorbars in their catalogs. 
These samples were weighted for incompleteness using the completeness
functions provided by the authors (Table~2). Galaxies were included
in the \oii computation if brighter than the closest available apparent magnitude 
limits corresponding to the absolute magnitude limits adopted
for EDisCS, as a function of redshift. The radial coverage for 
these clusters in units of $R_{200}$ are shown in Table~2.
We note that the spectroscopic sample of the POL01 cluster
was selected in rest-frame U-band, while the EDisCS and all other
clusters used in this analysis were selected at $\sim 5000 \pm 500$ \AA $\,$ 
rest frame.
Though the estimate of the \oii fraction in clusters in these external 
samples cannot be carried out in a way that is
fully homogeneous with the analysis performed on the EDisCS 
data, due to the slight differences in radial coverage, magnitude limit 
and so on, such differences are sufficiently small to allow an
interesting comparison with the EDisCS data: this will be presented
in \S5.

\section{The [O{\sc ii}] fractions at low redshift: Sloan}

In order to compare with clusters at low redshift, we constructed a
local comparison sample from the spectroscopic Sloan Digital Sky
Survey. Rather than trying to obtain a sample with the largest
possible number of clusters, we aimed to build a sample with selection
criteria similar to EDisCS. For simplicity, we used the Abell cluster
catalog. Its selection is based on (projected) overdensities of
galaxies, which can be regarded as being similar to the selection of
EDisCS clusters, that were chosen by their light excess over the
background.

Our Abell sample was built according to the following steps:

1) At the time of sample selection, the most comprehensive compilation of
properties of Abell clusters was by Struble \& Rood (1991). From this, we
selected clusters with a redshift estimate based on at least two
galaxies. This yields 774 clusters. 

2) Only clusters with $0.04<z<0.085$ were selected. This
reduces the sample to 227 clusters. The lower limit in redshift 
is chosen to reduce fibre aperture effects. At $z>0.04$, the Sloan 
fibres sample a significant fraction of the galaxy light, and the spectral
classification into star-forming and non--star-forming galaxies 
should not differ significantly from the integrated spectral class 
(Kewley, Jansen \& Geller 2005).
The upper redshift limit is imposed by the need to have spectra
for galaxies down to a sufficiently deep absolute magnitude, for comparison
with EDisCS. Above $z=0.085$, the Sloan spectroscopy samples
only the bright end of the galaxy luminosity function,
galaxy numbers per cluster become too small and errors on the \oii
fraction are too high to reach solid conclusions. 

3) For each cluster, we identify galaxies from the spectroscopic DR2 SDSS
catalog which lie within one Abell radius ($R_A = 1.7\arcmin/z$) from
the cluster center quoted by Struble \& Rood (1991). 
Only clusters with at least 20 matched galaxies
are retained (32 clusters).

4) 
  At this stage, the image of each cluster was 
  inspected interactively. We restricted
  the sample to clusters that are well separated from the survey
  boundaries, and we identified a BCG
  from the SDSS imaging data. 
  Redshift histograms were also inspected to verify the presence
  of a concentration of galaxies at the redshift given by
  Struble \& Rood (1991). 
  These constraints yield a sample of 24 clusters, to which
  we add two clusters with redshifts slightly
  lower than 0.04 and two with redshifts slightly
  higher than 0.085.
The final list of 28 clusters is presented in Table~3.

5)  As for the EDisCS sample, we rely on the biweight estimator of
  Beers et al. (1990) for determining the cluster redshift and
  velocity dispersion, as described in Appendix C.

Once our low--redshift comparison sample was selected,
a number of steps were taken to ensure a meaningful comparison with 
our high-z sample.

1) The Sloan spectroscopic target selection was performed
in the $r$ band ($r<17.7$).
In order to more closely approximate the rest-frame
EDisCS selection wavelength, we extracted a $g$-selected sample 
from the Sloan spectroscopic catalogs.
This corresponds to the subset of galaxies with $g \le 18$:
brighter than this limit, 99\% of the galaxies have $r<17.7$
and their $g$-magnitude distribution follows closely the distribution
in the whole $g$-band photometric sample. Galaxies
brighter than $g=12$ and $r=12$ were excluded, being brighter than any cluster
member of the clusters considered. Therefore, the Sloan spectroscopic sample 
we used is the subset of the Sloan catalogs with $12<r<17.7$ and $12<g<18.0$.

2) Each galaxy in the spectroscopic catalog was assigned
completeness weights as a function of magnitude and position
comparing, cluster by cluster, 
the number of galaxies in the Sloan spectroscopic and 
photometric ($g$-band) catalogs, as done for EDisCS (see Appendix A).
As in EDisCS, essentially all targeted galaxies (99.9\%) yield
a reliable redshift  (Strauss et al. 2002), thus no correction is 
required to account for the spectroscopic success rate.

3) The analysis of the fraction of [O{\sc ii}]-emitters was carried out
on the Sloan data as for EDisCS. 
The center of each cluster was assumed to coincide with the
BCG. Only galaxies within $R_{200}$ were
considered, down to an absolute V magnitude limit $= -19.8$, corresponding
to the limit at which spectroscopic incompleteness sets in 
at these redshifts in the Sloan sample. This limit 
was used to determine the absolute magnitude limit in the distant 
clusters, once passive evolution was taken into account. 
Galaxy absolute V magnitudes were obtained
from the absolute Petrosian 
magnitudes in the Sloan system 
using the transformation of Blanton  
(http://astro.physics.nyu.edu/\~{}mb144/kcorrect/linear.ps).

 \begin{figure}
 \vspace{-3cm}
 \centerline{\hspace{2cm}\includegraphics[width=12cm]{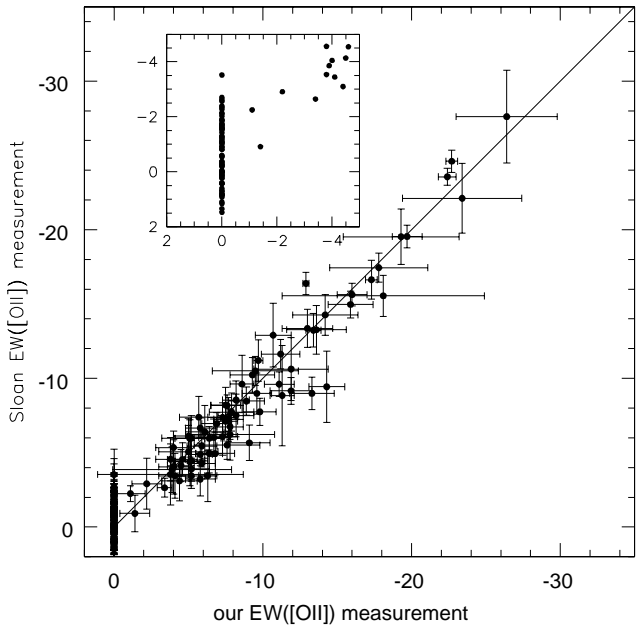}}
 \caption{EW(\oii)s of Sloan spectra in plate
 \#973. The measurement obtained with the Ediscs method
 is compared with the EW listed by Sloan. The inset is a blow-up of
 the lower left corner of the plot, in which errorbars are omitted
 for clarity. In this plot, 
 EDisCS EWs = 0 correspond to those spectra in which the line is not
 detected and the spectral fluctuations in that region are
 considered noise. Since the Sloan measurements are fully automatic,
 these cases can yield a non-null negative or positive
 EW value, that is however consistent
 with zero within the errorbar in most cases.  
 \label{sloanspectra}}
 \end{figure}

4) The star-forming fraction was computed as the fraction of galaxies
with EW(\oii)$< -3$ \AA $\,$ rest frame. 
The EWs(\oii) of the Sloan spectra were taken from 
Brinchmann et al. (2004a,b).  
Since these were measured with a different routine than the
one we used for EDisCS, we first
tested the consistency between the Sloan and our EW measurements on
all the 162 spectra of a random Sloan plate (plate \#973).
The comparison is shown in Fig.~\ref{sloanspectra}, where a
good agreement is visible. Since our purpose is to obtain
a meaningful comparison of the fraction of galaxies with EW(\oii)$< -3$
\AA, we are mainly interested in those eventual cases of
discrepancy between our and the Sloan measurement that could cause a
different classification (star-forming versus non-star-forming). 
We find that in 99\% of the cases 
our EW measurements agree with the Sloan EWs 
for the purpose of dividing galaxies into $>$ and $< -3$ 
\AA.\footnote{As visible in the inset of Fig.~\ref{sloanspectra},
only 2 cases of small discrepancy are found
among the 162 galaxies: galaxy \#506 with 
EW(Sloan)=-2.6$\pm$0.6 and EW(EDisCS)=-3.4$\pm$0.5 \AA,  and galaxy \#369
with EW(Sloan)=-3.5$\pm$1.1 and EW(EDisCS)=0, because no line appears 
to be present from a visual inspection of the spectrum.} 
We conclude that using the EWs by Brinchmann et al. (2004)
does not introduce any systematics in the comparison with the high
redshift clusters, and we use their EW measurements to compute the
\oii fraction in low-z clusters.
\oii fractions computed with and without completeness corrections,
together with redshifts, velocity dispersions, $R_{200}$ and number of
spectroscopically confirmed cluster members used to compute the
\oii fractions are listed for our Abell-Sloan clusters in Table~3.

\begin{table}
\begin{center}
\caption{Sloan clusters.\label{tbl3}}
\begin{tabular}{lclcccc}
&&&&&& \\
\tableline\tableline
Cluster & $z$ & $\sigma$ $\pm{\delta}_{\sigma}$ & $N_{\oii}$ & $R_{200}$ & $f_{\oii}$ &  $f_{\oii}^{uncorr}$ \\
\tableline
A2255  & 0.0801  &1151$\pm$64  & 106& 2.74 & 0.22$\pm$0.05 & 0.24 \\
A1767  & 0.0707  & 908$\pm$54  & 48 & 2.17 & 0.19$\pm$0.06 & 0.19 \\
A85    & 0.0555  & 861$\pm$48  & 46 & 2.07 & 0.16$\pm$0.06 & 0.17 \\
A160   & 0.0425  & 842$\pm$64  & 25 & 2.04 & 0.29$\pm$0.11 & 0.28 \\
A1066  & 0.0691  & 833$\pm$62  & 39 & 1.99 & 0.31$\pm$0.09 & 0.31 \\
A2670  & 0.0761  & 804$\pm$48  & 53 & 1.92 & 0.28$\pm$0.08 & 0.30 \\
A1650  & 0.0837  & 770$\pm$72  & 32 & 1.83 & 0.17$\pm$0.08 & 0.22 \\
A1809  & 0.0794  & 730$\pm$55  & 37 & 1.74 & 0.16$\pm$0.07 & 0.16 \\
A628   & 0.0838  & 667$\pm$67  & 24 & 1.58 & 0.54$\pm$0.14 & 0.50 \\
A1424  & 0.0755  & 664$\pm$53  & 33 & 1.58 & 0.21$\pm$0.08 & 0.21\\
A2593  & 0.0417  & 659$\pm$52  & 13 & 1.60 & 0.07$\pm$0.08 & 0.08 \\
A1564  & 0.0792  & 600$\pm$63  & 19 & 1.43 & 0.17$\pm$0.11 & 0.21 \\
A2197  & 0.0303  & 586$\pm$27  & 31 & 1.43 & 0.25$\pm$0.09 & 0.26 \\
A117   & 0.0551  & 570$\pm$46  & 17 & 1.37 & 0.30$\pm$0.13 & 0.29 \\
A1559  & 0.1056  & 541$\pm$86  & 8  & 1.27 & 0.31$\pm$0.18 & 0.25 \\
A933   & 0.0969  & 515$\pm$55  & 15 & 1.22 & 0.26$\pm$0.13 & 0.27 \\
A1780  & 0.0776  & 500$\pm$63  & 18 & 1.19 & 0.25$\pm$0.12 & 0.28 \\
A1452  & 0.0616  & 485$\pm$105 & 7  & 1.16 & 0.42$\pm$0.25 & 0.43 \\
A116   & 0.0667  & 471$\pm$71  & 4  & 1.13 & 0.47$\pm$0.35 & 0.50 \\
A2448  & 0.0820  & 466$\pm$81  & 11 & 1.11 & 0.28$\pm$0.16 & 0.27 \\
A1468  & 0.0850  & 464$\pm$105 & 16 & 1.10 & 0.38$\pm$0.15 & 0.38 \\
A1139  & 0.0393  & 436$\pm$46  & 10 & 1.06 & 0.38$\pm$0.20 & 0.40 \\
A1507  & 0.0599  & 419$\pm$47  & 11 & 1.01 & 0.54$\pm$0.22 & 0.55 \\
A2630  & 0.0669  & 402$\pm$69  & 9  & 0.96 & 0.32$\pm$0.19 & 0.30 \\
A1218  & 0.0801  & 365$\pm$77  & 7  & 0.87 & 0.55$\pm$0.29 & 0.57 \\
A1171  & 0.0748  & 352$\pm$51  & 5  & 0.84 & 0.67$\pm$0.35 & 0.60 \\
A1534  & 0.0699  & 333$\pm$39  & 11 & 0.80 & 0.18$\pm$0.13 & 0.18 \\
A1279  & 0.0544  & 192$\pm$37  & 3  & 0.46 & 1.00$\pm$0.58 & 1.00 \\
\tableline
\end{tabular}
\tablecomments{Col. (1): Cluster name.  
Col. (2) Cluster redshift. Col. (3) Cluster velocity dispersion. 
Col. (4) Number of members used for computing the \oii fraction.
Col. (5) $R_{200}$ in Mpc. Col. (6) \oii fraction corrected
for completeness. Col. (7) \oii fraction uncorrected for completeness.
}


\end{center}
\end{table}

To test our results on a larger control sample that, however, resembles 
less closely the EDisCS sample for selection and characteristics, we also
chose a second
cluster sample in the SDSS from the C4 catalog of Miller et al. (2005) 
at redshift $0.04 \le z \le 0.08$. 
The purity of the C4 sample is discussed in Miller et al. (2005).
Such a sample is more prone to be contaminated
by filaments, sheets and multiple structures 
yielding an overestimated $\sigma$.
To minimize the contamination
of structures with severely overestimated velocity dispersion,
we retained only clusters with $\sigma \le 1500 \, \rm km \, s^{-1}$, 
$\sigma \le {\sigma}_{C4} + 200 \, \rm km \, s^{-1}$ and a number of galaxies
within 3$\sigma$ from the cluster redshift and within $R_{200}$ equal or 
greater
than 7, where $\sigma$ was measured from the Sloan spectroscopic tables 
as described in Appendix C 
and ${\sigma}_{C4}$ is the velocity dispersion given by Miller 
et al. (2005).
Clusters with clear multiple peaks in the redshift histograms indicating
that the $\sigma$ is unreliable were excluded. 

Since all but one of the EDisCS structures have at least 8
members within these magnitude and radial limits ($N_{\oii}$ in
Table~1), we included only C4 clusters with at least 8 members usable
to compute the \oii fraction.  Our final C4-based sample consists of
88 clusters, whose centers were taken to coincide with the BCG listed
by Miller et al. (2005).\footnote{The BCG was chosen by Miller et
al. to be the brightest entry in the SDSS catalog 
within $500 \, h^{-1} \, kpc$
from the peak of the C4 density field, with 
EW($\rm H\alpha)>-4$ \AA $\,$ and within 4$\sigma$ from the cluster redshift, 
or with no spectroscopy but brighter than $m_r=19.6$ and $M_r=-19.8$
with colors lying within the cluster color-magnitude sequence and
no more than two magnitudes dimmer than the BCG identified based
on the 4$\sigma$ criterion.
}
The \oii fraction of these clusters was computed from a $g$-selected sample
of galaxies within $R_{200}$ and with $M_V < -19.8$, as it was done
for the Abell sample. For the C4-based sample we did not apply
completeness weights, given that these corrections did not affect
significantly the \oii fractions of the Abell clusters (see
Table~3).

 \begin{figure*}[t]
 \vspace{-3cm}
 \centerline{\hspace{2cm}\includegraphics[width=1.5\columnwidth]{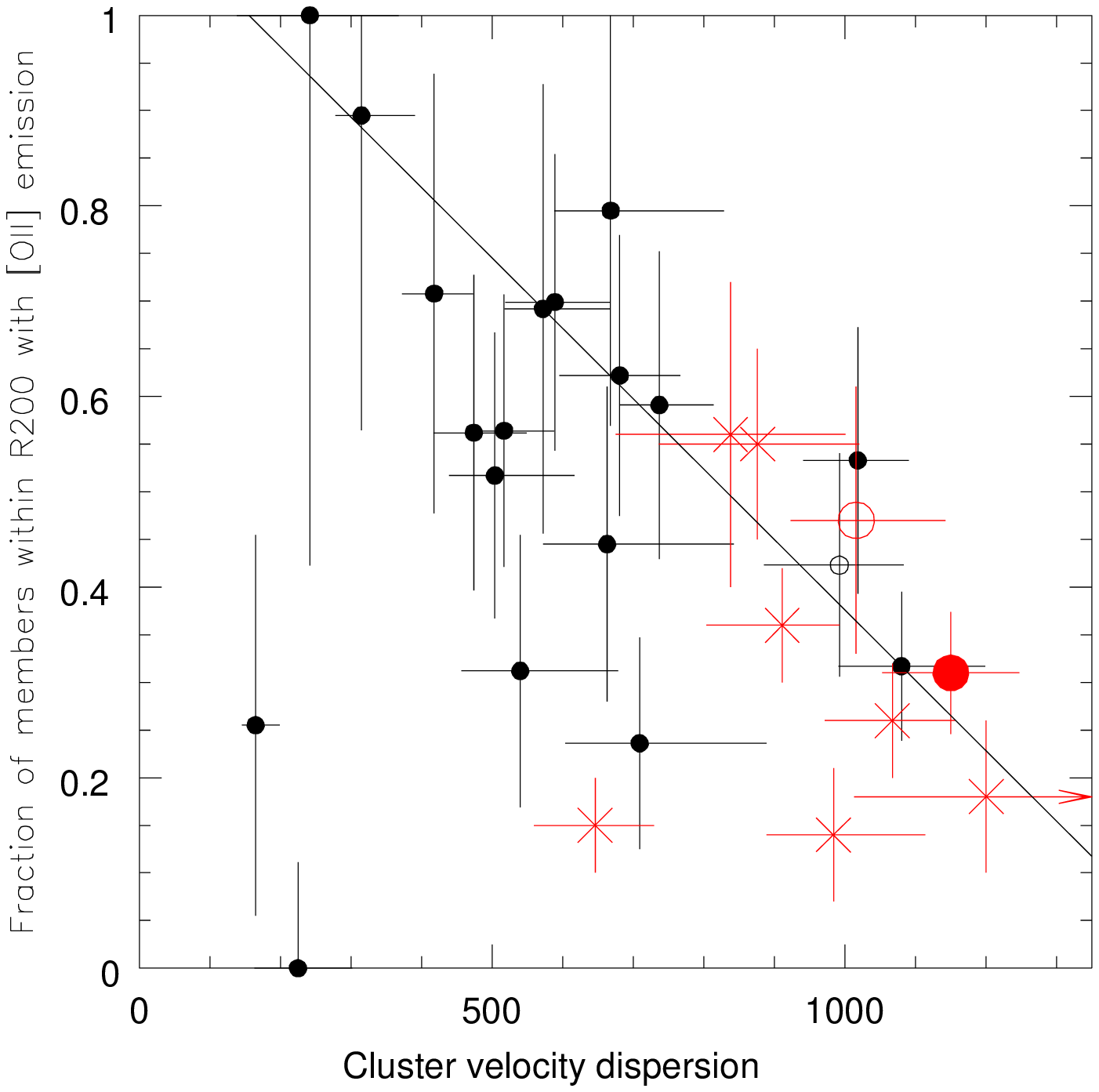}\hfill\hspace{-3cm}\includegraphics[width=1.5\columnwidth]{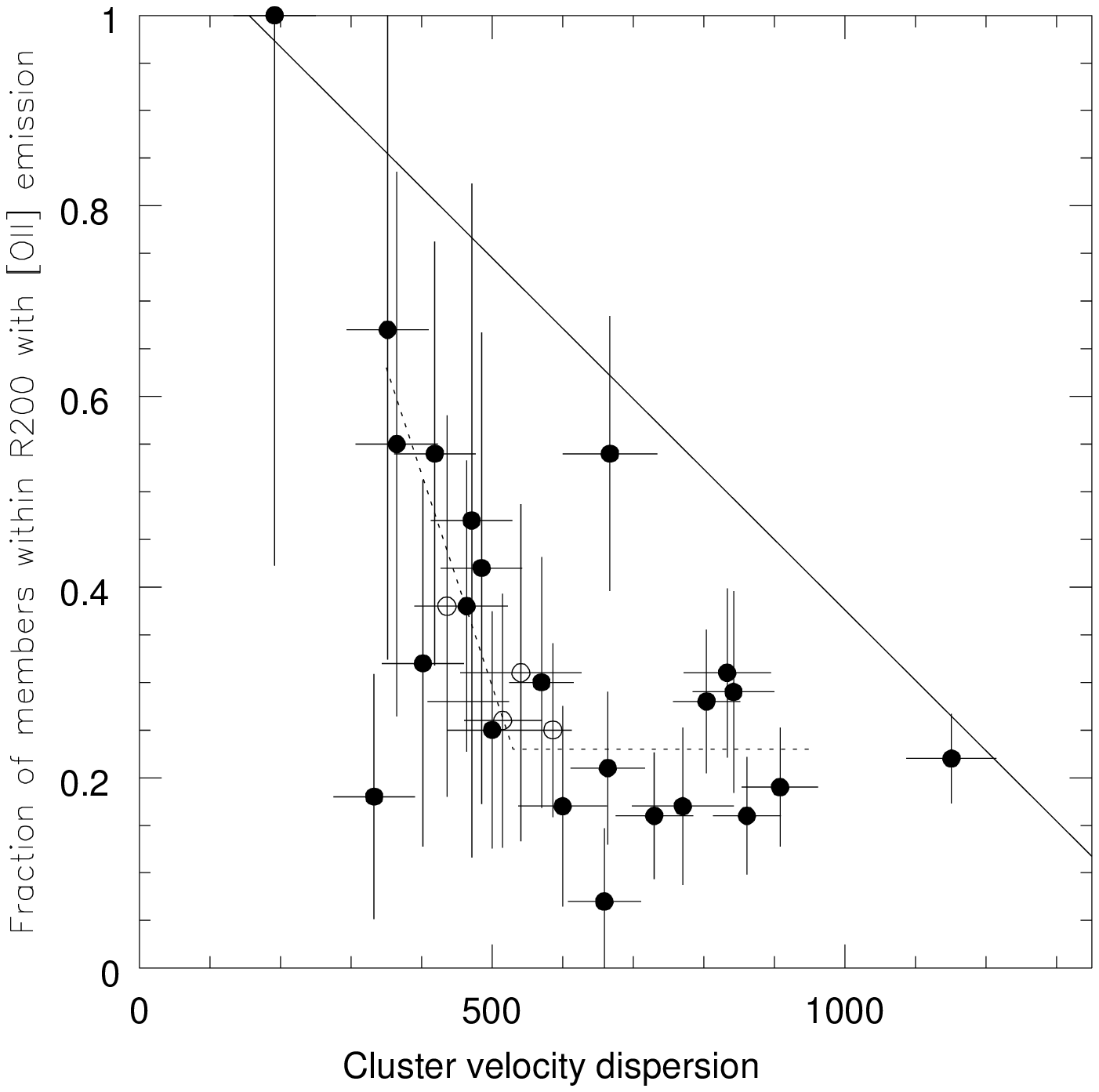}}
 \caption{Left. The \oii - $\sigma$ relation in clusters at $0.4 \le z \le 0.8$.
 The fraction of cluster members within $R_{200}$ with \oii emission
 is plotted versus the cluster velocity dispersion. Solid black
 dots represent EDisCS datapoints in
 each one of the 18 fields, as in Table~1. The red crosses are for MORPHS clusters;
 the red solid large circle is the cluster MS1054-03 from van Dokkum
 et al. and the red empty circle is Cl1324+3011 from Postman et al.
 Errorbars on the fractions
 are Poissonian.
 The solid line across the plot is taken as a description of 
 the upper envelope of the high-z points (see text).
 The same line is repeated in the right panel, to
 illustrate the differences. At high-z, the majority of systems group around 
 this line, while at low-z (right panel)
 the great majority fall well below the line. 
 Right. The \oii - $\sigma$ relation in low redshift clusters from Sloan
 described in \S4. Empty circles indicate those Sloan clusters slightly outside
 of the preferred redshift range. The two dotted lines
 represent the eye-fitted most heavily populated region in the plot.
 \label{main}}
 \end{figure*}

\section{Results}

\subsection{Star formation activity at high redshifts
as a function of the cluster mass}

Our main result is shown in Fig.~\ref{main}. The left panel presents
the fraction of \oii emitters as a function of cluster velocity
dispersion for EDisCS clusters (filled small circles), and for the
other clusters at z=0.4 to 0.8 whose data were taken from the
literature as described in Sec.~3.1.

Most datapoints occupy a stripe in this diagram, indicating that
most clusters follow a broad anticorrelation between the fraction of
star-forming galaxies and the cluster velocity dispersion: generally, more
massive clusters have a lower fraction of star-forming galaxies.  When
including all clusters, a Kendall test shows that an
anticorrelation between cluster velocity dispersion and star-forming
fraction is present with a 97.1\% probability.  
There are evident
outliers that do not follow the [O{\sc ii}]-$\sigma$ trend defined by
the majority of clusters. The most evident outliers are Cl\,1119 and
Cl\,1420, two groups with $\sigma < 400 \rm \, km \, s^{-1}$ that
will be discussed further in \S5.5.  The Kendall probability becomes
99.9\% when excluding these two outliers. The probabilities excluding
non-EDisCS datapoints become 48.0\% and 96.2\% when the two outliers are 
included and excluded, respectively.

Assuming that the cluster velocity dispersion is related to the mass
of the system\footnote{We note that two EDisCs clusters, Cl\,1216 and
Cl\,1232, show evidence for substructure and therefore in principle
their velocity dispersions may be a poor indicator of their masses
(Halliday et al. 2004). However, a subsequent weak lensing analysis of
EDisCS clusters has showed that substructure does not strongly affect
the spectroscopic measurement of their velocity dispersion, confirming
within the errors the spectroscopic estimate of $\sigma$ ($\sigma =
1152_{-78}^{+70}$ for Cl\,1216, and $\sigma = 948_{-55}^{+50}$ for
Cl\,1232, Clowe et al. 2005).}, the [O{\sc ii}]-$\sigma$ relation
shown in the left panel of Fig.~\ref{main} suggests that it is the
mass of the system, though with a significant scatter, that largely
determines what proportion of its member galaxies are forming stars at
$z \sim 0.6$.  Given the scatter and the outliers in this plot,
however, it is uncertain whether this is better described as a broad
relation between the \oii fraction and $\sigma$, or as an upper
envelope.  In fact, the most notable feature of this diagram is the
absence of datapoints in the upper right corner, above the most
populated stripe.  This 
envelope in the \oii fraction versus $\sigma$ plane seems to imply
that at $z=0.4-0.8$ a system of a given mass can have {\it at most} a
certain fraction of star-forming galaxies or, equivalently, must have
{\it at least} a given fraction of galaxies that are already passive
at this epoch. More massive systems have a lower maximum-allowed
fraction of star-forming galaxies or, equivalently, a higher
minimum-allowed fraction of passive galaxies.

The solid line in Fig.~4 is a "hand-drawn" description of this upper envelope\footnote{This line is taken to be equal to
a fit to the EDisCS points, after excluding
the two group outliers, using an M-estimate that minimizes absolute
deviations (Press et al., 1986).}:

\begin{equation}
f_{\oii} = -0.74 \times  (\frac{\sigma}{1000}) (\rm km \, s^{-1}) 
\, + 1.115
\end{equation}

 \begin{figure}
 \vspace{-3cm}
 \centerline{\hspace{2cm}\includegraphics[width=13cm]{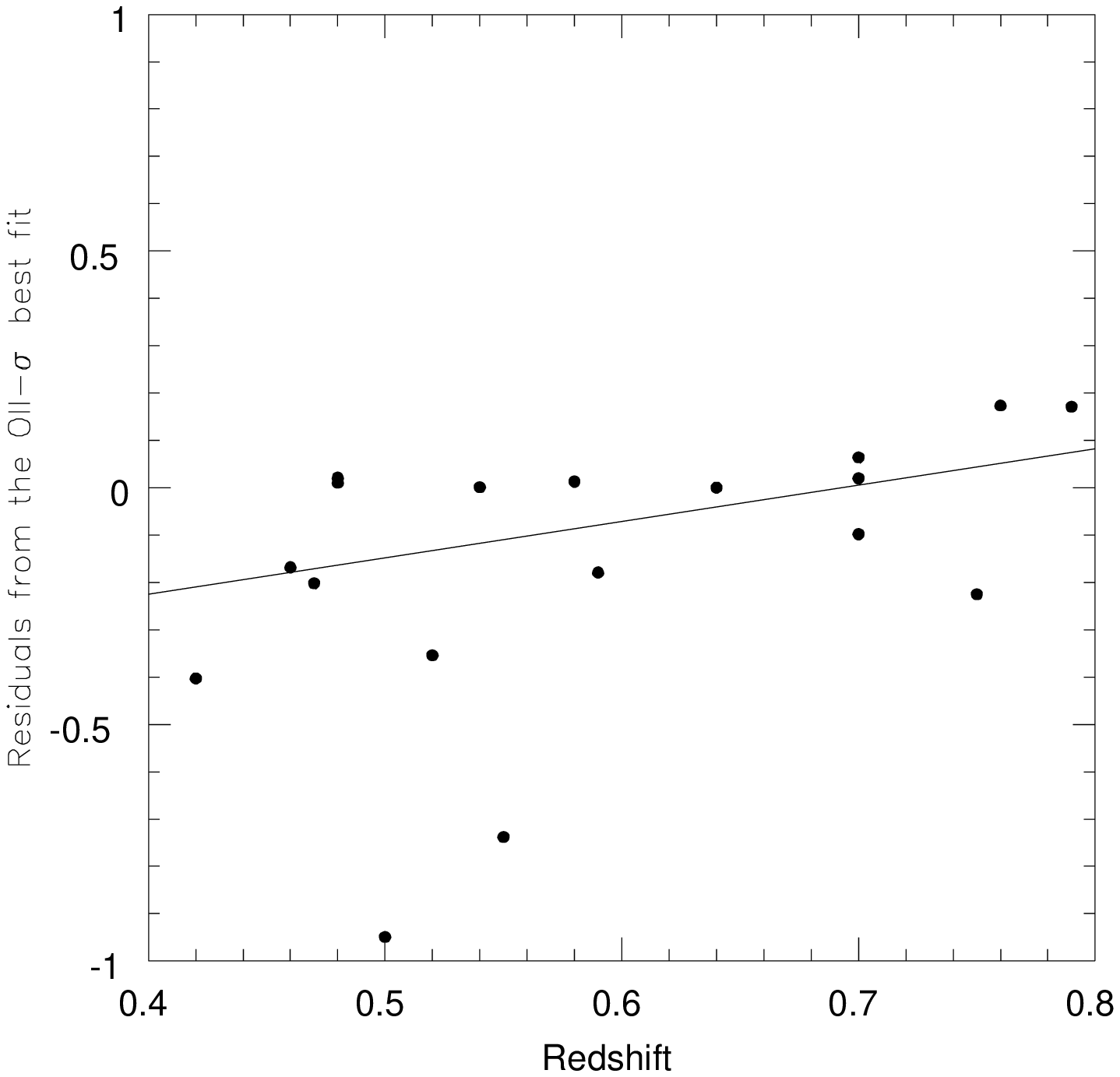}}
 \caption{Residuals of the \oii fraction of EDisCS
 clusters with respect to the line drawn in the \oii - $\sigma$
 relation at high-z shown in Fig.~\ref{main}, plotted versus
 cluster redshift. The solid line is the least squares fit having excluded 
 the two outliers (Cl\,1119 and Cl\,1420), which are the two points with 
 residuals $< -0.5$.
 \label{residuals}}
 \end{figure}

%
%
When plotting the residuals from this relation for EDisCS clusters as
a function of cluster redshift (Fig.~\ref{residuals}), there is a
trend of increasing \oii fraction towards higher redshifts in the
interval $z=0.4-0.8$.
Since within our sample there is no correlation between cluster
redshift and velocity dispersion (see Table~1), 
Fig.~\ref{main} and Fig.~\ref{residuals} 
indicate that both a trend of the \oii fraction with
$\sigma$ and one with redshift are present.  In the next subsection we
investigate the evolution with redshift in more detail, comparing
these results with our Sloan cluster sample.

\subsection{Evolution of the {\oii} - $\sigma$ relation in clusters}

The [O{\sc ii}]-$\sigma$ trend observed in distant clusters can be
compared with nearby clusters to quantify the evolution with redshift
of the star-forming fraction in clusters as a function of the system
velocity dispersion. 
The right panel of Fig.~\ref{main} shows that Sloan Abell clusters at
$z \sim 0.04-0.08$
have significantly lower fractions of star-forming galaxies than 
clusters at $z \sim 0.4-0.8$.
The solid line in this diagram is the same as the solid line in the left 
panel and corresponds to the line following
the high-z datapoints.  
While at high-z most clusters 
fall around this line (left panel), at low-z the great majority of clusters
fall well below this line (right panel).

At low-z, an approximate description of the datapoints
in the \oii - $\sigma$ diagram is given by the two
dotted lines in Fig.~\ref{main} (right panel), 
following:

\begin{equation}
\begin{array}{ll}
f_{\oii}= -0.0022 \, \sigma + 1.408 & \rm for \, \sigma < 530 \\
f_{\oii}=0.23 & \rm for \, \sigma > 530 \\
\end{array}
\end{equation}

In fact, the average \oii - $\sigma$ relation is flat for $\sigma >550
\rm km \, s^{-1}$; no clear trend seems to be present above this
velocity dispersion. In contrast, a trend is visible at $\sigma < 500
\rm km \, s^{-1}$, with the \oii fraction rising for most systems
towards lower velocity dispersions.  The average $f_{\oii}$ in 3 bins
of velocity dispersion for $\sigma < 550 \rm \, km \, s^{-1}$
increases from 0.31, to 0.41, to 0.54 going to lower $\sigma$.
A Kendall test shows the trend below $600 \rm \, km \, s^{-1}$
to be significant at the 98.5\% level.

\begin{figure}
 \centerline{\includegraphics[width=10cm]{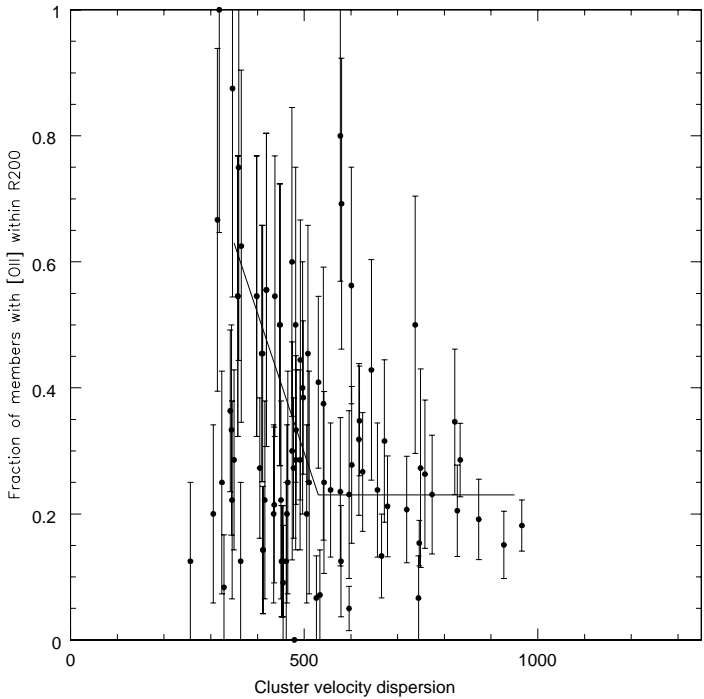}}
 \caption{The \oii - $\sigma$ relation in Sloan clusters at $0.04 < z < 0.08$
 taken from the C4 catalog (Miller
 et al. 2005). Errorbars on the $\sigma$ have been omitted for clarity. The two lines
 indicate the trend followed by the Abell sample, as the dotted lines in the right panel
 of Fig.~\ref{main}.
 \label{c4}}
 \end{figure}

The rather small low--z sample we use here, being 
quality-controlled and comparable to EDisCS, show significant
differences with respect to the high-z clusters.  It is interesting to compare
the results obtained for this sample with those obtained for the C4
control sample.  The results for C4 clusters are shown in
Fig.~\ref{c4}. These clusters confirm the main trends observed in the
Abell sample: for $\sigma > 550 \rm \, km \, s^{-1}$, the \oii
fraction does not seem to depend on $\sigma$ and is smaller than 0.3
for the great majority of clusters. At $\sigma \le 500 \rm \, km \,
s^{-1}$, many systems have an $f_{\oii}$ higher than 0.3, though there
are also systems with low $\sigma$ and low $f_{\oii}$.  While from the
Abell sample there appears to be a trend of increasing average \oii
fraction towards lower velocity dispersions, for the C4 clusters it is
unclear whether there is a trend at $\sigma < 500 \rm \, km \,
s^{-1}$, or simply a large scatter in the \oii fraction at low
$\sigma$. We note that the trend of rising average $f_{\oii}$
observed in the Abell sample is in excellent agreement with the
results of Martinez et al. (2002) who found the average fraction of
emission-line galaxies to decrease monotonically with virial mass for
groups ($10^{12} - 2 \times 10^{14} \, M_{\odot}$) from the 2dF Galaxy
Redshift Survey. The result from Martinez et al. (2002) parallels the
trend of increasing early-type fractions towards higher velocity
dispersion in the poor groups studied by Zabludoff \& Mulchaey (1998),
who noted that the early-type fractions in the most massive groups of
their sample were comparable to those found in rich clusters.
A rising fraction of ``late-type galaxies'' (defined on the basis of
their color and star formation from emission lines) towards lower velocity 
dispersions is also found below $500 \, \rm km \, s^{-1}$ in Sloan groups
by Weinmann et al. (2006). 
Our low-z trends also agree with the fraction of passive galaxies in 2dF
groups from Wilman et al. (2005b, see their Fig.~7).\footnote{For
clusters, Biviano et al. (1997) also found a trend with $\sigma$
in the fraction of emission-line galaxies in the ESO Nearby Abell
Cluster Survey between 400 and 1100 $\rm \, km \, s^{-1}$. A direct
comparison with our results cannot be carried out, due to the very
different threshold of line strength adopted for identifying
emission-line galaxies, but they found that on average the mean
emission-line fraction in three bins of $\sigma$ decreased towards higher
$\sigma$. If we analyzed the \oii fraction for our Sloan clusters
in three similarly wide bins of $\sigma$, instead of presenting
the results cluster by cluster, we would obtain a similar trend.}
Keeping in mind the caveat of a larger scatter in the C4 sample
at low $\sigma$, in the following we will adopt as reference the rising
trend observed in the Abell sample. As we will see, this does not 
influence our main conclusions.

In fact, the most relevant and striking aspect of the Abell and C4 comparison
is that both samples show a break in the behaviour of the \oii fraction with
$\sigma$ at the same velocity dispersion ($\sim 500 \rm \, km \,
s^{-1}$) and that most clusters above this $\sigma$ have a fraction
of star-forming galaxies around 20\%.  
As we will discuss later, this critical threshold in
$\sigma$ is an important observational landmark for
inferring the effects of the environment on galaxy star
formation histories.
\footnote{We note that our 
low-z samples are composed for the great majority of
structures with $\sigma$ between 350 and 900 $\rm km \, s^{-1}$,
therefore they do not allow an exploration of the properties of the
most massive and the least massive systems.}

 \begin{figure}
 \vspace{-3cm}
 \centerline{\hspace{2cm}\includegraphics[width=13cm]{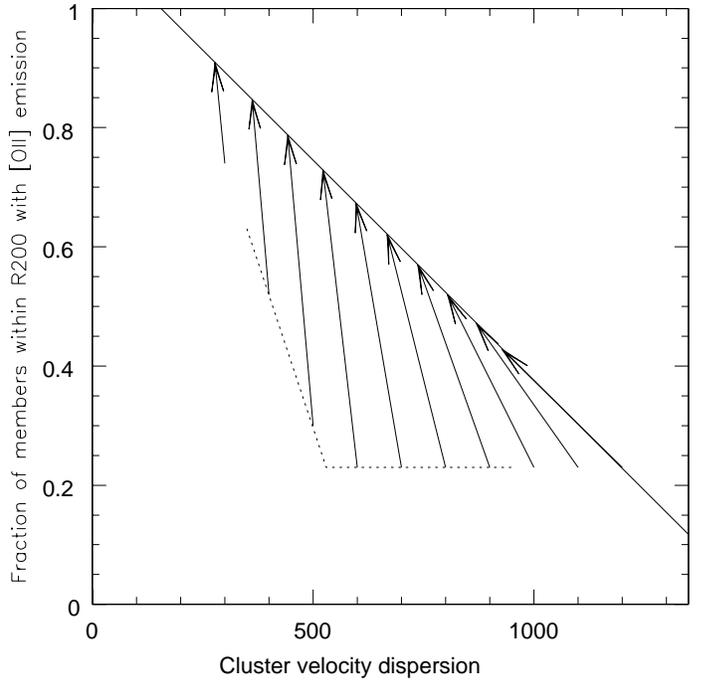}}
 \caption{Evolution of the \oii - $\sigma$ relation inferred from 
observations at different redshifts.
 The solid line and the two dotted lines represent the observed 
 relations followed
 by the majority of clusters at high- and low-z, respectively. These are the 
 same lines shown in the left and right panels
 of Fig.~\ref{main}. The arrows identify the average progenitors at $z=0.6$
 of clusters of different mass at $z=0$ as expected from simulations, 
 considering the fact that the cluster mass, hence its 
 velocity dispersion, evolves too (see text for details). 
 If $\sigma$ did not evolve with time, 
 the arrows would be vertical in this plot.\label{model}}
 \end{figure}

A schematic view of the evolution in the \oii -- $\sigma$
diagram is presented in
Fig.~\ref{model}.  The solid line and the two dotted lines illustrate
the position of the most densely populated regions at high and low
redshift, respectively, reproducing the average observed trends
followed by most clusters in the two panels of Fig.~\ref{main}.

The arrows in Fig.~\ref{model} indicate the average velocity dispersion
of $z=0.6$ progenitors of systems of different masses at $z=0$.
This was computed from the assembly history given by
high--resolution N-body simulations by Wechsler et al. (2002),
adopting concentration parameters as in Bullock et al. (2001) and
computing the relation between mass of the system and $\sigma$ as in
Finn et al. (2005):

\begin{equation}
M^{sys} = 1.2 \times 10^{15} \, {(\frac{\sigma}{1000 \, \rm km \, s^{-1}})}^3 \,
{\frac{1}{\sqrt{{\Omega}_{\Lambda} + {\Omega}_{0}(1+z)^3}}} \, h^{-1} \, \rm M_{\odot}
\end{equation}

\begin{figure}
 \centerline{\includegraphics[width=9cm]{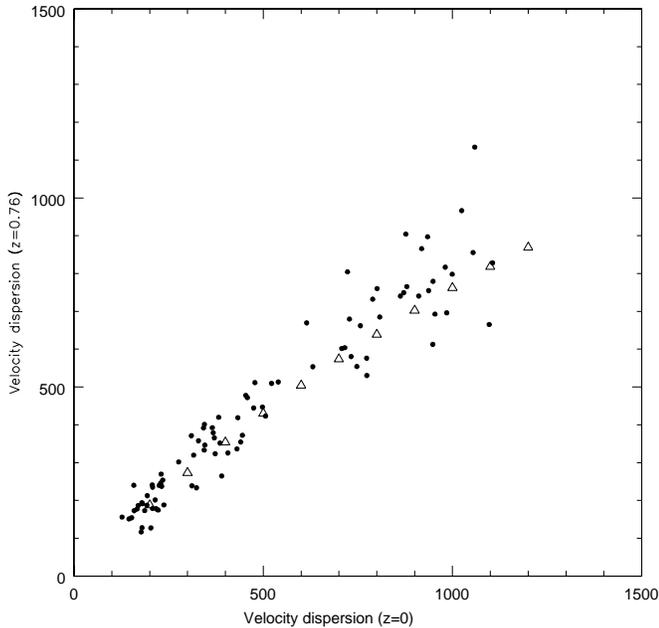}}
 \caption{Relation between the velocity dispersion of systems at $z=0$
 and at z=$0.76$. Empty triangles are average values derived analytically from 
 Wechsler et al. 
 (2002) and eqn.4, as described in \S5.2.
 Filled circles are results for 90 haloes from the Millennium Run
 simulation (Springel et al. 2005), as described in \S5.2. 
 \label{sigsig}}
 \end{figure}

In Fig.~\ref{sigsig}, the mean 
change of $\sigma$ between $z=0$ and $z=0.76$ derived in
this way (empty symbols)
is compared with the evolution obtained from the Millennium
Simulation (hereafter MS; Springel et al. 2005).
Filled symbols represent results for 90 haloes at $z=0$
extracted from the MS and followed back in time by tracking at each
previous redshift their most massive progenitor. The MS
follows $N= 2160^3$ particles of mass $8.6\times10^{8}\,h^{-1}{\rm
M}_{\odot}$ within a comoving box of size $500\, h^{-1}$Mpc on a side
and with a spatial resolution of $5\, h^{-1}$kpc.  We extracted 90
haloes within the simulation box, uniformly distributed in log(mass)
between $5\times10^{12}\,{\rm M}_{\odot}$ and $5\times10^{15}\,{\rm
M}_{\odot}$.  Dark matter haloes were populated using the
semi-analytic model presented in De Lucia et al. (2005) (see also
Croton et al. 2005).  For each halo, we considered all galaxies
within 2$R_{200}$ from the central galaxy to compute the projected
velocity dispersion along the x, y, and z axis. In Fig.~\ref{sigsig}
we have plotted the mean of these projected velocity dispersions.

Fig.~\ref{sigsig} shows a very good agreement between the evolution
of the velocity dispersions estimated from the two independent simulations.
In the following, we use the average evolution of $\sigma$ derived
from Wechsler et al. (2002) and eqn.(4) to establish
the evolutionary link between low-z and high-z structures.\footnote{We 
note that this is done selecting haloes at $z=0$ and computing the
average projected velocity dispersion 
of their ``main progenitor'' at $z=0.6$. 
We have verified that selecting haloes at
$z=0.6$ and following their descendants to $z=0$ gives similar results,
albeit the amount of evolution in $\sigma$ turns out to be slightly smaller
in this case. The differences found with the two selection methods
increase with the halo velocity dispersion ranging from less than  
$10 \rm \, km \, s^{-1}$ for systems up to $ 500 \rm 
\, km \, s^{-1}$ to about  
$100 \rm \, km \, s^{-1}$ for systems with $1000 \rm \,  km \, s^{-1}$.}
In Table~4, we list the average $f_{\oii}$ fractions observed
at high-- and low--redshift as a function of the cluster velocity dispersion
and mass at $z=0$, the corresponding average $\sigma$ and mass 
of that structure at $z=0.6$,
and the difference $\Delta f$ in $f_{\oii}$ between the two 
redshifts.\footnote{We stress that Table~4
extends to lower and higher masses than those probed by our
sample at low redshift. Therefore, for systems with $\sigma$ below 350 
and above 1000 $\rm km \, s^{-1}$ at $z=0$ the values listed are {\it
extrapolations} of the observed trends, not confirmed by any
observational evidence.  According to these extrapolations, no change
in \oii fraction with redshift would be observed for systems below
$200 \, \rm km \, s^{-1}$.  At these low masses also the evolution of
$\sigma$(mass) is negligible (Table~4).}  

\begin{table}
\begin{center}
\caption{Mean evolution between z=0 and z=0.6 
of the velocity dispersion $\sigma$, the mass of the system $M^{sys}$ 
and the \oii fraction.\label{tbl4}}
\begin{tabular}{ccccccc}
\tableline\tableline
$\sigma_{z=0}$ & $M^{sys}_{z=0}$ & ${f_{\oii}}_{z=0}$ & $\sigma_{z=0.6}$ & $M^{sys}_{z=0.6}$ & ${f_{\oii}}_{z=0.6}$ & $\Delta f$\\ 
 $ \rm km \, s^{-1}$  &   $M_{\odot}/h$  &  & $ \rm km \, s^{-1}$ &  $M_{\odot}/h$  & & \\
\tableline
1200 & $2.1 \times 10^{15}$ & 0.23 & 928 & $6.9 \times 10^{14}$ & 0.45 &  0.22 \\
1100 & $1.6 \times 10^{15}$ & 0.23 & 869 & $5.7 \times 10^{14}$ & 0.49 &  0.26 \\
1000 & $1.2 \times 10^{15}$ & 0.23 & 805 & $4.5 \times 10^{14}$ & 0.54 &  0.31 \\
900  & $8.8 \times 10^{14}$ & 0.23 & 737 & $3.5 \times 10^{14}$ & 0.58 &  0.35 \\
800  & $6.1 \times 10^{14}$ & 0.23 & 668 & $2.6 \times 10^{14}$ & 0.63 &  0.40 \\
700  & $4.1 \times 10^{14}$ & 0.23 & 597 & $1.8 \times 10^{14}$ & 0.68 &  0.45 \\
600  & $2.6 \times 10^{14}$ & 0.23 & 522 & $1.2 \times 10^{14}$ & 0.73 &  0.50 \\
500  & $1.5 \times 10^{14}$ & 0.30 & 443 & $7.5 \times 10^{13}$ & 0.79 &  0.49 \\
400  & $7.7 \times 10^{13}$ & 0.52 & 362 & $4.1 \times 10^{13}$ & 0.85 &  0.33 \\
300  & $3.2 \times 10^{13}$ & 0.74 & 278 & $1.9 \times 10^{13}$ & 0.91 &  0.17 \\
200  & $9.6 \times 10^{12}$ & 0.96 & 190 & $5.9 \times 10^{12}$ & 0.97 &  0.01 \\
100  & $1.2 \times 10^{12}$ & 1.00 & 98  & $0.8 \times 10^{12}$ & 1.00 &  0.00 \\
\tableline
\end{tabular}


\tablecomments{Col. (1) Velocity dispersion of a system at z=0. 
Col. (2) Corresponding mass of the system at z=0 
in units of solar masses/$h$. The relation between ${\sigma}_{z=0}$
and the mass is computed according to eqn.(4). Col. (3) Average \oii fraction 
at z=0 for a system of ${\sigma}_{z=0}$, derived from eqn.(3).
Col. (4) Mean velocity dispersion at z=0.6 of a system with ${\sigma}_{z=0}$
as in col.1. ${\sigma}_{z=0.6}$ is computed from Col.5 according to eqn.(4).  
Col. (5) Mean mass at z=0.6 of a system with ${M}_{z=0}$
as in col.2. The relation between mean mass at z=0 and mean mass at z=0.6
is computed according to Wechsler et al. (2002) (see \S5.2). Col. (6) Average \oii fraction
at z=0.6 for a system of ${\sigma}_{z=0.6}$ as in col.4, derived from eqn.(2).
Col. (7) Difference between the \oii fractions at z=0.6 and z=0 (col.6 - col.3).}
\end{center}
\end{table}

Fig.~\ref{model} and Table~4 show that
the strongest evolution in mass and in velocity dispersion is expected for the
most massive structures. The same figure and table also illustrate that
the observed change
in star--forming fraction between $z=0.6$ and $z=0$ is {\it maximum instead
for intermediate-mass structures}, those  with $\sigma \sim
5-600 \, \rm km \, s^{-1}$ at $z=0$ and $\sim 450-500 \, \rm km \, s^{-1}$ 
at $z=0.6$. The evolution 
is smaller at both higher and lower masses, but 
is still significant even for the most massive systems in the Sloan 
sample.\footnote{It is worth noting that the star-forming fractions in 
clusters with $\sigma > 1000 \rm \, 
km \, s^{-1}$ on average are quite similar at $z=0.6$ and $z=0$ (compare
the left panel in Fig.~\ref{main} with the right panel in the same figure
and with Fig.~\ref{c4}). 
However, as shown in
Fig.~\ref{model} and Table~4, clusters with $\sigma > 1000 \rm \, km \, s^{-1}$
at $z=0$ are those whose mass evolved the most between the two
redshifts. Their progenitors at $z=0.6$ were  
$\sim 800 \rm \, km \, s^{-1}$ systems, whose
star-forming fraction was higher than that of their descendants at $z=0$.
Thus, also the most massive systems at $z=0$ on average have experienced
a significant evolution of their fraction of star-forming galaxies.}
In the mass range we 
observe, the change in the average star--forming fraction $\Delta f$
ranges between 20-30\% and 50\%.

It is essential to keep in mind that the trends in Fig.~\ref{model}
and Table~4 depict an average evolution apparently followed by
the majority of systems, but a large scatter is present in the
observed \oii - $\sigma$ relation at all redshifts, and the
evolution of the cluster masses is expected to proceed with a significant
scatter too, as shown in Fig.~\ref{sigsig}. 

In this section we have shown that the fraction
of star--forming (and, conversely, passive) galaxies in clusters
has evolved significantly between $z \sim 0.6$ and $z=0$. For the first time,
we can quantify how the average evolution varies with
the mass of the system. 
Why the proportion of passive/star--forming galaxies broadly 
correlates/anticorrelates
with the velocity dispersion of the system for most clusters
at high-z, why this proportion evolves with redshift, why
the evolution is maximum for intermediate-mass systems and
why there is no clear trend with $\sigma$
at z=0 for systems more massive than
$\sim 550 \rm \, km \, s^{-1}$ are questions that will be
addressed in \S6.

\subsection{Other environments at high-z}

About 35\% of the galaxies in the 
EDisCS spectroscopic catalog reside in the 18 structures 
that have been discussed so far. The other 
spectra can be used to investigate other environments at high redshift,
such as groups and the ``field''.

We have identified other structures in our spectroscopic sample as
associations in redshift space. 
Given our selection criteria for spectroscopic targets (\S2 and Halliday et al. 2004),
we can treat the spectroscopic catalog in a redshift slice close
to the redshift targeted in each field as a purely I-band selected sample
(Milvang-Jensen et al. 2006). To ensure that no selection bias can be present,
we choose conservative redshift limits and
we only consider galaxies  within $\pm 0.1$ in $z$ from the
cluster targeted in each field. 
With these criteria we can study two other clusters ($\sigma > 400 \rm \, km \, s^{-1}$), listed
in Table~5, and several groups ($\sigma < 400 \rm \, km \, s^{-1}$).

Two types of groups have been isolated. 
Six structures have at least 7 member galaxies with spectroscopic
redshifts. For these, velocity dispersions have been computed. In the following
we will refer to these as ``groups''. 
These are listed in Table~5 and are among the groups
studied in detail in Halliday et al. (2006).  

\begin{table}
\begin{center}
\caption{Other EDisCS structures.\label{tbl5}}
\begin{tabular}{lcccccc}
\tableline\tableline
&&&&&& \\
Cluster & z & $\sigma$ $\pm{\delta}_{\sigma}$ & $N_{mem}$ & $N_{\oii}$ & $f_{\oii}$ & $Dist/{\sigma}$ \\ 
\tableline
C2\_1138   & 0.4549 & 529$_{-54}^{+109}$ & 11 & 8  &  0.38$\pm$0.22 & 7  \\
C2\_1227   & 0.5822 & 432$_{-81}^{+225}$ & 11 & 8  &  0.88$\pm$0.33 & 17 \\
G1\_1301   & 0.3971 & 393$_{-46}^{+84}$  & 17 & 13 &  0.38$\pm$0.17 & 25 \\
G1\_1103\tablenotemark{a}   & 0.6258 & 329$_{-28}^{+50}$  & 14 & 10 &  0.50$\pm$0.22 & 43 \\
G1\_1040   & 0.7798 & 259$_{-52}^{+91}$  &  8 & 6  &  0.83$\pm$0.37 & 32 \\
G1\_105411 & 0.6130 & 227$_{-28}^{+72}$  &  8 & 7  &  1.00$\pm$0.38 & 25 \\
G1\_105412 & 0.7305 & 182$_{-69}^{+58}$  & 10 & 10 &  0.50$\pm$0.22 & 7  \\
G2\_1040   & 0.6316 & 179$_{-26}^{+40}$  & 11 & 8  &  1.00$\pm$0.35 & 31 \\
\tableline
\end{tabular}
\tablenotetext{a}{named Cl\,1103.7-1245a in White et al. (2005).}
\tablecomments{Col. (1) Name of the structure. 
Col. (2) Redshift. Col. (3) Velocity dispersion.
Col. (4) Number of spectroscopically confirmed members.
Col. (5) Number of galaxies
members of the cluster used for the calculation of the \oii fraction,
hence brighter than the adopted magnitude limit.
Col. (6) \oii fraction. Col. (7) Distance from another structure in units
of $\sigma$ of the most massive of the two.
}
\end{center}
\end{table}

\begin{figure}
 \vspace{-1cm}
 \centerline{\includegraphics[width=7cm]{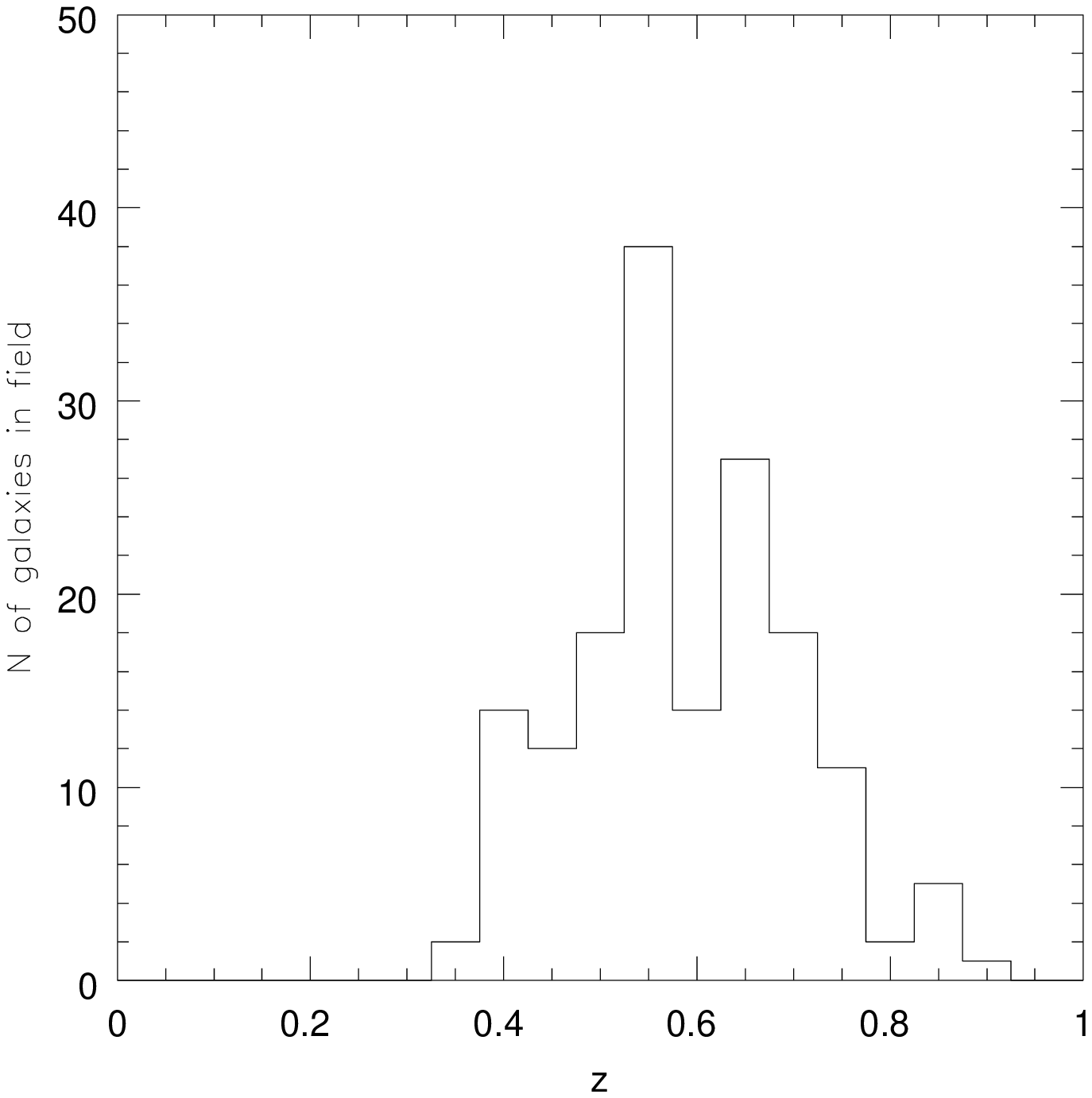}\hfill\hspace{-2cm}\includegraphics[width=7cm]{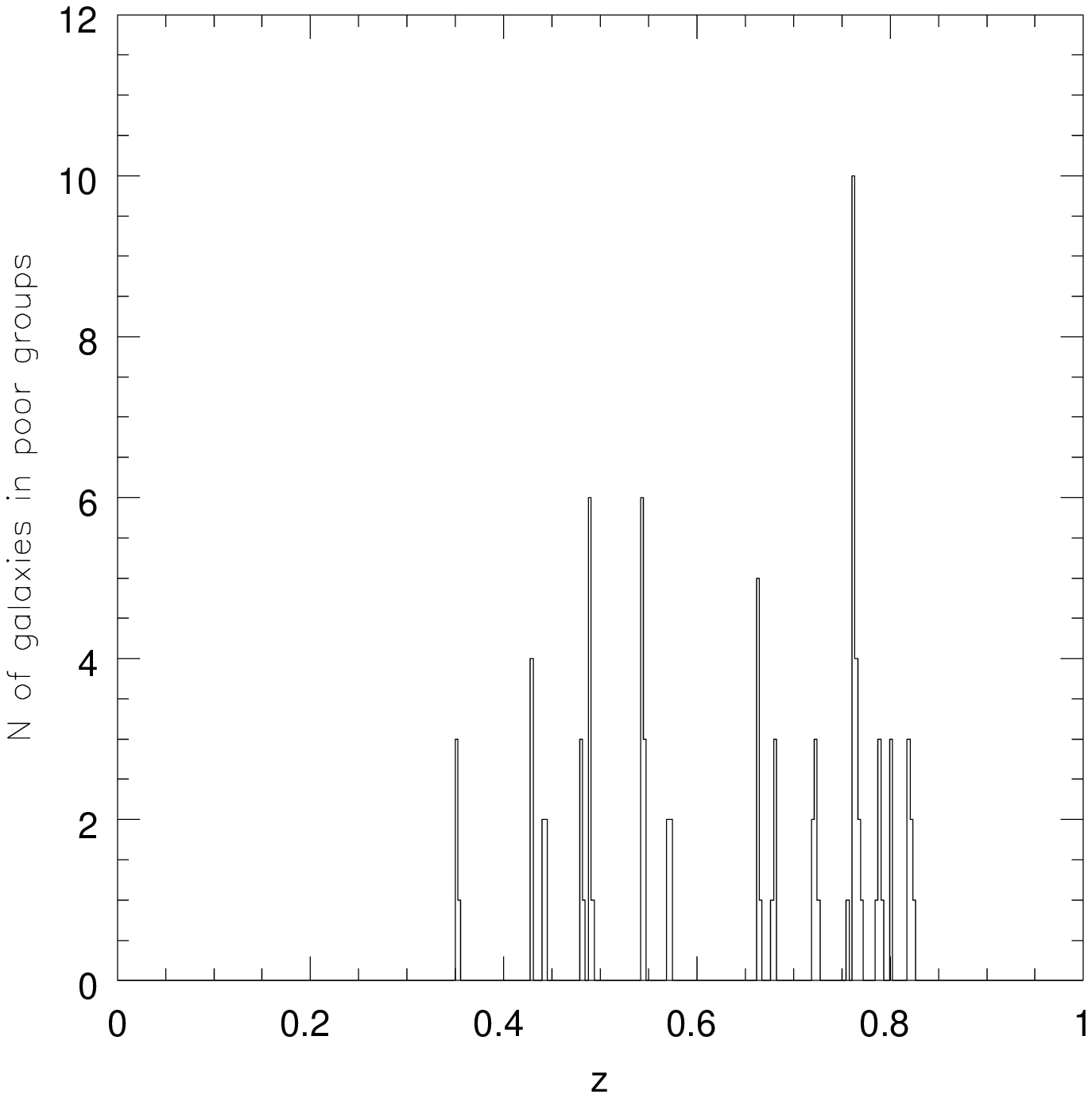}}
 \caption{Left. Redshift distribution of our field spectroscopic sample.
 Right. Redshift distribution of galaxies in our poor groups.\label{zfiegro}
 }
 \end{figure}

Associations in redshift space with between 3 and 6 galaxies in our 
spectroscopic catalog
have been treated separately, and hereafter will be referred to as
``poor groups''.  In total our poor groups comprise 84 galaxies,
whose redshift distribution
is shown in Fig.~\ref{zfiegro}.  We did not attempt to derive velocity
dispersions for these systems given the small number of
redshifts. 

Any other galaxy in our spectroscopic catalog
that was not a member of any of our clusters, groups or poor
group associations, will be hereafter named a ``field'' galaxy. 
Our field sample is
composed of 162 galaxies, whose redshift distribution is shown in
Fig.~\ref{zfiegro}. 
Our ``field'' sample should be dominated by galaxies in regions less populated 
than the clusters/groups/poor groups we isolated, but will 
also contain galaxies belonging
to structures that were not detected in our spectroscopic catalog.

The \oii fractions in these additional structures and in the field
have been computed for galaxies with absolute magnitudes brighter than
the limit adopted for the main 18 structures described in \S3. No radial
distance criterion has been introduced for the additional structures
(those presented in Table~5 and the poor groups) and the field. The lack of 
a radial criterion, however, does not significantly affect
the \oii fractions derived for these systems.

 \begin{figure}
 \vspace{-2cm}
 \centerline{\hspace{2cm}\includegraphics[width=12cm]{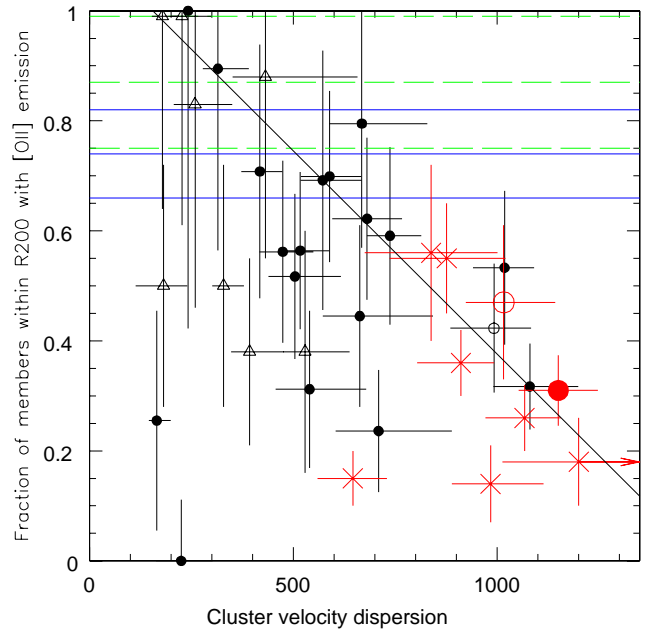}}
 \vspace{-1cm}
 \caption{Same as Fig.~\ref{main}, but now including also other
 environments. Empty triangles are other EDisCS structures with
 at least 7 spectroscopically confirmed members as in Table~5. The average \oii
 fraction and corresponding errorbars in poor groups (those between 3
 and 6 spectroscopic members) are shown with green dashed 
 lines. The \oii fraction among field galaxies (see text) and its
 errorbars are shown as blue solid lines.\label{ediscsall}}
 \end{figure}

The fraction of \oii emitters we find in the ``field'' is
$f_{\oii} = 0.74$ and is shown in
Fig.~\ref{ediscsall} as a solid blue horizontal line with its errorbars.  
The \oii fraction in
poor groups is even higher (87\%, green dashed horizontal line in
Fig.~\ref{ediscsall}), though still compatible with the fraction in
the field within the errors.

The other structures for which we derived a velocity dispersion are
shown in Fig.~\ref{ediscsall} as empty triangles.  Among these,
there are 4 systems with very high \oii fractions ranging between 83 and
100\%. In redshift space, these systems are all far from any known
cluster, always at least 17 ${\sigma}_{clu}$ away (see Table~5).
The other 4 additional
structures in Fig.~\ref{ediscsall} have relatively low \oii
fractions, around 40-50\%.  Two of these are quite close to
other massive structures (7 ${\sigma}_{clu}$ from the most massive
cluster in that field, see Table~5).

To summarize, both the field and the poor groups contain a high
proportion of star-forming galaxies (70 to 100\%), 
comparable to that observed in
more than half of the systems with $\sigma < 400 \rm \, km \, s^{-1}$,
and in agreement with the line that traces the relation between
$f_{\oii}$ and $\sigma$ up to the most massive systems. 
In addition, there are 5 other
groups with $\sigma < 400 \rm \, km \, s^{-1}$ (including Cl\,1119 and
Cl\,1420, previously discussed) that have \oii fractions significantly
lower than the rest of the groups and the field.  Their \oii fractions
are $ \le 50$\%. We will come back to discussing 
these systems and describe their
properties and those of their galaxy populations in \S5.5.

\subsection{EW([O{\sc ii}]) distributions}

As shown above, the fraction of star-forming galaxies in high-redshift clusters
depends on cluster velocity dispersion. It is interesting to investigate whether
also the star formation activity in star-forming galaxies depends on 
$\sigma$ and, more generally, on environment. 
In this section we analyze how the distributions of EW([O{\sc ii}]) 
for star--forming EDisCS galaxies vary in the different environments.
We do not attempt any comparison with the low-redshift sample, given
that a quantitative comparison of the EW strength is affected by
the uncertainties related to aperture effects between
$z=0.8$ and $z=0$.

We consider the ``environments'' defined in \S5.3, namely clusters, groups, poor
groups and field. Clusters have been further subdivided into more and
less massive clusters. We define as ``massive clusters'' 
all clusters in Table~1 with $\sigma > 800 \rm \, km \, s^{-1}$ 
(Cl\,1216, Cl\,1232), and
as ``less massive clusters'' those 
with $400 < \sigma < 800 \, \rm km \, s^{-1}$
(Cl\,1138, Cl\,1411, Cl\,1301, Cl\,1354, Cl\,1353, Cl\,1054-11,
 Cl\,1227, Cl\,1202, Cl\,1059, Cl\,1054-12, Cl\,1018, Cl\,1040).
Among the groups, we consider separately groups with high
and low \oii fractions as discussed in the previous section.  
``Groups with low-emission'' are those groups with
$\sigma < 400 \, \rm km \, s^{-1}$ and \oii fractions significantly below
the line in Fig.~\ref{main}
(Cl\,1420, Cl\,1119, G1\_105412, G1\_1301, G1\_1103).
``Groups with high-emission'' are those groups with 
$\sigma < 400 \, \rm km \, s^{-1}$ and \oii fractions that roughly follow
the line in Fig.~\ref{main} (Cl\,1037, Cl\,1103,
G1\_1040, G2\_1040, G1\_1054-11).
The same radial, magnitude and EW limits and completeness weights used for computing
the \oii fractions have been applied to the EW distributions.

 \begin{figure}
 \vspace{-2cm}
 \centerline{\includegraphics[width=12cm]{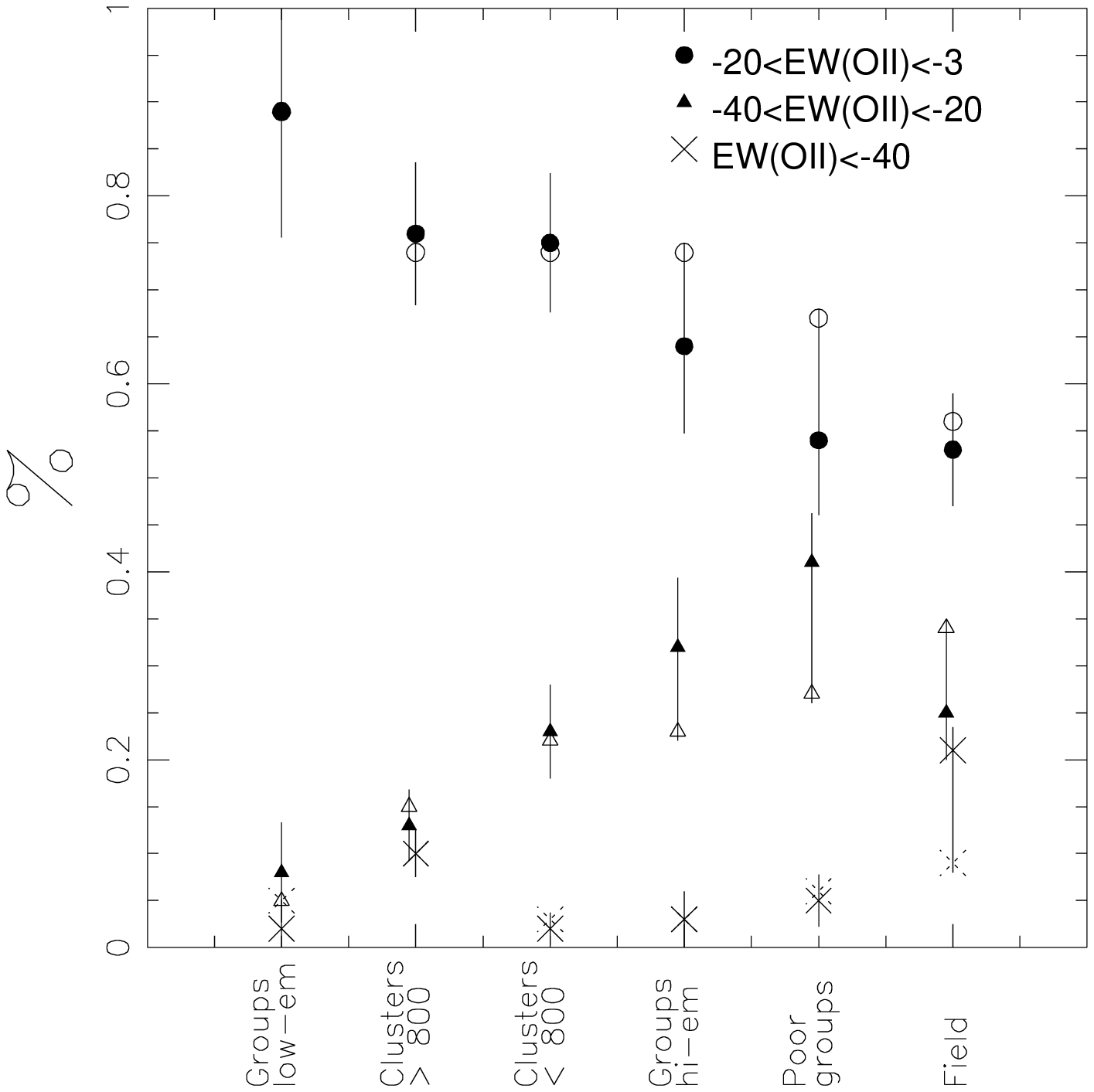}}
 \caption{Fraction of star--forming galaxies with a strong, moderate
 and relatively weak equivalent width of \oii in different environments. 
 Environments are located on the x-axis in order of increasing
 average \oii fraction as from Fig.~\ref{main}. The fraction
 of galaxies with relatively low EW ($-20<$EW$<-3$ \AA, circles) decreases 
 going to environments with higher \oii fraction.
 The fraction of galaxies with
 intermediate-strength EW (triangles) follows the opposite
 trend. The fraction of galaxies with strong EW (crosses, EW$<-40$ \AA) 
 is higher in the field than in any other environment, and a 
 possible excess
 is observed in massive clusters compared to the other environments.
 Filled symbols and solid line crosses refer to values corrected for
 incompleteness, while the uncorrected values are shown as empty symbols
 and dotted crosses.
 \label{ewfra}}
 \end{figure}

We have studied the proportion of star--forming galaxies with strong ($EW<-40$ \AA), intermediate
($-40<EW<-20$ \AA) and relatively modest equivalent width
of (\oii) ($-20<EW<-3$ \AA).\footnote{Note that 
the equivalent width measures the strength of the \oii line relative to the
underlying continuum. As such, it is not proportional to the SFR in 
solar masses per year, but to the current star formation 
rate per unit of galaxy luminosity at $\sim 3700$ \AA.} 
These proportions are shown in Fig.~\ref{ewfra} as a 
function of environment.
When ordering the environments in order of decreasing \oii fraction, 
there is a progressive trend also in equivalent width strength. 
The fraction of galaxies with relatively weak 
EW ($-20<$EW$<-3$ \AA, circles) decreases 
going to environments with higher \oii fraction.
The fraction of galaxies with
intermediate-strength EW (triangles) follows the opposite
trend. The fraction of galaxies with strong EW (crosses, EW$<-40$ \AA) 
appears higher in the field than in any other environment, and a
possible excess is observed in massive clusters compared to the other
environments.\footnote{The presence of some galaxies with very high
EWs in massive clusters is consistent with the detection of
some spirals with enhanced star formation in rich clusters
(see e.g. Milvang-Jensen et al. 2003, Bamford et al. 2005 and references
therein).} 
These trends are visible both in the completeness-weighted and unweighted
distributions (solid and empty symbols in Fig.~\ref{ewfra}), except
for the fraction of galaxies with strong EWs in the field
which is significantly reduced when the completeness correction is ignored.
The minimum errorbar for each point in Fig.~\ref{ewfra} is computed from
Poissonian statistics, while total errorbars have been increased
to take into account the difference between weighted and unweighted
values. Total errorbars are plotted in Fig.~\ref{ewfra}.

Thus, it is not only the proportion of galaxies with active star
formation that changes as a function of environment,
but also the star
formation properties in star-forming galaxies. The two things are closely
related to each other, following a parallel progressive trend: the
distribution of EW(\oii) is more skewed towards high values for
environments with higher \oii fractions. 
This seems to be at odds with the uniformity of the EW($\rm H\alpha$) 
distribution as a function of environment found by Balogh et al. (2004) 
for local samples, although
a study analogous to ours has not been carried out at low-z.

\begin{figure*}[t]
 \centerline{\includegraphics[width=0.8\columnwidth]{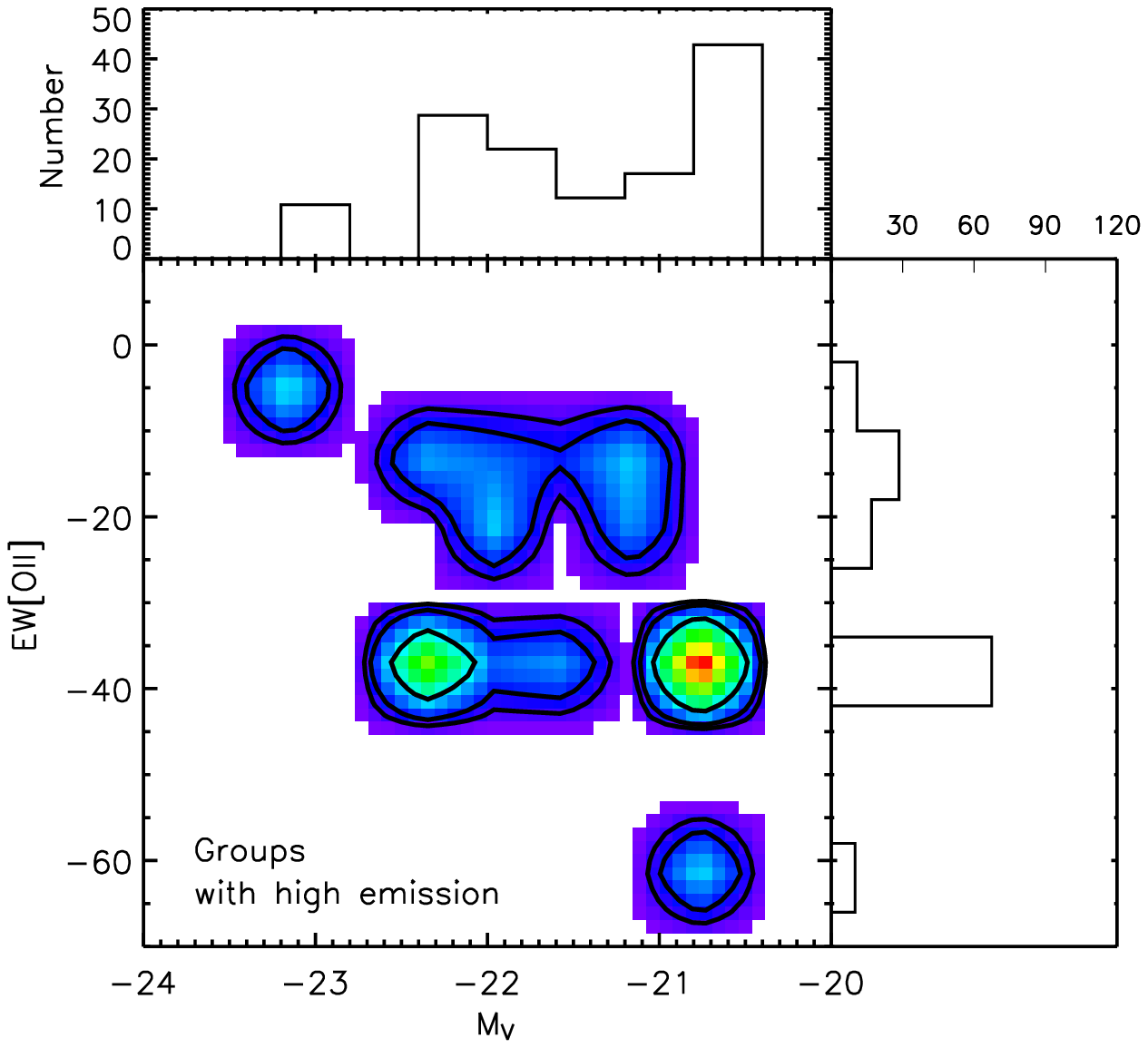}\hfill\hspace{-1cm}\includegraphics[width=0.8\columnwidth]{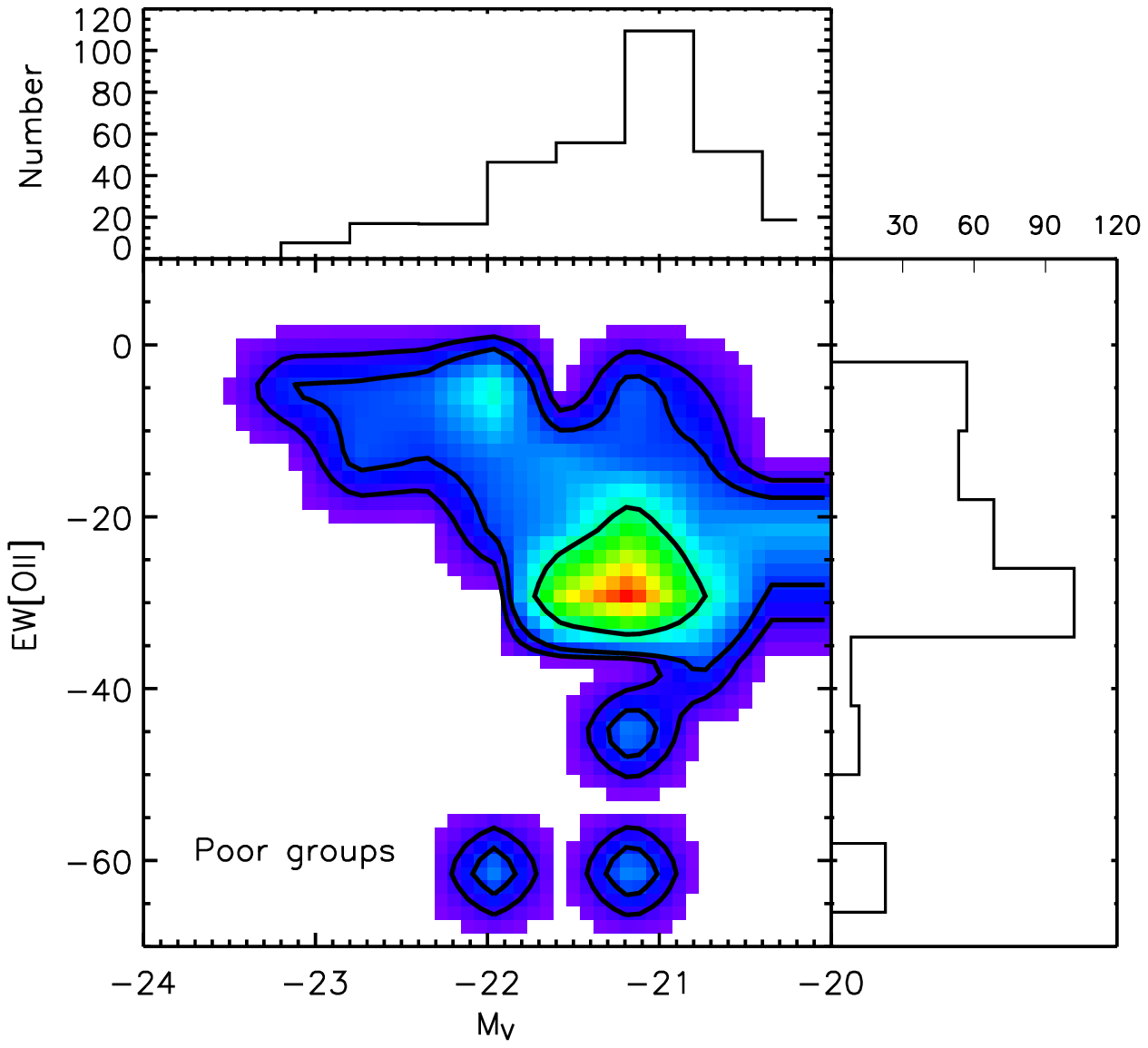}\hfill\hspace{-1cm}\includegraphics[width=0.8\columnwidth]{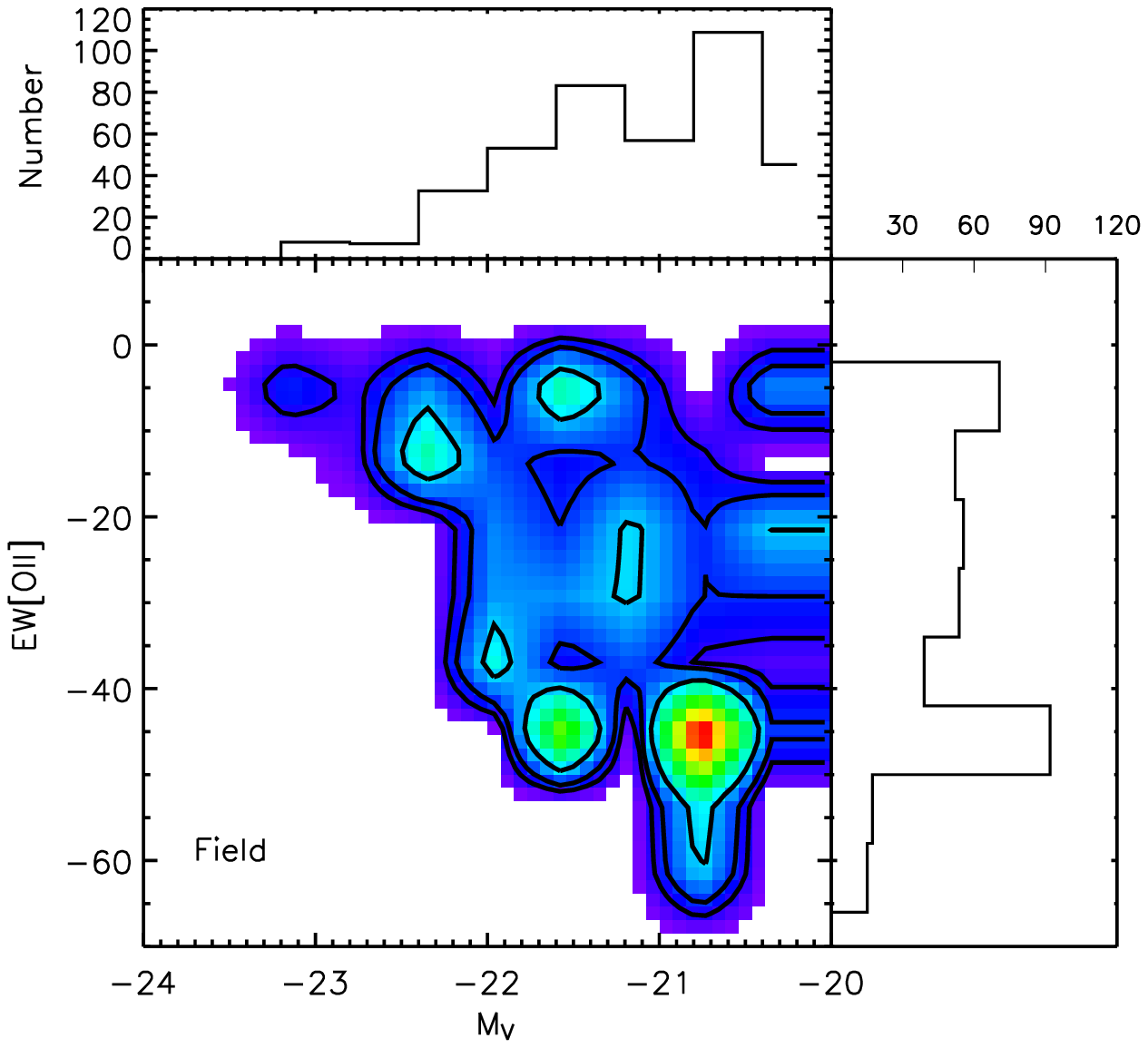}}
 \centerline{\includegraphics[width=0.8\columnwidth]{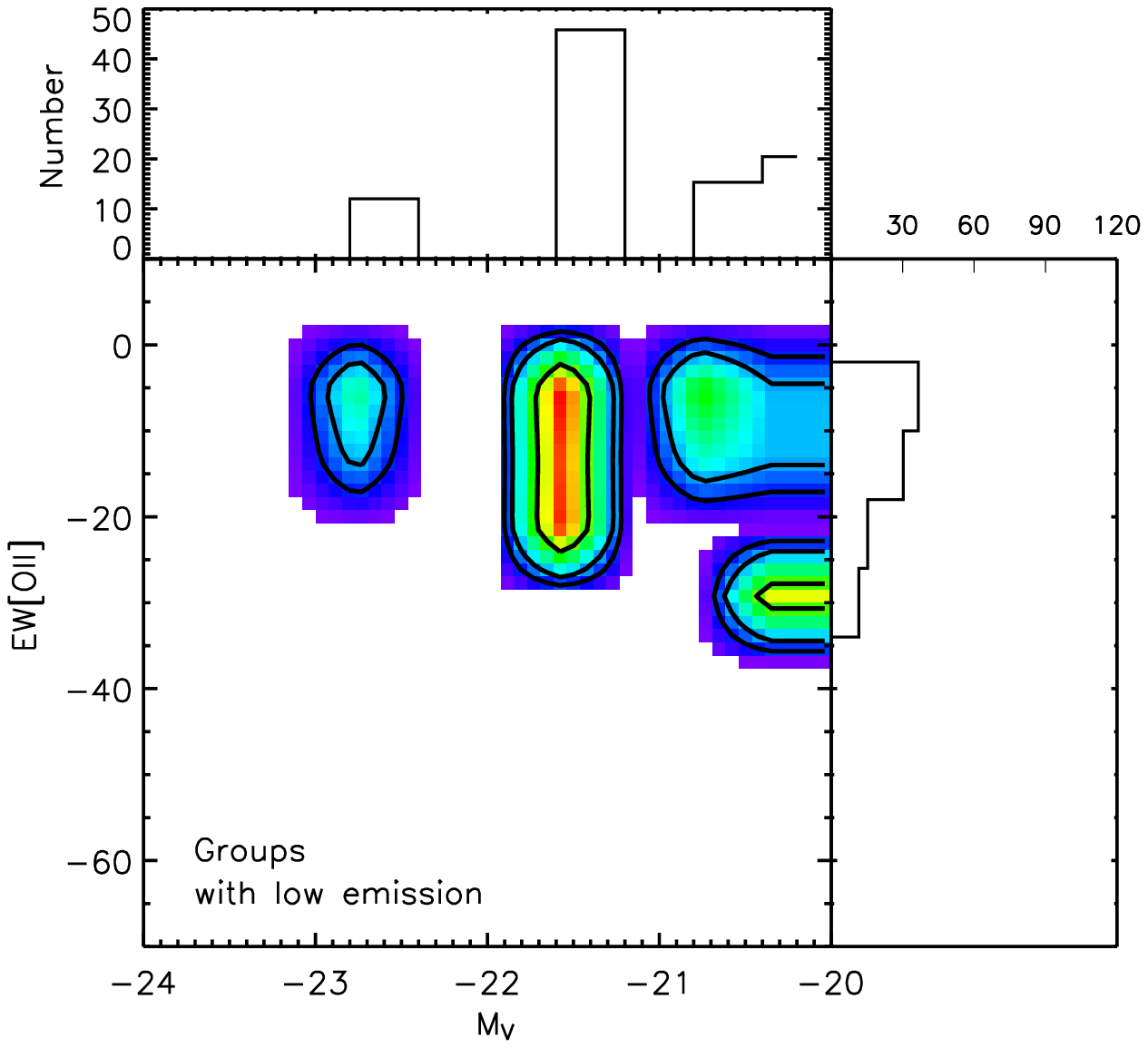}\hfill\hspace{-1cm}\includegraphics[width=0.8\columnwidth]{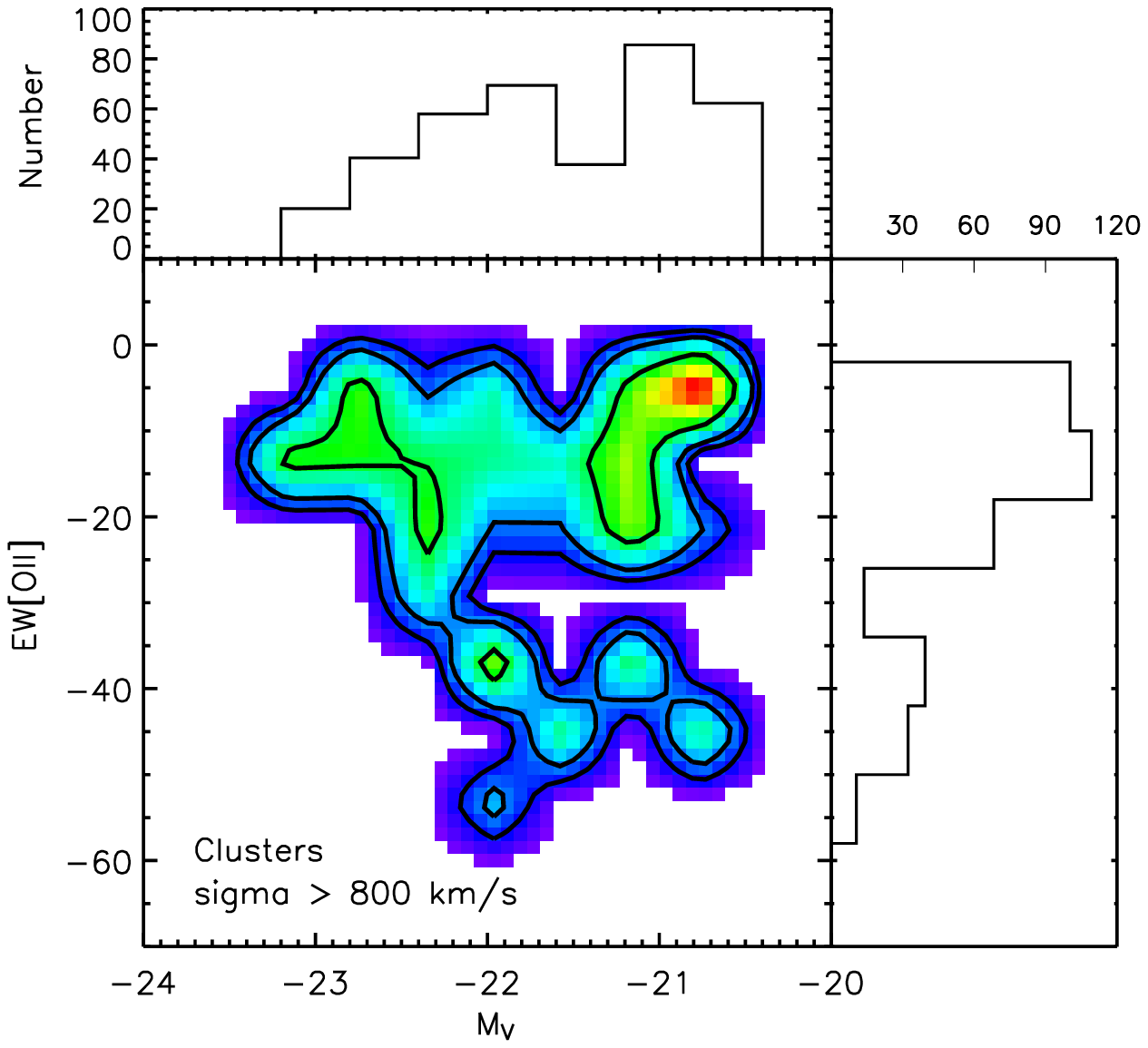}\hfill\hspace{-1cm}\includegraphics[width=0.8\columnwidth]{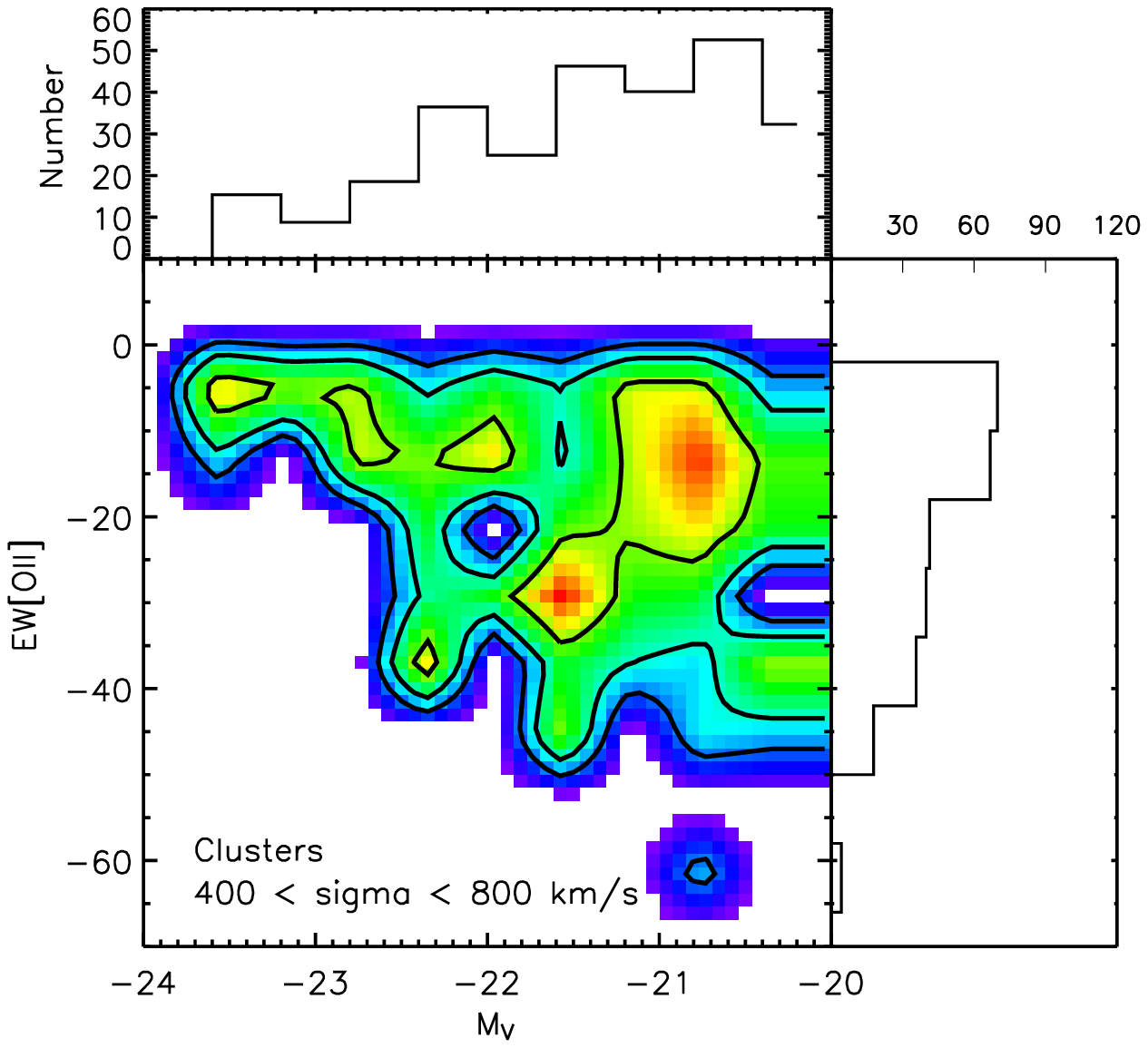}}
 \caption{Number density maps of star--forming galaxies in different environments as a function of galaxy absolute V magnitude and EW(\oii). Contours represent
 25, 75 and 90 percentiles, while the color scale is meant to guide the eye. 
 The histograms on
 top and on the right present the number of galaxies in bins
 of absolute magnitude and EW(\oii), respectively. Numbers in this figure 
 have been corrected for spectroscopic incompleteness. 
 The raw (uncorrected) numbers of galaxies are 39 (massive clusters), 126
 (less massive clusters), 31 (groups with high emission), 19 (groups
 with low emission), 51 (poor groups) and 75 (field). 
 \label{maps}}
 \end{figure*}

In principle, a different EW distribution could simply reflect a
different luminosity distribution of galaxies in the different
environments, since lower luminosity star--forming galaxies are
known to have on average higher EWs than bright star--forming
galaxies. 
To investigate whether the differences in the EW(\oii) distributions 
are due to a different luminosity distribution with
environment, or whether it is the SF in galaxies of similar luminosity
that changes with environment, we plot in Fig.~\ref{maps} the number
density distribution of star--forming galaxies in a two-dimensional
space of galaxy absolute V magnitude and EW(\oii). 

Following a well known trend, galaxies occupy a characteristic
triangular region in this diagram: in all environments, there are no
or at most a few bright galaxies with EW(\oii) stronger than $-20$
\AA, while going to fainter magnitudes the EW distribution extends to
stronger and stronger EWs.  In fact, fainter galaxies span a wide
range of EWs, while the brightest galaxies are confined to
small/moderate EWs.

While these are the general trends, the maps also show some
interesting variations with environment.  Groups with low \oii
fractions (bottom left panel) lack a significant population of faint
galaxies with moderate and strong EWs that {\it are} present in the
other environments. In these groups, star--forming galaxies have weak
EWs, regardless of their luminosity.  According to a KS test, considering
only star-forming galaxies with $M_V=-20$ to $-21.5$,
the EW distribution in groups with low \oii emission is different from 
the summed distribution in all other environments at the 99.9\% confidence 
level (99.4\% if the EW distributions are not weighted for completeness).

Inspecting the histograms of EW(\oii) (on the right side of each
panel), the field and poor group environments (middle and right top
panels) display an excess of galaxies with intermediate and strong EWs
compared to both massive and less massive clusters (middle and right
bottom panels).  This is the same effect seen in Fig.~\ref{ewfra},
that shows that the fraction of galaxies with intermediate+strong EWs
increases going from clusters to poor groups and the field. The
distributions of absolute magnitudes of star-forming galaxies are
presented in the top histogram of each panel. These show that the
magnitude distribution in the field and poor groups is skewed towards
fainter average magnitudes than in clusters of all masses.  In fact,
the absolute magnitude distribution in the field+poor groups differ
from that in clusters at the 99.9\% confidence level. In this case,
the difference is not significant if the distributions are unweighted.
Moreover, at a given faint luminosity, galaxies in the field and poor
groups tend to have stronger EWs on average than similarly luminous
star-forming galaxies in clusters (see the variation of the Y position
of the red/yellow highest peak at faint magnitudes in the maps of
field/poor-groups versus clusters).  The difference in the EW
distribution for faint galaxies ($M_V=-20$ to $-21.5$) in field+poor
groups compared to massive+less massive clusters is significant at the
99.9\% level (96.0\% if unweighted).

Thus, the behaviour of the EW(\oii) distributions with environment
seen in Fig.~\ref{ewfra} appears to be the result of a combination
of different EWs at a given faint luminosity (stronger EWs in
environments with higher star-forming fractions) and different luminosity
distributions of star-forming galaxies (a higher faint-to-bright galaxy 
number ratio in environments with higher star-forming fractions).

\subsection{Outliers in the \oii - $\sigma$ relation}

It is worth analyzing separately the properties of outliers from the
\oii -- $\sigma$ relation. Here we consider as outliers those EDisCS
structures with an \oii fraction significantly lower than the
fractions of the majority of structures of similar velocity
dispersion.

The two most outstanding outliers are two groups, Cl\,1119 and
Cl\,1420, with $\sigma = 165 \, \rm and \, 225 \, km \, s^{-1}$,
respectively, and an \oii fraction $<30$\%.  Both of these structures
have quite a high number of spectroscopically confirmed bright members
($>20$) within a small velocity dispersion.  For Cl\,1119, the
non-detection in the weak lensing analysis and the corresponding upper
limit on $\sigma$ confirm the low-mass nature of this system, while
for Cl\,1420 a meaningful comparison cannot be performed because the
lensing-based estimate is likely contaminated by other mass
structures (Clowe et al. 2005).  Thus, at least for Cl\,1119 this
seems to rule out the possibility that the system deviates from the
general trend due to its $\sigma$ measurement strongly underestimating
its mass.  Moreover, as shown in column 12 of Table~1, these two
systems remain outliers also if we relax the $R_{200}$ criterion and
compute the \oii fraction over a larger field.\footnote{The \oii fraction of Cl\,1420 changes from 0 within
$R_{200}$ to 40\% over the whole FORS2 field. Even adopting this latter
value, however, this system remains an outlier for its velocity
dispersion.}

Interestingly, besides the low \oii fractions, these two systems also
stand out for their unsually high fraction of early-type galaxies for
their velocity dispersion, as found by the analysis of galaxy
structural parameters based on 2D bulge+disk decomposition (Simard 
et al. 2006), and for their low 
fractions of blue galaxies (De Lucia
et al. 2006).  The evidence for the peculiarity of these outliers is
further reinforced by the analysis of the \oii equivalent width
distributions.  We have seen in \S5.4 that the EW distribution 
in star-forming galaxies of the groups with low \oii
fractions (including Cl\,1119 and C\,1420) is different from that in
any other environment.
The population of faint
galaxies with strong EWs, common in other environments, is absent in
these groups.  All of these findings (low \oii and blue fractions,
high early-type fractions for their
velocity dispersion, as well as peculiar equivalent width distribution
of \oii) suggest that these groups are intrinsically peculiar when
compared to the majority of other structures. 
In fact, the properties
of their galaxies resemble those of galaxies in the core of much
more massive clusters, as if they were ``bare'' massive--cluster cores
lacking the less centrally-concentrated population of
star--forming galaxies.

Another point worth stressing concerns those systems that are
relatively close in redshift to a cluster (within 10 ${\sigma}_{clu}$)
without being part of it (at velocities $> 3 {\sigma}_{clu}$).  There
are two such systems in our sample, G1\_105412 and C2\_1138 in
Table~5.  Both of these systems have a low star--forming fraction for
their velocity dispersion.  An intriguing hypothesis is that systems
close to more massive structures, thus embedded in a massive
superstructure, have a different galactic content than completely
isolated systems of similar mass. On the other hand, the two other groups
with a low \oii fraction (G1\_1301 and G1\_1103) are much further
away from any other structure detected within the field we observed
(26 and 47 $\sigma$, respectively). 

The \oii -- $\sigma$ trend observed at high redshift is suggesting
that the fraction of star--forming galaxies in clusters 5-7 Gyr ago
depends on cluster mass -- or on something that is closely related
to the cluster mass.  The existence and characteristics of the
outliers, as well as the fact that the two systems close to other
structures possess a low \oii fraction, seems to suggest that the
driving factor might be density (mass per unit volume), instead of
mass.\footnote{Interestingly, as shown by Gray et al. (2004) for the
Abell 901/902 supercluster, the local dark matter mass density
measured from weak gravitational lensing correlates with local galaxy
number density, though with considerable scatter.}  Density and mass
will be closely related for most systems, and the outliers might be
those systems of unusually high density for their mass, i.e.  those
regions that were very dense at high redshift but failed to acquire
star-forming galaxies at later times, possibly due to the
characteristics of their surrounding supercluster environment.  

\subsection{Star formation versus galaxy morphologies}

As shown in \S5.2, the relation between the \oii fraction and
the cluster velocity dispersion changes significantly between $z=0.8$
and $z=0$. Over the same redshift range, also galaxy morphologies have been
observed to evolve in clusters.  Distant clusters generally contain a
higher proportion of spirals, and a correspondingly lower proportion
of S0 galaxies, than low-z clusters (Dressler et al. 1997, Couch et al. 1998,
Fasano et al. 2000, Treu et al. 2003, Smith et al. 2005, Postman et
al. 2005). Low S0 fractions and high spiral fractions are found also
in our sample and we refer to Desai et al. (2006) and Simard et
al. (2006) for a detailed analysis of the morphological content of
EDisCs clusters.  It is then interesting to investigate whether galaxy
morphologies evolve with redshift in the same way as the star-forming
fraction does, and how the star formation histories of galaxies are
related to the Hubble type.

 \begin{figure}
 \vspace{-2cm}
 \centerline{\hspace{2cm}\includegraphics[width=12cm]{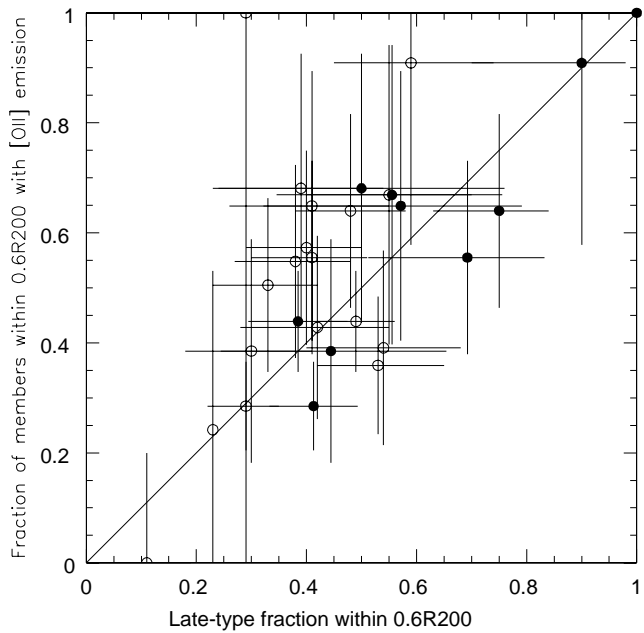}}
 \caption{\oii fraction of EDisCS clusters versus the fraction of late-type
 galaxies, both computed within $0.6R_{200}$ and down to $M_V=-20.0$. 
 The late-type fraction is computed as the fraction of spectroscopically
 confirmed members. Empty dots represent the fraction of late-type galaxies 
 obtained from structural parameters derived from 2D bulge+disk
 decompositions of VLT images (Simard et al., 2006). 
 Filled dots represent the fraction of late-type galaxies 
 (all galaxies excluded ellipticals and S0s) derived
 from visual morphological classifications of HST images
 (Desai et al., 2006). The 1:1 line is shown for comparison.
 \label{oiispirals}}
 \end{figure}

Fig.~\ref{oiispirals} presents the \oii fraction versus the fraction
of late-type galaxies (spirals+irregulars) for EDisCS clusters.  The
fraction of late-type galaxies has been derived with two different
methods: for clusters with HST imaging, from visual morphological
classifications (Desai et al. 2006), and for all 18 systems using
structural parameters derived from 2D bulge+disk decompositions of VLT
images (Simard et al. 2006).  A comparison of visual and automated
classifications can be found in Simard et al. (2006).  The plot shows
that the proportions of late-type galaxies are roughly consistent with
the star-forming fractions we find in this paper. We note for example
that the two systems with the lowest \oii fractions (Cl\,1119 and
Cl\,1420) are also those with the lowest late-type fractions.

We now compare the evolution of the star--forming fraction with the
evolution of the morphological types.  In massive clusters ($\sigma =
800 - 1100 \rm \, km \, s^{-1}$) the fraction of late-type
galaxies evolves from $30-50$\% at
high-z to $\sim 20$\% at $z=0$ (Desai et al. 2006 and Fasano et
al. 2000).\footnote{The late-type fractions are computed for
galaxies brighter than $M_V = -20.7$.}  
This corresponds to a comparable increase of the S0 galaxy
fraction.  For clusters of this velocity dispersion we find that the
star--forming fraction changes on average from 30-50\% at $z=0.6$ to
20\% at $z=0$ (Fig.~\ref{main}). In massive clusters, the change in the 
star-forming fraction is therefore similar to the observed evolution 
of the late-type population.

In clusters with $\sigma = 400-700 \rm km \, s^{-1}$, both the late-type
and the star-forming fractions range from about 40 to 80\% at high
redshift (Desai et al. 2006, and this paper).  Unfortunately, a
detailed study of the elliptical/S0/spiral fractions as a function of
the cluster velocity dispersion is not available at low redshift for
comparison.  If the evolution of the star--forming fraction between
$z=0.6$ and $z=0$ reflects the evolution of spirals into S0s (and
{\it vice versa}), our results on the \oii evolution as a function of
the system mass would imply that the evolution of the S0
population should be maximum in intermediate-mass clusters, those with
$\sim 600 \rm \, km \, s^{-1}$ at $z=0$.

The agreement between the evolution of the star-forming and S0 fractions
suggests that star-forming late-type galaxies are being transformed
into passive S0 galaxies.
However, it is necessary to stress that morphology and star formation
history can be partly decoupled in clusters: several of the cluster
spirals at all redshifts do not have emission lines in their spectra,
and both their spectra and their colors indicate a lack of current
star formation activity (e.g. Poggianti et al. 1999, Couch et
al. 2001, Goto et al. 2003).  These passive spirals are believed to be
an intermediate stage in the transformation from star-forming spirals
to passive S0 galaxies.  The existence of passive spirals, and the
fact that most of the poststarburst galaxies in distant clusters have
spiral morphologies (Dressler et al. 1999, Poggianti et al. 1999) are
strong indications that the timescale for morphological transformation
is longer than the timescale over which the spectrophotometric
signature of recent star formation disappers: galaxies are first
quenched, and then eventually their morphology changes on a longer
average timescale (Poggianti et al. 1999).

\begin{figure}
 \centerline{\includegraphics[width=8cm]{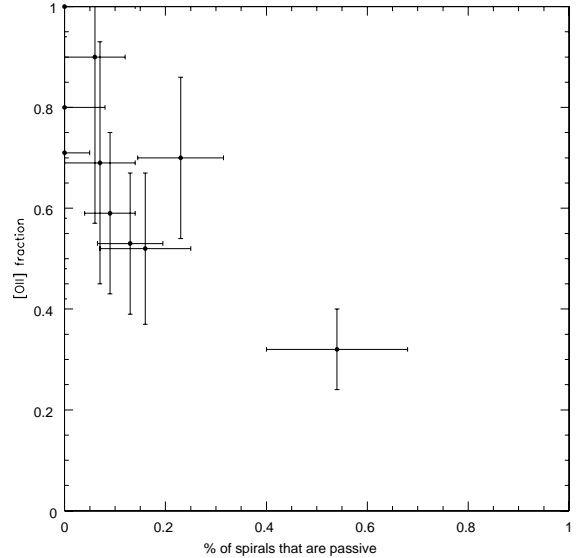}}
 \caption{Fraction of late-type galaxies (spirals+irregulars from HST images) 
 that are passive
 (lack emission lines in their spectra) versus \oii fraction for EDisCS
 systems.\label{passpi}}
 \end{figure}

Also in EDisCS clusters the population of star-forming(=emission-line)
and late-type galaxies do not fully coincide.  On average over all
clusters, we find that 15\% of the star-forming galaxies are
classified as ellipticals or S0 galaxies, and conversely that 13\% of
the morphologically late-type galaxies do not show any sign of ongoing
star formation. A detailed one-to-one comparison between galaxy
morphologies and star formation histories in EDisCS clusters is
deferred to a later paper, but for the purposes of this paper we plot
the fraction of spirals that are passive versus the \oii fractions in
Fig.~\ref{passpi}.  There is a hint that clusters with a lower \oii
fraction also might have a higher proportion of their spirals that are
passive, though this conclusion is mostly based on one cluster
(Cl\,1232) in which $>50$\% of the spirals are passive.

The decoupling of morphologies and star formation, however, involves
only a relatively modest proportion of galaxies in most clusters.  In
addition, those clusters with the highest fraction of spirals that are
passive, also tend to be those with the lowest fraction of spirals,
therefore the decoupling is not strongly affecting the global
morphological budget. It remains true that the correspondence between the
morphological evolution and the evolution in the star-forming fraction
supports the hypothesis that the evolution observed between $z=0.8$ and
$z=0$ concerns star-forming late-type galaxies evolving into passive S0
galaxies.

\section{Discussion}

The results found in this paper provide
for the first time a quantitative description of the evolution of
the star-forming galaxy population in clusters as a function of
redshift and $\sigma$.  These results highlight the need to study
the evolution of the star--forming galaxy fraction as a function of
system mass.  Ignoring this dependence can lead to incorrect 
conclusions regarding the evolution.  
Low-z surveys lack large numbers of massive clusters with
$\sigma \ge 1000 \rm \, km \, s^{-1}$ (due to their rarity) while most
high-z samples include only the most massive clusters. If the cluster
mass dependence is not taken into account, 
$1000 \rm \, km \, s^{-1}$ high-z clusters  end up being
compared with $ 500-700 \rm \,
km \, s^{-1}$ low-z clusters, making the evolution harder to 
detect.

Our results also provide a likely explanation of 
why it has been so difficult to observe trends with cluster mass/$\sigma$.
The way the data are distributed in Fig.~\ref{main} already
shows that
finding the general
trends of star-forming fraction
with system mass requires: a) a large number of clusters, with data as
homogeneous as possible; b) a sample covering a wide range
of cluster masses; c) 
a high quality spectroscopic dataset, both in terms of the
number of spectra per cluster and of the quality of the spectra
themselves; and d) reliable and thoroughly controlled $\sigma$'s when 
using velocity dispersion as a proxy for mass.  The most crucial 
requirement is the range of
cluster/group masses that needs to be explored. For example,
from Fig.~\ref{main} it is evident that sampling at high-z 
only half of the range in
$\sigma$ (only systems with $\sigma >$ or $< 700 \rm \,
km \, s^{-1}$) would result in the trend being buried in the scatter and
unrecognizable. At low z, no trend can be observed when including
only systems with $\sigma > 500 \rm \, km \, s^{-1}$.

This might explain at least some of the contrasting results that have
been found in the literature regarding the presence or absence 
of a relation between galaxy properties and system ``mass'' as
determined from velocity dispersion, X-ray luminosity or other global
cluster properties.  For example, the relatively limited range in
cluster mass explored by X-ray selected samples might be the reason
why several works could not find a trend of blue fraction with cluster
X-ray luminosity (Smail et al. 1998, Andreon \& Ettori 1999, Ellingson
et al. 2001, Fairley et al. 2002). Only sampling the whole mass range,
from groups to massive clusters, trends at high-- and low--z become
recognizable.  In this respect, the fact that the mass distribution of
EDisCs clusters differs significantly from the distribution of X-ray
selected samples at high redshift, extending to much lower masses
(Clowe et al. 2005), is an advantage.  Moreover, the range of masses
sampled by EDisCS, when evolved to z=0, matches significantly better
the mass distribution of nearby clusters than X-ray high-z samples do,
as the latter only contain the progenitors of the highest 
mass tip of the low redshift cluster mass distribution. 
It is also true that very low-mass/low-richness non-centrally concentrated 
systems could be under-represented in our sample, given the selection method
of EDisCS. This type of systems are 
generally those with the highest incidence of star-forming galaxies.
Hence, though our
selection criteria could not be responsible for the \oii -- $\sigma$ trend
observed, they might influence the observed density of points in
the \oii -- $\sigma$ diagram. If anything, there should be more
high-\oii low mass groups and the \oii -- $\sigma$ relation would then
be even stronger than we have observed.

\subsection{A possible scenario for the trends of the star formation
activity as a function of environment}

Understanding the origin of the trends of star formation with velocity
dispersion would represent a significant step forward towards
comprehending the link between galaxy evolution and environment.  
If galaxy properties depend on the mass of the system where they reside
or have resided during their evolution, there should be a connection
between the trends observed and the way cosmological structures have
grown in mass with redshift. 

In this section we investigate whether 
the trends in star formation activity correspond in some way to the 
growth history of structure.
To quantify the evolution of cosmological structures, we adopt two
different approaches. Using a Press-Schechter formalism (hereafter, PS;
Bower 1991, Lacey \& Cole 1993), we analyze the growth history of
systems of different mass. In particular, we study what fraction of the
system mass was already in massive structures at different
redshifts.  In addition, we use the Millennium Simulation (Springel et
al. 2005) to study the growth history in terms of number fraction of
galaxies instead of mass, to assess what fraction of the galaxies in
systems of a given mass were already in massive structures at
different redshifts.  This was computed by populating dark matter
haloes with galaxies by means of semi-analytic models (De Lucia et
al. 2005, Croton et al. 2005).  We have chosen to employ both
approaches because it is important to examine the results both in
terms of mass and of number of galaxies. The evolution of the mass of
cosmological structures is totally independent of assumptions
regarding galaxy formation and evolution, therefore it is not affected
by all the uncertainties inherent to these assumptions.  At the same
time, it is important to ascertain whether the evolution in the number
of galaxies follows the mass evolution, given that the former quantity
is the one that is directly observed.

In the following we will name ``clusters'' systems with masses
$>10^{14}$ $M_{\odot}$ and ``groups'' systems with masses between $3
\times 10^{12} < M <10^{14}$ $M_{\odot}$. These mass limits
approximately correspond to the velocity dispersion limits we have
adopted in this paper for defining clusters and groups ($>$ and $< 400
\, \rm km \, s^{-1}$, respectively).

In the comparison between observations and theory we are guided
by four considerations:

1) So far, we have focused on the fraction of star-forming galaxies
$f_{\oii}$. At each epoch and in each environment, the fraction of
galaxies {\it with no} ongoing star formation is simply (1-$f_{\oii}$),
and we will refer to these as ``passive galaxies''. 
Observational studies of clusters suggest that there may
be two distinct families of passive galaxies.

\begin{itemize}
\item The first family is composed of galaxies
whose stars all formed at very high redshift ($z>2$) over a short
timescale, that have been observed in clusters up to and beyond $z=1$
(Bower et al. 1992, Ellis et al. 1997, Barger et al. 1998, Kodama et
al. 1998, van Dokkum et al. 2000,2001, Blakeslee et al. 2003, De Lucia
et al. 2004, Barrientos et al., 2004, Holden et al. 2005).  
This family is largely 
composed of luminous cluster ellipticals (e.g. Ellis et al. 1997). 
%
\item The second family corresponds
to passive galaxies that have had a more extended period of star 
formation activity (with a longer star formation timescale). Star formation
in these galaxies has been quenched when they were accreted into the 
dense environment
(Dressler \& Gunn 1983, 1992, Couch \& Sharples 1987, Balogh et
al. 1997, Poggianti et al. 1999, Ellingson et al. 2001, Kodama \&
Bower 2001, van Dokkum \& Franx 2001, Tran et al. 2003, 
Poggianti et
al. 2004, Wilman et al. 2005b). Most of these galaxies are spirals
up to at least 1 Gyr after the star formation is quenched (Dressler et al. 
1999, Poggianti et al. 1999, Tran et al. 2003).
The passive nature of these galaxies is considered to be a consequence 
of the interaction with their environment.
\end{itemize}
In the following we will refer to
these two families as ``primordial passive galaxies'' and ``quenched
galaxies'', respectively. As the growth of cosmological structures
proceeds and clusters and groups accrete more galaxies, we should 
expect the relative proportion of the two types of passive
galaxies to change. In those environments that efficiently quench star
formation, quenched galaxies should progressively become a larger part of the 
passive population (going to lower redshifts) while
primordial passive galaxies should dominate the passive population in
systems at high redshift.

2) Observationally, primordial passive galaxies are preferentially
located in the densest, more massive structures at all redshifts.  At
the epoch when they formed their stars ($z \ge 2.5$), essentially no
system more massive than $10^{14}$ $M_{\odot}$ existed according to
current hierarchical theories. The most massive structures at $z=2.5$
had masses similar to those of systems that at low redshift
we would call ``groups'' ($> 3 \times 10^{12}$).  Thus,
when they had just completed their star formation, primordial passive 
galaxies were in systems of masses comparable to groups today.

3) Considering cluster crossing times (typically 1 Gyr),
timescales associated with the various physical processes that might
lead to the truncation of the star formation activity
(e.g. harassment, ram pressure, strangulation, mergers - 1-2 Gyr at
most) and the spectrophometric timescale for the evolution of the
\oii signature ($\sim 5 \times 10^7$ yr), a few Gyr should be sufficient
for suppressing the \oii emission in most quenched galaxies.  We will
then consider a timescale of 3 Gyr as a reasonable upper limit for the
time required to totally extinguish star formation in newly accreted
galaxies.

4) The existence of a break-point ($\sim 550 \rm \, km \, s^{-1}$) in
the \oii -- $\sigma$ relation observed at low redshift, above which
essentially every system has a low \oii fraction regardless of its
mass, suggests that systems above this mass are highly efficient at quenching
star formation in galaxies falling into them. A system with a
velocity dispersion around $\sim 500 \rm \, km \, s^{-1}$ at $z=0$
approximately corresponds to a system that 3 Gyr ago (see point (3))
had a mass $\sim 1-2 \times 10^{14}$ $M_{\odot}$ (see Table~4). As a
working hypothesis it is then natural to adopt $10^{14}$ $M_{\odot}$
as the reference mass for efficiently quenching star formation, i.e.
the mass above which the quenching is a widespread phenomenon affecting
sooner or later (within 3 Gyr according to point 3) all accreted galaxies.

 \begin{figure*}[t]
 \vspace{-12truecm}
 \centerline{\includegraphics[width=2.5\columnwidth]{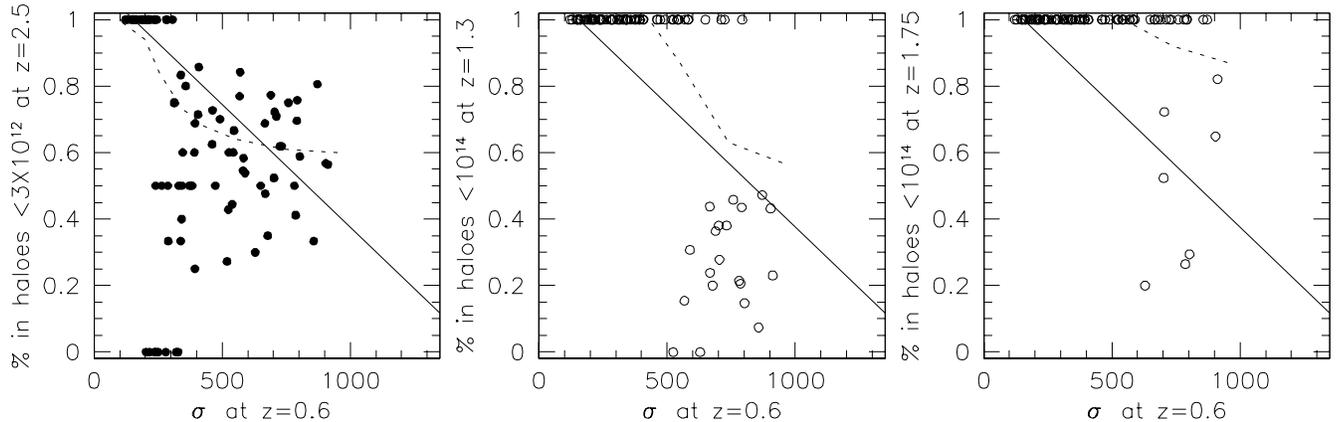}}
 \vspace{-2truecm}
 \caption{Clusters at high redshift.
 The solid line is the line drawn in the \oii -- $\sigma$ diagram 
 observed at high redshift. 
 Left. Dots represent the {\it fraction of galaxies} in 90 haloes selected at 
$z=0.6$ in the MS
 simulation that were in haloes with mass $< 3 \times 10^{12}$ at z=2.5.
 The dotted line is the prediction
 of the average {\it fraction of mass} in haloes $< 3 \times 10^{12}$ at z=2.5 
 derived from the Press-Schecter formalism.
 Middle. Empty dots represent the {\it fraction of galaxies} in the MS 
 simulation
 that were in haloes with mass $ < 10^{14}$ $\sim 3$ Gyr prior to $z=0.6$,
 thus at z=1.3. The dotted line is the prediction
 of the {\it fraction of mass} in haloes $< 10^{14}$ at $z=1.3$ 
derived from the PS.
 Right. Empty dots represent the {\it fraction of galaxies} in the MS 
 simulation
 that were in haloes with mass $ < 10^{14}$ $\sim 3$ Gyr prior to $z=0.8$,
 thus at z=1.75. The dotted line is the prediction
 of the {\it fraction of mass} in haloes $< 10^{14}$ at $z=1.75$ 
derived from the PS.
 \label{mr2}}
 \end{figure*}

We first consider the family of primordial passive galaxies.
In Fig.~\ref{mr2} we compare the \oii observations at $z=0.4-0.8$ with
the theoretical expectations for the growth history.  The solid line
(in both panels) is the line drawn in the \oii -- $\sigma$ diagram
observed at $z=0.4-0.8$ (Fig.~\ref{main}).  In the left panel, the
dotted line is the fraction of mass of systems at $z=0.6$ that was in
systems with $M_{sys}<3 \times 10^{12}$ $M_{\odot}$ at z=2.5 as
derived from the Press-Schechter formalism, averaged over 100
haloes. Solid dots represent the fraction of galaxies within $R_{200}$
and with $M_V$ limits as for EDisCS that were in systems with
$M_{sys}<3 \times 10^{12}$ $M_{\odot}$ at z=2.5 obtained for 90 haloes
in the Millennium Simulation. Haloes were in this case selected at
$z=0.6$ from the MS, similarly to how it was done for haloes at $z=0$
in \S5.2.

Both the PS results and the MS upper envelope trace remarkably well
the \oii -- $\sigma$ relation observed at high
redshift. The scatter of the MS points illustrates that for systems of
any given $\sigma$ at $z \sim 0.6$ there is a range in the fraction of
galaxies that were already in groups at $z=2.5$.  This scatter indeed
resembles the scatter of the datapoints in the observed \oii--$\sigma$
diagram of distant clusters (see the left panel of
Fig.~\ref{main}). This figure shows that {\it the fraction of passive
galaxies observed in $z=0.4-0.8$ clusters of a given $\sigma$/mass is
comparable with the fraction of its galaxies (or its mass) that was
already in dense environments (=groups) at $z = 2.5$.}  We tentatively
identify the latter with the population of primordial passive galaxies
as described in point 2) above.  Identifying primordial passive
galaxies with galaxies already in groups at $z=2.5$ implies that the
great majority of galaxies belonging to environments more massive then
$3 \times 10^{12} \, M_{\odot}$ at $z=2.5$ completed their star
formation activity at high redshift, and, {\it vice versa}, that those
galaxies that completed their star formation at high redshift are
mostly galaxies that were in environments more massive than $3 \times
10^{12} \, M_{\odot}$ at $z=2.5$.

We note that among the 90 haloes extracted from the MS there are
also a few ``outliers'' located in the lower left region of
Fig.~\ref{mr2}.  This means that a large fraction of their galaxies
resided in haloes of masses $M_{sys}> 3 \times 10^{12}$ at $z=2.5$.
The comparison between Fig.~\ref{mr2} and Fig.~\ref{main} then
suggests that \oii outliers at high redshift might represent 
systems that had a high fraction of their mass already in groups at
$z=2.5$ and did not accrete a large population of ``field'' galaxies
between $z=2.5$ and $z \sim 0.6$.

In the middle and right hand panels of Fig.~\ref{mr2}, we contrast the
observed fractions of passive galaxies with the trends expected for
quenched galaxies. Open circles in the middle panel show the fraction
of galaxies in $z=0.6$ systems in the MS that were in haloes with mass
$ < 10^{14}$ $\sim 3$ Gyr prior to $z=0.6$, thus at $z=1.3$ (see
points 3) and 4) above).  For most systems with $\sigma < 700 \rm \,
km \, s^{-1}$ the fraction of galaxies that experienced the cluster
environment for at least 3 Gyr is zero, and the predicted trend is
inconsistent with the observational results.  {In these systems, the
passive galaxy population is not consistent with the fraction of
galaxies/mass that was already in systems more massive than $10^{14}
\, M_{\odot}$ 3 Gyr prior to $z=0.6.$} In the most massive systems
($\sigma > 700 \rm \, km \, s^{-1}$), the middle panel of
Fig.~\ref{mr2} shows that there is already a considerable fraction of 
galaxies at $z=0.6$ that have resided in a clusters at least since $z=1.3$. 
We find that only $\sim 50$\% of these galaxies were in groups at z=2.5,
therefore there is a non-negligible proportion of galaxies that have
experienced the cluster environment (=have been quenched according to point
4)) without being ``primordial'' passive galaxies. This suggests
that, while the lower mass systems at $z=0.6$ contain essentially no
quenched galaxies, in more massive systems at these redshifts quenched
galaxies can already account for more than 1/3 of the passive
population.  

In the right panel of Fig.~\ref{mr2}, open circles show the fraction
of galaxies in MS systems that were in haloes with mass $ < 10^{14}$
$\sim 3$ Gyr prior to $z=0.8$ (the highest redshift in the sample we
study here), thus at $z=1.75$. Few galaxies have experienced prolonged
exposure to cluster-like environments since $z=1.75$. The comparison
of the middle and right panel of Fig.~\ref{mr2} shows that between
$z=1.75$ and $z=1.3$ in massive systems there is a dramatic change in
the fraction of mass/galaxies that have experienced environments more
massive than $10^{14} \, M_{\odot}$, hence $z \sim 1.5$ is an important epoch
for the build-up of clusters and the beginning of the quenching process.

\begin{figure*}[t]
 \vspace{-10truecm}
 \centerline{\includegraphics[width=2.2\columnwidth]{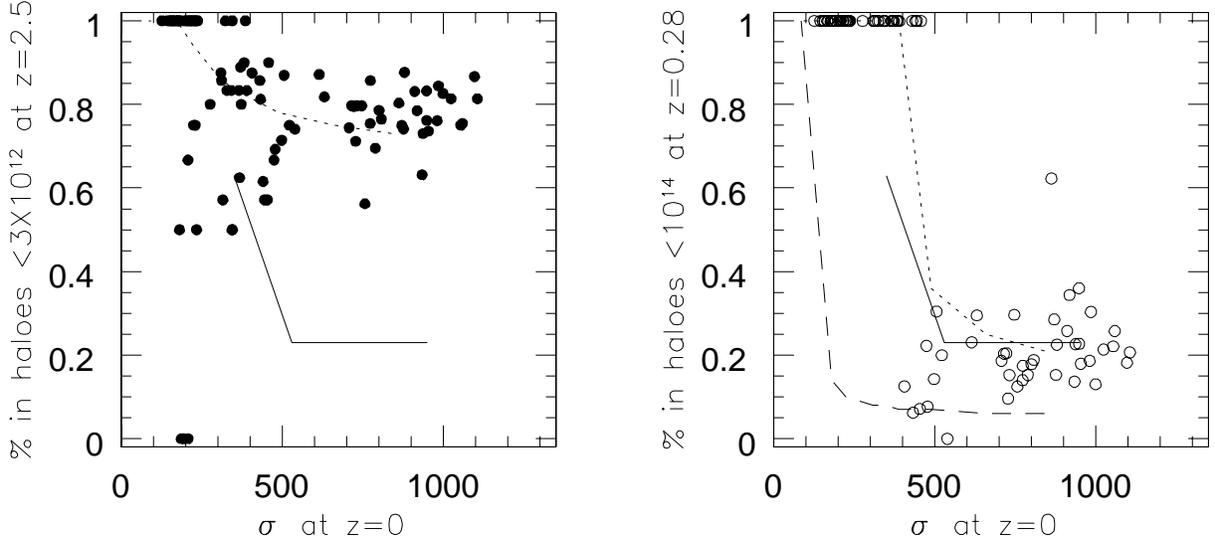}}
 \vspace{-1.7truecm}
 \caption{Clusters at low redshift.
 Comparison between the \oii -- $\sigma$
 relation observed at low redshift (solid broken line) and results from 
 the Millennium Simulation (dots) and from the
 Press-Schecter formalism (dotted lines).
 Left. The solid broken line traces the \oii -- $\sigma$ relation
 observed at low redshift. 
 Dots represent the {\it fraction of galaxies} in 90 haloes in
 the MS simulation
 that were in haloes with mass $< 3 \times 10^{12}$ at z=2.5.
 The dotted line is the prediction
 of the {\it fraction of mass} in haloes $< 3 \times 10^{12}$ at z=2.5 
 for a system of a given $\sigma$ at z=0.0 
 derived from the Press-Schecter
 formalism.
 Right. The solid broken line is repeated from the left panel.
 Empty dots represent the {\it fraction of galaxies} in 90 haloes
 from the MS simulation
 that were in haloes with mass $< 10^{14}$ $\sim 3$ Gyr prior to observations,
 thus at z=0.28. The short dashed line is the prediction
 of the {\it fraction of mass} in haloes $< 10^{14}$ at z=0.28 
 for a system of a given $\sigma$ at z=0.0 
 derived from the Press-Schecter
 formalism. The long dashed line is the prediction
 of the {\it fraction of mass} in haloes $< 3 \times 10^{12}$ at z=0.28
 for a system of a given $\sigma$ at z=0.0 
 derived from the Press-Schecter
 formalism. 
 \label{mr1}}
 \end{figure*}

Turning to low redshifts, in Fig.~\ref{mr1} we compare the trends
observed in Sloan clusters (solid broken line) with the fraction of
mass and galaxies in massive environments at previous redshifts. The
left panel shows the fractions of mass and number of galaxies in systems
at $z=0$ that
were in systems with $M_{sys}<3 \times 10^{12}$ $M_{\odot}$ at z=2.5
from the PS and MS (dotted line and filled dots), respectively. For
the MS, only galaxies within $R_{200}$ and with $M_V$ limit as for
Sloan are considered.  {\it 
In contrast to the high-z clusters, the fraction of
passive galaxies in systems at z=0 does not agree well with the fraction of
galaxies residing in groups already at high redshift. While 80\%
of galaxies in massive systems at z=0 are passive, only 20\% 
were in groups at z=2.5.}

The right panel in Fig.~\ref{mr1} shows the fractions of mass and
galaxies that were in systems of mass $M_{sys}< 10^{14}$ $M_{\odot}$
3 Gyr prior to the observations (corresponding to z=0.28), from the PS
and MS (dotted line and empty dots), respectively. In this case, the
agreement between the observations of Sloan clusters
(solid broken line) and the PS and MS results is
remarkable.   
This shows that {\it the observed fraction of passive galaxies
in systems with $\sigma > 500 \, \rm km \, s^{-1}$ at z=0 is
compatible with the fraction of galaxies that have resided in a {\it
cluster} ($M_{sys}> 10^{14}$ $M_{\odot}$) for at least 3 Gyr, and
therefore have had the time to have their star formation switched
off} (see points 3) and 4) above). 
The passive population in clusters at z=0 amounts
to about 80\% of all galaxies\footnote{Within $R_{200}$ and for
magnitudes brighter than $M_V=-19.8$.}, {\it of which} 20\% (left
panel in Fig.~\ref{mr1}) are ``primordial'' passive galaxies that have
evolved passively since $z=2.5$ and 60\% are galaxies which are 
``quenched'' at
lower redshift.\footnote{This is the case because we find that $>90$\%
of the galaxies that were in groups (haloes with masses $M_{sys}> 3
\times 10^{12}$) at z=2.5 end up being in clusters (haloes with
$M_{sys}> 10^{14}$) at z=0.28.  Thus, the population of galaxies in
clusters at z=0.28 (right panel of Fig.~\ref{mr1}) essentially {\it
contains} the galaxy population that was in groups at z=2.5 (left
panel of Fig.~\ref{mr1}).}
Also in this case the scatter in the growth history of MS haloes is
similar to the observed scatter in the fraction of star-forming
galaxies (compare the right panels of Fig.~\ref{mr1} and
Fig.~\ref{main}). The scatter observed in the \oii -- $\sigma$ relation
at all redshifts probably simply reflects the scatter in the growth
histories of systems of any given mass.

According to this discussion, while the passive galaxy populations of
the distant clusters are predominantly composed of primordial passive
galaxies, the populations of lower redshift clusters are dominated by
quenched galaxies.  We considered whether it is possible to obtain an
agreement between the fraction of quenched galaxies and the high-z
observations by choosing a lower reference mass for quenching star
formation.  However, the reference mass of $ \sim 10^{14}$ $M_{\odot}$
is set by the mass (3 Gyr ago) of a system with a velocity dispersion
at $z=0$ corresponding to the break observed at $\sim 500 \rm \, km \,
s^{-1}$ in Sloan clusters.  If the minimum mass of a system
efficiently quenching star formation were much
lower, such as for example $3 \times 10^{12}$ $M_{\odot}$, the
fraction of passive galaxies would be too high (and the fraction of
star-forming galaxies too low) compared to the low-z observations, as
shown by the long dashed line in the right panel of Fig.~\ref{mr1}.
Thus, under the assumption that physical processes operate
at $z=0.6$ as they do at $z=0$ and adopting the same quenching
reference mass at all redshifts, both the primordial and the quenched
channels are required to simultaneously match the observed trends at
high and low redshift.

A key point to note from this discussion is that the behaviour of the
\oii fraction with $\sigma$ at low redshift appears to rule out the
possibility that the group environment universally quenches star
formation.  If the quenching was a widespread phenomenon in ``groups''
(systems with masses significantly lower than $ \sim 10^{14}$
$M_{\odot}$), then all groups and clusters at low redshift should
contain much lower fractions of star-forming galaxies than is
observed. We note that this does not exclude that star formation might
be quenched in {\it some} galaxies in groups or all galaxies
in some of the
groups, but a truncation affecting all galaxies within 3 Gyr from
infall into systems with masses $ \ll 10^{14}$
$M_{\odot}$ cannot be reconciled with the observations. Conversely,
the $10^{14}$ $M_{\odot}$ reference mass indicates that the quenching
of the star formation is not limited to very massive clusters,
but is highly efficient also in low-mass clusters.

Adopting the scenario depicted above as our working hypothesis,
we can address the questions raised in \S5.2:

a) Why does the proportion of passive/star-forming galaxies
correlate/anticorrelate (with a large scatter) with the velocity
dispersion of the system for the majority of clusters at z=0.4-0.8?

Our previous discussion shows that the observed fractions of passive
galaxies at $z=0.4-0.8$ roughly agree with the fraction of
mass/galaxies that were already in groups at $z=2.5$. Primordial
passive galaxies make up most of the passive population observed at $z
\sim 0.6$, but in systems more massive than $700 \rm \, km \, s^{-1}$
the proportion of quenched galaxies is already significant.  The
anticorrelation observed arises because more massive systems tend to
have a higher fraction of their mass/galaxies that were already in
groups at z=2.5, and massive systems also have a significant population
of quenched galaxies.


b) Why does the proportion of passive/star-forming galaxies evolve with
redshift in the way observed? In other words, why is there a
Butcher-Oemler effect?  

At any redshift, the star-forming population is made up of
galaxies that were not in groups at $z>2.5$ {\it and} were not in
clusters in the last few Gyrs. In this scenario the proportion of
star-forming galaxies varies with redshift because the proportion
that was in groups at $z>2.5$ and the proportion in clusters
during the last 3 Gyr change according to the growth history, the sum
of the two growing towards lower redshifts.

c) Why is there no clear trend with $\sigma$ at $z=0$ for systems more
massive than $500 \rm km \, s^{-1}$?

In clusters with $\sigma > 500 \rm km \, s^{-1}$ at $z=0$, about 80\%
of all galaxies are passive and have resided in clusters for at least
3 Gyr. Of these, 20\% are primordial passive galaxies that formed in
groups at $z>2.5$ and 60\% are quenched galaxies.  At $z=0$, both the
proportion of galaxies that were in groups at $z=2.5$ and the
proportion of galaxies that were quenched is flat as a function of the
system mass, as shown by Fig.~\ref{mr1}, and this gives rise to the
observed plateau at $\sigma > 500 \rm km \, s^{-1}$. In systems less
massive than $400-500 \rm \, km \, s^{-1}$, that are not as efficient as
more massive systems in quenching star formation in galaxies infalling
into them, the passive population could in some cases still largely
coincide with the population of primordial passive galaxies formed at
$z>2.5$. However, if all the passive population in low $\sigma$ systems
originated as primordial passive galaxies, systems at low-z on average
would have {\it higher} starforming fractions than similar systems at high-z
(compare the left panels of Fig.~\ref{mr2} and  Fig.~\ref{mr1}).
Hence, it is probable that either the same process active
in clusters (but with a lower efficiency) and/or other mechanisms are
at work suppressing star formation in some of the galaxies in groups,
or in some of the groups. As discussed previously, if the most important factor
were density instead of mass, the large scatter of the \oii fractions
at low $\sigma$ could be due to variations of density for haloes of similarly
low masses. As discussed in the next section, the existence 
of S0 galaxies in groups may be suggesting that star formation is indeed
truncated also in groups under certain circumstances.


The consistency between the observations and the theoretical scheme
outlined above does not constitute a definite proof of the validity of this
scenario; this should be further tested by additional observations,
especially at redshifts even higher than those considered here.  It is
suggestive, however, to find that the observed star formation trends 
follow both qualitatively and quantitatively the growth history of
structure.  If the scenario we have proposed above approximates the real
situation, its implications are far-reaching, as discussed in the next section.

\subsection{Implications}

In the scenario outlined in the previous section there are two
channels that ``produce'' passive galaxies in dense environments.
``Primordial'' passive galaxies form all of their stars at $z>2$ and
it is reasonable to largely identify them with ellipticals, while
``quenched'' galaxies have their star formation truncated at much
later times, when infalling into an environment that can cause a
truncation in the star formation activity, and we tentatively identify
them with the population of spirals evolving into S0s. 
Each one of these two channels seems to correspond to a different {\it typical
mass} of the system. While primordial passive galaxies are related to
systems with masses typical of groups at $z>2.5$, quenched galaxies
appear to be a universal phenomenon in clusters, i.e.  
systems with masses $M_{sys}>
10^{14}$ $M_{\odot}$.  

We also note that the galaxy mass and luminosity distributions of
primordial passive galaxies and quenched galaxies are expected to
differ. Since the star formation activity in galaxies proceeds in a
downsizing fashion, both in clusters and in the field (Cowie et
al. 1996, Smail et al. 1998, Kodama \& Bower 2001, Poggianti et
al. 2001, Gavazzi et al. 2002, Kauffmann et al. 2003, De Lucia et
al. 2004, Poggianti et al. 2004), galaxies terminating their star
formation at higher redshift (e.g. primordial passive galaxies) will
be on average more massive/luminous than galaxies with a more
protracted star formation activity that are quenched at later
epochs when they are accreted in the dense environment.  As a
consequence, quenched galaxies will be on average less massive/fainter
than primordial passive galaxies.

Our results show that galaxy properties could be directly linked with
the growth history of DM structure: 
as shown in Fig.~\ref{mr2} and
~\ref{mr1}, the history of the mass of structures is reflected in the
star-forming fraction we observe. 
This suggests that, even without
using galaxy formation and evolution models, we can use our knowledge
of the growth of structure to explain the trends of galaxy properties
in clusters.
We have only used two pieces of information, namely how much
mass/how many galaxies in a system of a given mass at a given redshift
were in dense regions at $z>2.5$, and how much mass/how many galaxies
experienced the cluster environment for at least a few
Gyrs. If this extremely simple, double-channel picture is generally correct, it
represents a very powerful recipe for interpreting the environmental
trends observed.

If this scenario approximates the real situation, it can also serve as a 
key to understand
the evolution of galaxy morphologies.  We have seen that the observed
environmental trends of galaxy properties originate in two ways. The
proportion of ``primordial'' passive galaxies tends to increase 
with system mass in high-z systems. Systems with proportionally
more massive seeds at $z>2.5$ formed more
``primordial'' passive galaxies (mostly ellipticals). In massive systems,
other galaxies are added (as S0s) to the passive
population as time goes by. These are galaxies that would have
continued forming stars had they not been acquired by the dense
environment that has switched off their star formation activity.  From
a morphological point of view, it is reasonable to associate the
ellipticals with the primordial component and {\it some} of the S0
galaxies with the component quenched at $z<1$.\footnote{As discussed
in many previous studies, probably not all S0 galaxies originate from
the quenching of spirals at $z<1$ (e.g. Dressler et al.  1997). The
0-20\% of S0s observed in clusters at $z=0.4-0.8$ might have originated from
spirals evolving into S0s at $z>1$, or by some other mechanism. The
existence of S0s in groups (e.g. Hickson, Kindl \& Auman 1989)
shows that this type of galaxies
can be produced also in systems less massive than $400-500 \rm \, km \, s^{-1}$.}
The fact that in some low-z clusters ellipticals have been found to have
only old stellar populations, while a significant fraction of the S0s show
signs of a more recent star formation activity
is consistent with this scenario
(Kuntscher \& Davies 1998, Poggianti et al. 2001, Smail et al. 2001,
Terlevich et al. 2001, Thomas 2002).  This would also explain why some
local clusters are dominated by S0 galaxies and some others by
ellipticals. Oemler (1974) suggested that elliptical-rich and S0-rich
clusters are not two evolutionary stages in cluster evolution, but
intrinsically different types of clusters in which the abundance of
ellipticals was established at high redshifts. This suggestion was
supported by the findings of Fasano et al. (2000) that clusters at
$z\sim 0.1-0.2$ have a low (high) S0/E number ratio if they display
(lack) a strong concentration of elliptical galaxies towards the
cluster center.  In the scenario we outline above, elliptical-rich
clusters would be those with the highest incidence of primordial
passive galaxies, and S0-rich clusters those in which quenched
galaxies represent a dominant portion of the passive population.

Let us compare spectroscopic and morphological evolution in more
detail. The average fraction of ellipticals in clusters at z=0 from
Dressler (1980) is $\sim 20$\% (see Fasano et al. 2000 or Desai et
al. 2006).  In agreement with this , at z=0 the fraction of
galaxies in haloes with
masses $> 3 \times 10^{12} \, \rm M_{\odot}$ at $z>2.5$, is $\sim
20$\% for the majority of systems with $\sigma > 300 \, \rm km \,
s^{-1}$ (left panel of Fig.~\ref{mr1}).  In total, the
early-type population (ellipticals+S0s) reaches $\sim 80$\% at z=0, in
agreement with the fraction of passive (non-star-forming) galaxies observed at
z=0 (right panel in Fig.~\ref{main}) and with the fraction of
``passive'' galaxies (primordial+quenched) of haloes at z=0 that were
in haloes $> 10^{14} \, M_{\odot}$ since z=0.28.
The observed late-type fraction at all redshifts is in rough agreement
with both the observed fraction of star-forming galaxies and the predicted
fraction of galaxies that were not in groups at $z=2.5$ {\it and} did
not reside in a cluster for at least a few Gyrs. All these three
quantities are roughly equal to $\sim 20$\% in clusters at
z=0 (compare Fig.~3 in Desai et al. (2006), the right panel of
Fig.~\ref{main} and Fig.~\ref{mr1}). All of these three quantities also
show a trend with $\sigma$ at high-z (compare Fig.~7 in
Desai et al. (2006), the left panel of Fig.~\ref{main} and
Fig.~\ref{mr2}).

There is, however, one
inconsistency when grossly identifying the population of
late-type galaxies with the population of star-forming galaxies.
While there is no clear trend in the star-forming fraction with $\sigma$
at z=0 above $500 \, \rm km \, s^{-1}$ (Fig.~\ref{main}), 
the percentage of spiral galaxies in nearby clusters has been
shown to anticorrelate with the X-ray luminosity in clusters of
$\sigma \sim 700-1000 \rm \, km \, s^{-1}$ (Bahcall 1977, 
Edge \& Stewart 1991). Although the available
morphological studies in X-ray clusters at $z=0$
were not done in a similar way to ours (for selection
of members, radial coverage, morphological classifications etc.)
and notwithstanding the fact that passive spirals could play an even more
important role at low than at high z according to the quenching
scenario, this remains an unsolved issue.

Interestingly, significant morphological evolution seems to have taken
place in clusters for a large number of galaxies only at $z \le 0.4$.
In fact, as discussed by Desai et al. (2006), the S0 and spirals fractions
appear to flatten out at $z>0.45$. A tentative explanation for this
behaviour can be found in the scenario proposed.  At redshifts higher
than $\sim 0.6$, the population of passive galaxies is generally
dominated by primordial passive galaxies (mostly ellipticals, but also
the few S0 galaxies present at high redshift). This is supported by
the fact that in clusters at redshifts $z \sim 0.6$ (and even more so
at $z=0.8$) the passive
fraction can largely be accounted for by 
the fraction of galaxies that were in dense
environments at $z>2.5$.  Only at $z<0.6$ does the quenched galaxy
population become a dominant part of the passive population. Given
the delay between the truncation of the star formation and the
morphological evolution (Poggianti et al. 1999), 
this might translate into a morphological
evolution observable for a large number of galaxies only at
$z<0.4$.\footnote{Passive spirals may be galaxies that are caught in 
the transition phase of this transformation. Moreover, 
this might also explain why the morphology-density
relation does not evolve much (except in the very highest density bin)
between z=0.5 and z=1 (Smith et al. 2005), see below.  At z=1, the MD
(and SFD) relation observed is mostly the ``primordial'' relation as
established at very high redshift.}  Thus, the epoch where we can
observe the quenching of star formation for a significant fraction of
galaxies in clusters is only at $z \le 0.8$, while the epoch where
morphological transformations have taken place for a significant
fraction of the cluster galaxies is only at $z \le 0.4$. The redshift
range in which these transformations are observable is due to how
the relative infall rate 
(fraction of system mass/galaxies)
from low-mass/low-density regions onto clusters/groups changes with
redshift, as shown in Fig.~\ref{mr2} and Fig.~\ref{mr1}.
Interestingly, $z \sim 0.6$ seems to be a special epoch also for the
evolution of quasars in rich environment (Yee \& Ellingson 1993).

So far, we have considered the existence of a trend of the \oii
fraction with $\sigma$ as a sign of a relation between the star
formation of galaxies and the ``global environment'' (mass of the
system) in which galaxies reside. However, it is possible
that this is a secondary relation induced by the fact that
mass and density are closely linked. In fact, density at early and
later times might be the driving factor.

The main galaxy properties (star formation activity and
morphology) are observed to vary with the ``local''
environment in a systematic way.  The most emblematic way to describe
these systematic variations is the morphology-density relation (MD),
that is the observed correlation between the frequency of the various
Hubble types and the local galaxy density, normally defined as the
projected number density of galaxies within an area including its
closest 10 neighbours. In clusters in the local Universe, the
existence of this relation has been known for a long time: ellipticals
are frequent in high density regions, while the fraction of spirals is
high in low density regions (Oemler 1974, Dressler 1980).  An MD
relation qualitatively similar to the one observed in the local
Universe has been observed up to z=1 (as it is logical to expect,
galaxy properties correlate with environment at all redshifts), but
this relation is {\it quantitatively} strongly evolving between z=0
and z=0.5: in distant clusters the frequency of S0 galaxies is lower,
and the frequency of spirals higher, at all densities (Dressler et
al. 1997).  Interestingly, first results at $z=0.7-1.3$ seem to
indicate that between z=0.5 and z=1 what changes in the MD relation is
only the occurrence of early-type galaxies in the very highest density
regions (Smith et al.  2005), and that the frequency of ellipticals at any
given local density is the same at z=1 and at z=0 (Postman et
al. 2005).

In parallel to the MD relation, there is a star formation-density
relation (SFD). For a very long time it has been known that in the
nearby Universe also the average star formation activity correlates
with the local density: in higher density regions, the mean
star formation rate per galaxy is lower.  This is not surprising, given
the existence of the MD relation: the highest density regions have
proportionally more early-type galaxies devoid of current star
formation. The correlation between mean SF and local density extends
to very low local densities, comparable to those found at the virial
radius of clusters, and such a correlation exists also outside of
clusters (e.g. Lewis et al. 2002, Gomez et al. 2003, Kauffmann
et al. 2004).  Again, this
seems to parallel the fact that an MD relation is probably existing in
all environments, and it has been observed in clusters of all types
(Dressler et al. 1980), groups (Postman \& Geller 1984) and cluster
outskirts (Treu et al. 2003) -- though the MD relation is {\it not
quantitatively the same} in all environments, being different in
concentrated vs. irregular clusters, and high- vs. low-$L_X$ clusters
(Dressler 1980, Balogh et al. 2002).

Rephrasing our picture in terms of density instead of mass, both the
MD and SFD relations should have a ``primordial'' component and an
``evolved'' component, and both of these components should depend on
the environment, but in a different way.  In this scenario, the MD
relation and the SFD relation are {\it established} at very high
redshift at the moment the first stars formed in galaxies, and they
exist due to the close link between the initial star formation
activity of galaxies and the ``primordial'' local density of their
environment (ellipticals formed and have always resided in the highest
density regions of the Universe).  Thus it could be the ``primordial
local density'' at very high redshift that determines the properties
(star formation history, morphology - and probably mass, see Steidel
et al. 2005) of galaxies
formed in that region. Primordial local density and primordial mass of
the cluster seed are probably closely related, and the relation we
observe with the fraction of mass in massive environments at $z>2.5$ could
reflect a relation between the primordial local density and the type
of galaxy formed in that region.\footnote{Outliers in the
\oii--mass($\sigma$) relation (groups with low \oii fraction and low
masses) could be systems with high primordial local density that have
grown in mass much less than the average system with similar
primordial density.}  Therefore, the {\it origin} of the MD and SF
relations should be ``primordial'', in the sense that a relation
between galaxy properties and environment must have been in place at
$z>3$. In fact, a morphological and star formation segregation is an
outcome of CDM simulations of large scale structure and semianalytic
models because the local density of galaxies and DM is related to the
epoch of initial collapse (Bower et al. 1991, Kauffmann 1995a, 1995b,
Kauffmann et al. 1999, Benson et al. 2001, Diaferio et al. 2001,
Springel et al. 2001): the most massive structures at any epoch are
the earliest to collapse. A morphological segregation is
built-in at a very fundamental level in hierarchical theories of
galaxy formation.

\begin{table}
\begin{center}
\caption{KS-test probabilities for the color distributions
of spectroscopic and photometric catalogs
of EDisCS fields to be indistinguishable.\label{tbl6}}
\begin{tabular}{lrrrrrrrr}
\tableline\tableline
Cluster & ${p}_{corre}^{KS}$ & ${p}_{uncor}^{KS}$ \\ 
\tableline
 Cl\,1018     & 0.368 & 0.250 \\
 Cl\,1037     & 0.846 & 0.996 \\
 Cl\,1040     & 0.845 & 0.845 \\
 Cl\,1054-11  & 0.364 & 0.363 \\
 Cl\,1054-12  & 0.246 & 0.158 \\
 Cl\,1059     & 0.371 & 0.251 \\
 Cl\,1103     & 0.515 & 0.515 \\
 Cl\,1119     & 0.251 & 0.164 \\
 Cl\,1138     & 1.000 & 1.000 \\
 Cl\,1202     & 0.018 & 0.058 \\
 Cl\,1216     & 0.245 & 0.245 \\
 Cl\,1227     & 0.686 & 0.685 \\
 Cl\,1232     & 0.249 & 0.162 \\
 Cl\,1301     & 0.685 & 0.515 \\
 Cl\,1353     & 0.249 & 0.250 \\
 Cl\,1354     & 0.246 & 0.247 \\
 Cl\,1411     & 0.518 & 0.368 \\
 Cl\,1420     & 0.518 & 0.250 \\
\tableline
\end{tabular}


\end{center}
\end{table}

However, in addition to this, the MD and SF relations {\it evolve}
with redshift in a way that {\it depends on environment}. In those
environments that are effective in quenching star formation, galaxies
coming from lower mass/density environments are transformed by
environmental effects when they enter the denser region.  In fact, all
models so far have failed to reproduce the S0 population (which, it is
worth remembering, represents $>$40\% of the galaxies in some
rich clusters at z=0), recognizing that additional processes seem to
be required (Diaferio et al. 2001, Springel et al. 2001, Okamoto \&
Nagashima 2001, 2003, Benson et al. in prep.).  

Unfortunately, ``trends with environment'' have often been confused
with ``environmental effects'', where the latter is used as a
synonym for a physical mechanism switching off star formation in
infalling galaxies. Thus, for example, the fact that star formation
trends exist down to very low local densities and outside of clusters
has often been interpreted in the sense that also such low density
environments must somehow ``suppress'' star formation in
galaxies. This is not necessarily the case, as discussed at length
above.  A trend with environment could be ``imprinted'' very early on
simply due for example to the amount of galaxies with a short star
formation timescale that were able to form at high redshift in that
region.  To fully comprehend why galaxy properties depend on
environment in the way it is observed, it is necessary to disentangle
high-z ``imprinting'' of the initial conditions from
``proper'' environmental effects acting on galaxies when they
experience a dense environment for the first time.  The dependence of
galaxy properties on environment does not necessarily arise from a
``suppression'' of star formation: depending on the density/mass of
the environment, the relative importance of ``primordial'' and
``quenched'' passive galaxies can vary significantly.


Two main challenges remain at this point. Observationally, the
physical mechanism responsible for quenching the star formation still
needs to be identified. The characteristic mass of $M_{sys}> 10^{14}$
$M_{\odot}$ (500 $\rm km \, s^{-1}$) suggested by this work may help
in discriminating among the various processes, but still does not
uniquely pick out a culprit. Our knowledge of how the efficiency of
the various physical mechanisms proposed (e.g. harassment, ram pressure,
strangulation) depends on the mass of the system is still too poor to
draw solid conclusions and discriminate between them.  
From a theoretical point of view,
one of the most useful pieces of information that can come from
state-of-the-art simulations is the link between mass and density at
primordial and successive times, as well as the relation between the
density experienced by a galaxy at different epochs, to assess whether
the relations observed with mass are simply the mirror of relations
with density.

\section{Summary}

In this paper we have studied the fraction of galaxies with ongoing
star formation as a function of environment at z=0.4-0.8, comparing
the results with those at z=0. As a signature for the presence of
ongoing star formation we have used the \oii line in emission with an
equivalent width stronger than 3 \AA.

Our dataset is based on 16 high-z clusters with a velocity dispersion
$\sigma > 400 \rm \, km \, s^{-1}$, 10 groups with $160 < \sigma < 400
\rm \, km \, s^{-1}$ and another 250 galaxies in poorer groups and the
field with high quality spectroscopy from the ESO Distant Cluster
Survey, plus 9 massive clusters at the same redshifts from
previous spectroscopic surveys.  As a local comparison, we have
selected samples of structures from the Sloan Digital Sky Survey at
$0.04< z < 0.08$.

We have presented how the fraction of star-forming galaxies, measured
within $R_{200}$, depends on the velocity dispersion of the cluster/group,
both at high and low redshift. We have discussed how the
evolution of the fraction of star-forming galaxies compares
with the evolution of galaxy morphologies found by previous
authors and by our own survey. We propose a simple scenario
that is able to account for the origin and the evolution
of the observed trends.

In more detail, our results can be summarized as follows:

1) At z=0.4-0.8, most systems follow a broad anticorrelation 
with significant scatter between the fraction of star-forming galaxies and
velocity dispersion: generally, more massive clusters have a
lower fraction of star-forming galaxies.  This [O{\sc ii}]-$\sigma$
relation suggests that the mass of the system, though with a
significant scatter, largely determines what proportion of
galaxies are forming stars at these redshifts.

2) The most evident feature in the \oii fraction versus $\sigma$
diagram observed at high redshift is the presence of a "ridge", or
upper envelope, delimiting a region of the diagram where no datapoint
is found.  This envelope implies that a system of a given mass at this
redshift has {\it at most} a certain fraction of star-forming galaxies
or, equivalently, has {\it at least} a given fraction of galaxies that
are already passive at this epoch. More massive systems have a lower
maximum-allowed fraction of star-forming galaxies or, equivalently, a
higher minimum-allowed fraction of passive galaxies.

3) We find that at z=0.4-0.8 the field and the poor groups in our
sample contain a high proportion of star-forming galaxies (70 to 100\%)
comparable to that observed in more than half of the systems
with $\sigma < 400 \rm \, km \, s^{-1}$.  There are, however, groups
with significantly lower \oii fractions ($<50$\%). These are outliers
that do not follow the [O{\sc ii}]-$\sigma$ trend defined by the
majority of clusters at z=0.4-0.8, and that stand out also for 
other properties of their galaxies, resembling those of galaxies in
the core of much more massive clusters. The existence and characteristics
of the outliers, as well as the fact that the two systems close
to other structures possess a low \oii fraction for their velocity
dispersion, suggest that the factor driving the observed trends might
be density, instead of mass. The behaviour of the \oii fraction with
$\sigma$ might be a secondary relation induced by the fact that
mass and density are closely linked.

4) In addition to the fraction of star-forming galaxies, also the
star-formation properties in star-forming galaxies vary systematically
with environment. Environments with higher \oii fractions have on
average stronger values of equivalent widths of \oii among star-forming
galaxies. This is due to the fact that both the 
equivalent width strengths at a given luminosity and the luminosity
distribution of star-forming galaxies vary with environment.

5) Sloan clusters at $z=0.04-0.08$, analyzed in the same way as
EDisCS, show significantly lower fractions of star-forming galaxies
than clusters at $z \sim 0.4-0.8$.  Moreover, Sloan clusters show the
existence of a plateau for $\sigma$'s above a critical velocity
dispersion, equal to $\sim 550 \rm \, km \, s^{-1}$, above which the
\oii fraction does not vary systematically with velocity dispersion
and remains below 30\% for most clusters.  A trend still might be
present at lower $\sigma$'s, with the average \oii fraction rising
towards lower velocity dispersions.

6) Using N-body simulations to quantify how the mass and velocity
dispersion of a cluster or group evolve on average between $z=0.6$ and
$z=0$, we infer the evolutionary connection between systems at $z=0$
and their progenitors at z=$0.6$ and quantify the average evolution of
the star-forming fraction as a function of velocity dispersion.  While
the strongest evolution in mass is expected for the most massive
structures, the observed evolution of the star-forming fraction is
strongest in intermediate-mass systems, those with $\sigma = 500-600 \rm
\, km \, s^{-1}$ at z=0.  The evolution is lower in higher and lower
mass systems, but is still significant even for the most massive
systems at z=0. The change in star-forming fractions between z=0.4-0.8
and z=0 ranges between 20-30\% and 50\%.

7) We compare the proportions of star-forming galaxies with the
incidence of late Hubble types (spirals and irregulars). Although in
our as in other samples star formation activity and morphologies are
partly decoupled, we find good agreement between the
morphological evolution and the evolution in the star-forming fraction,
consistent with the hypothesis that the evolution observed between
$z \sim 0.6$ and $z=0$ mostly concerns late-type star-forming galaxies evolving
into passive S0 galaxies.

8) Our results quantify the evolution of the star-forming galaxy
populations in clusters and groups, for the first time as a function
of the mass of the system.  Therefore, they are a quantitative
description of the Butcher-Oemler effect in its most general sense.
The way datapoints are distributed in the \oii fraction versus
velocity dispersion at high-- and low--z provide a likely explanation
of why it has been so difficult to observe trends with cluster
mass/$\sigma$ and sometimes even with redshift.  Only when sampling a very
wide range of system masses with large cluster+group samples, do the trends
become recognizable.

9) To understand the origin of the observed trends between galaxy
properties and mass of the environment, we use the Press-Schechter
formalism and the Millennium Simulation to investigate whether galaxy
star formation histories are related to the growth history of the
structures where galaxies reside and have resided during their
evolution. 
We consider a scenario in which the population of passive
galaxies (those devoid of ongoing star formation at the time they are
observed) consists of two different components: ``primordial'' passive
galaxies whose stars all formed at $z>2.5$ and ``quenched'' galaxies
whose star formation has been truncated due to the dense environment
at later times. We find that the observed trend of the fraction of
passive galaxies with velocity dispersion at $z=0.4-0.8$ 
follows the fraction (in mass and in number of galaxies) that is
expected to have been in groups ($M_{sys} > 3 \times 10^{12} \,
M_{\odot}$) already at $z=2.5$ and that we identify with the
primordial passive population, though in the most massive systems 
quenched galaxies should already represent a non-negligible fraction 
of the passive population.  At $z=0$, on the other hand, the observed
fraction of passive galaxies in clusters (80\%) resembles the fraction
(in mass and in number of galaxies) that has resided in clusters
($M_{sys} > 10^{14} \, M_{\odot}$) during at least the last 3 Gyr,
including 20\% of primordial passive galaxies and 60\% of quenched
galaxies.  This scheme is able to interpret the observed relations
between \oii and $\sigma$, thus providing a viable quantitative
explanation for the evolution of the star formation activity in dense
environments, and a possible explanation for the origin of the
Butcher-Oemler effect.  If this scenario approximates the real situation, 
galaxy star
formation histories are closely linked with the galaxy ``environmental
history'', and this link is actually extremely simple to predict.

 \begin{figure*}[t]
 \centerline{\includegraphics[width=8cm]{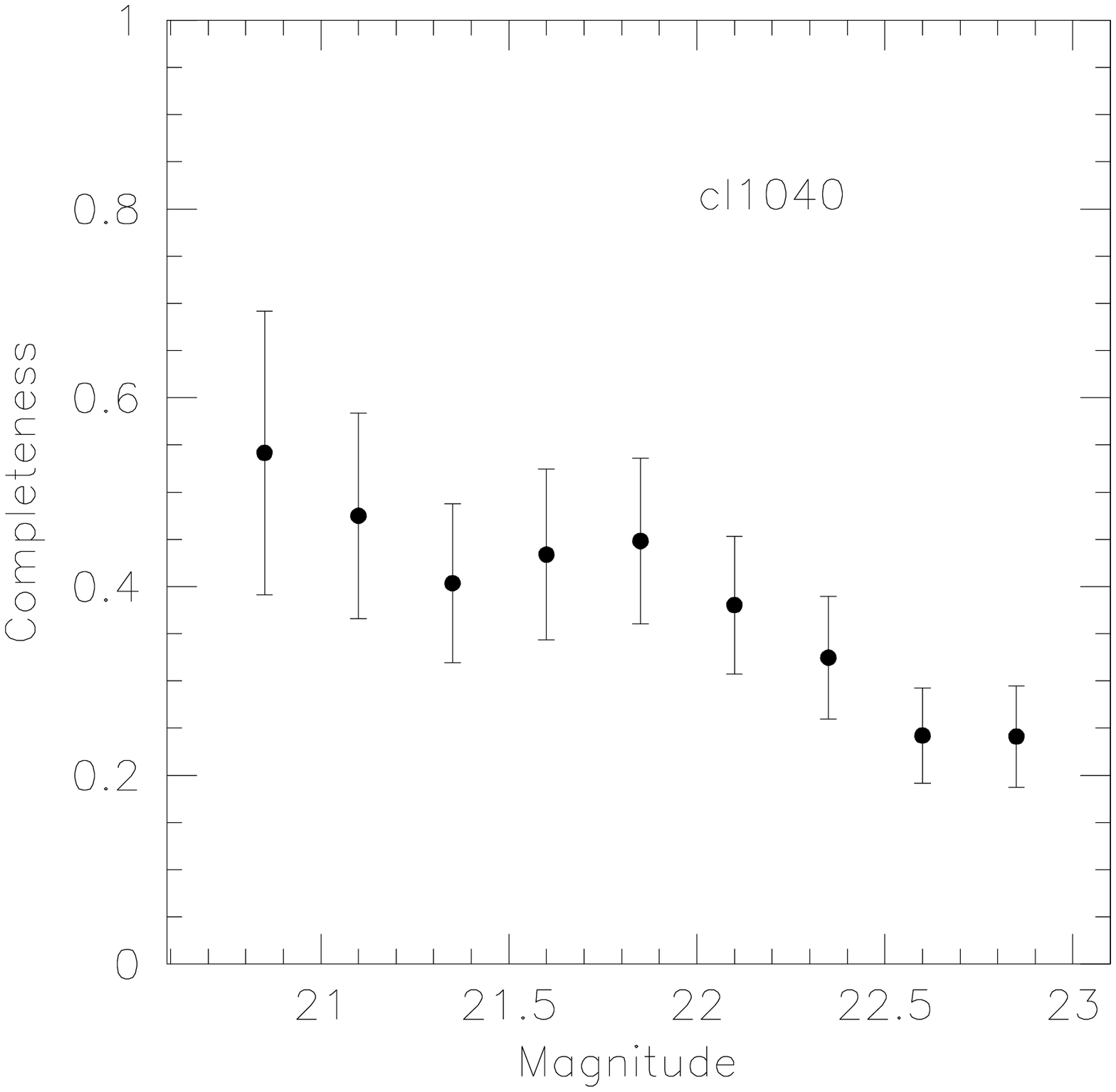}\hfill\includegraphics[width=8cm]{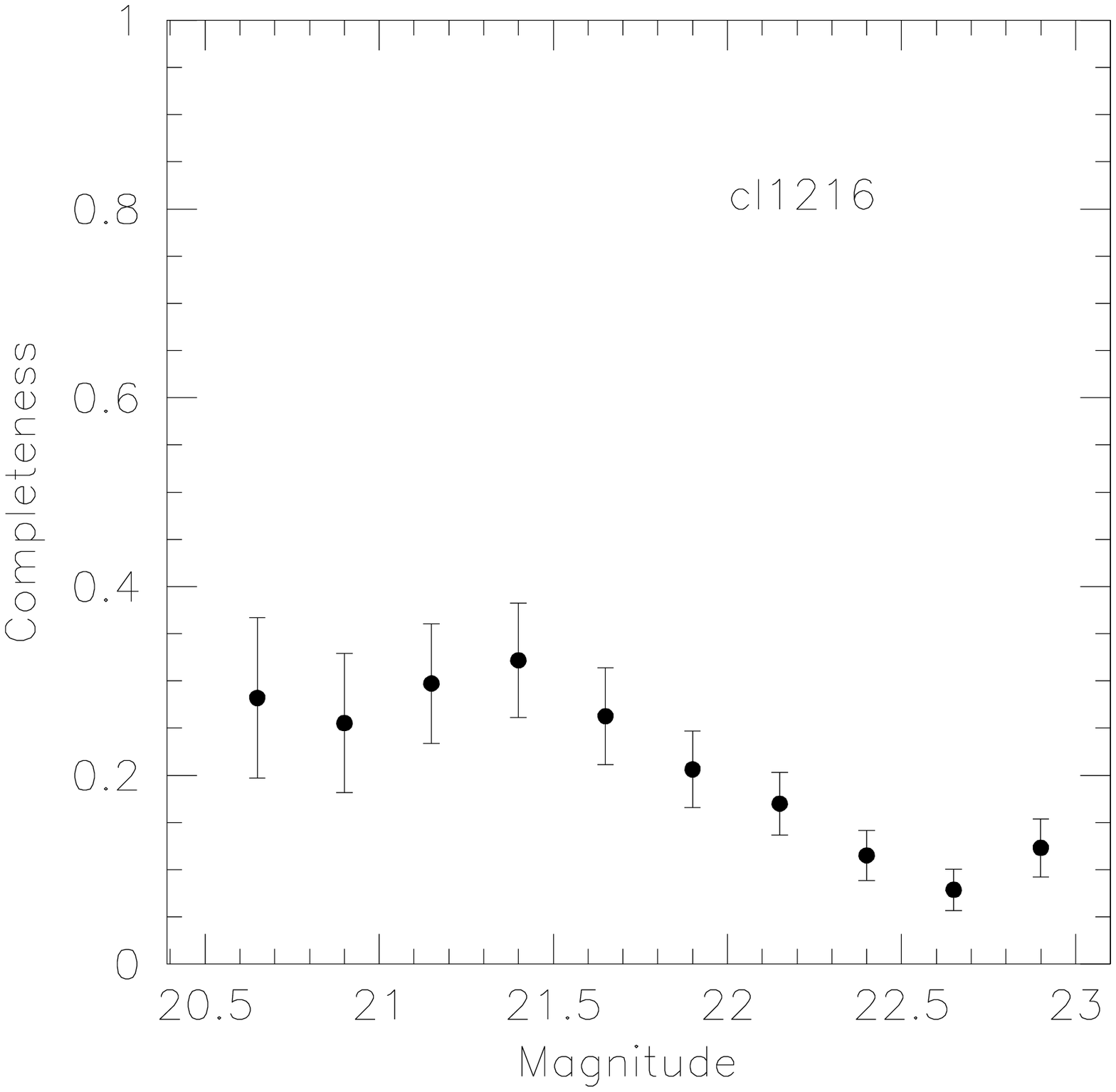}}
 \centerline{\includegraphics[width=8cm]{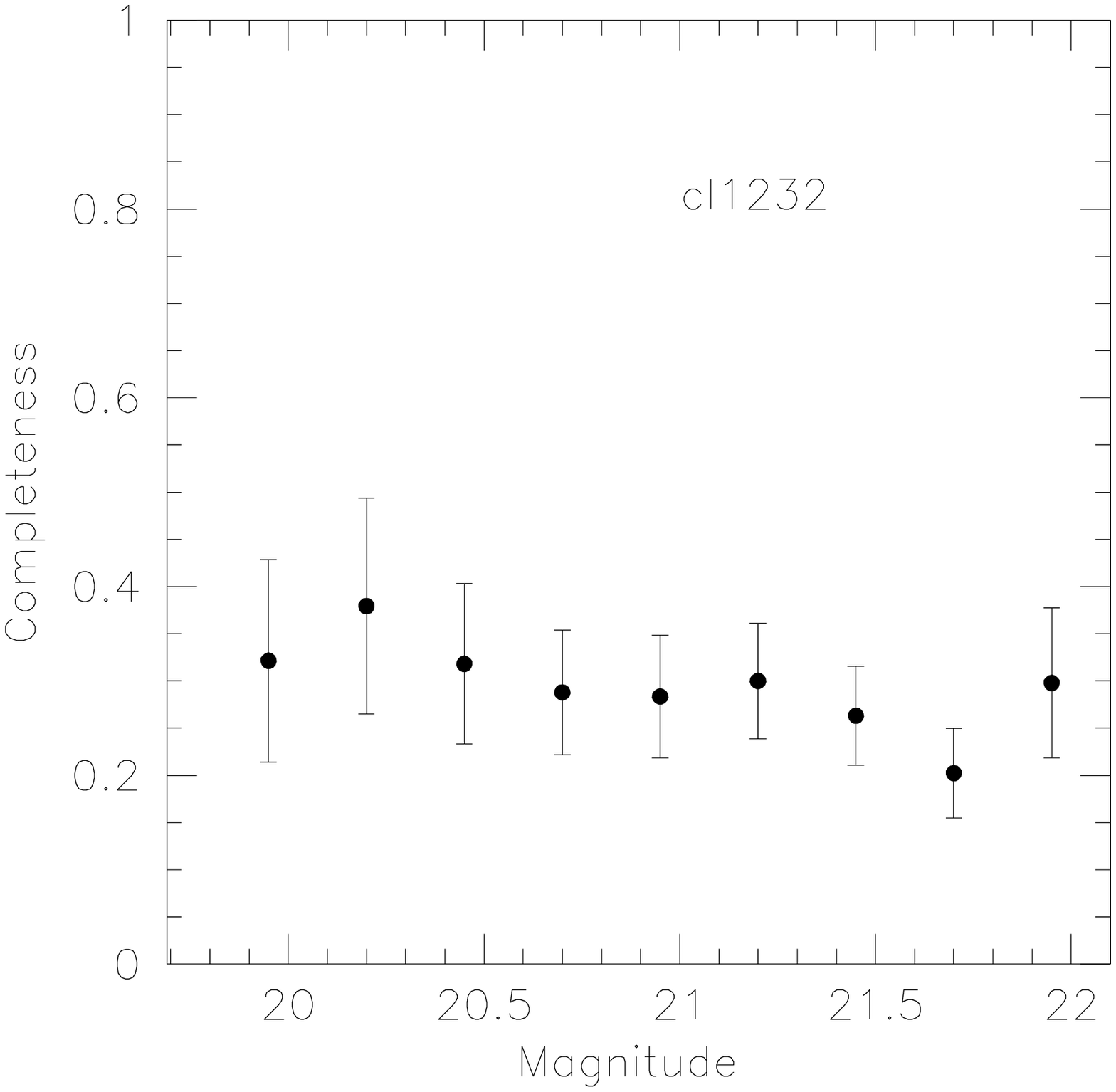}\hfill\includegraphics[width=8cm]{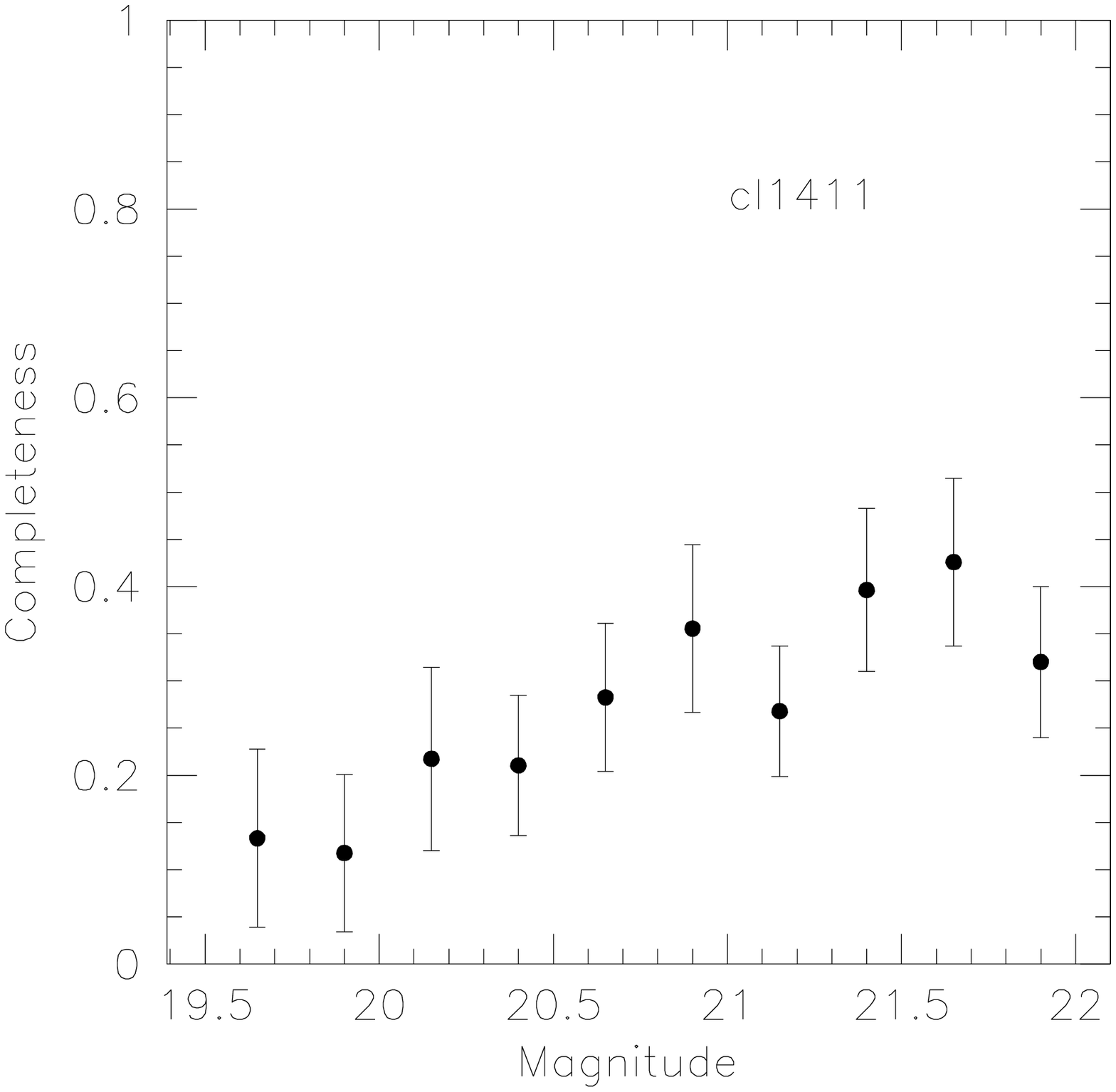}}
 \caption{Completeness functions versus galaxy magnitudes
 for four EDisCS clusters.\label{completeness}}
 \end{figure*}

10) The behaviour of the \oii fraction with $\sigma$ at z=0 appears to
rule out the hypothesis that the group environment efficiently and
universally quenches star formation. 
In fact, the existence of a plateau in the
\oii fraction for $\sigma > 500-550 \rm \, km \, s^{-1}$ at z=0 
suggests that only systems more massive than about $450 - 500
\rm \, km \, s^{-1}$ are highly efficient at truncating star formation
in galaxies infalling into them. If the quenching of star formation
was a widespread phenomenon also in less massive systems, the fractions of
star-forming galaxies in clusters and groups at low redshift should be
much lower than is observed. Conversely, the observed reference mass
indicates that the quenching of star formation is not limited to 
very massive clusters, but is efficient also in clusters of modest
mass.

 \begin{figure*}[t]
 \vspace{-9cm}
 \centerline{\hspace{6cm}\includegraphics[width=20cm]{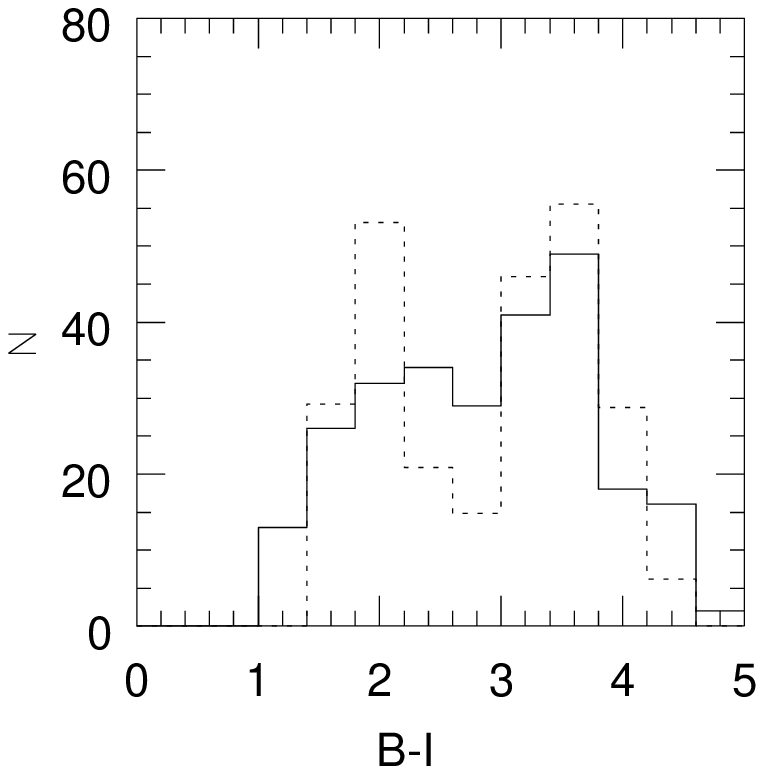}}
 \vspace{-1.5cm}
 \caption{Color distributions 
 of the parent photometric catalog
 (solid line) and of the spectroscopic catalog corrected for
 incompleteness (dashed line) for the cluster C\,1202.
 \label{cl1202}}
 \end{figure*}

 \begin{figure*}[t]
 \vspace{-9cm}
 \centerline{\hspace{8cm}\includegraphics[width=20cm]{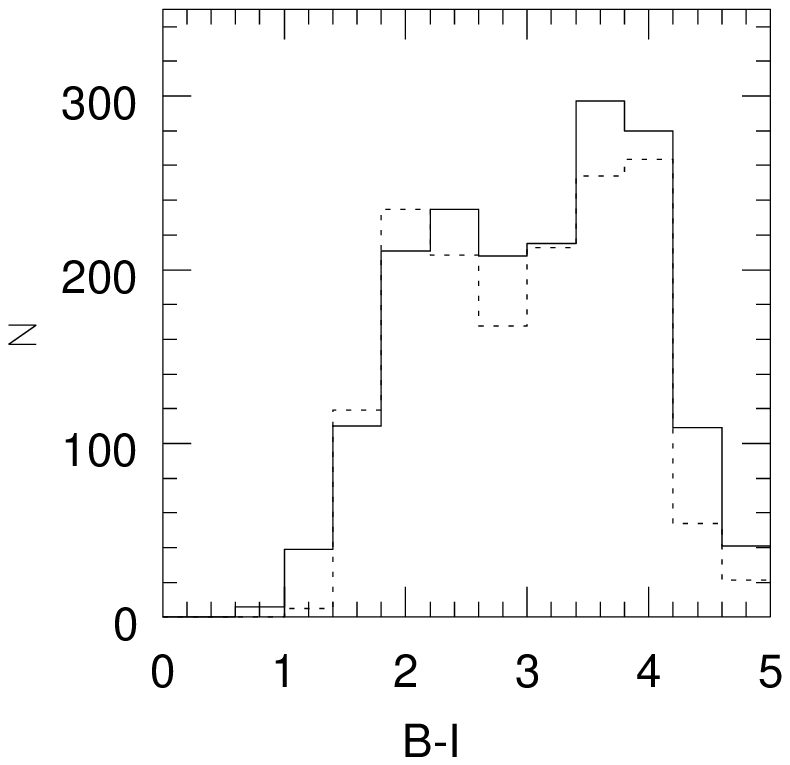}\hfill\hspace{-12cm}\includegraphics[width=20cm]{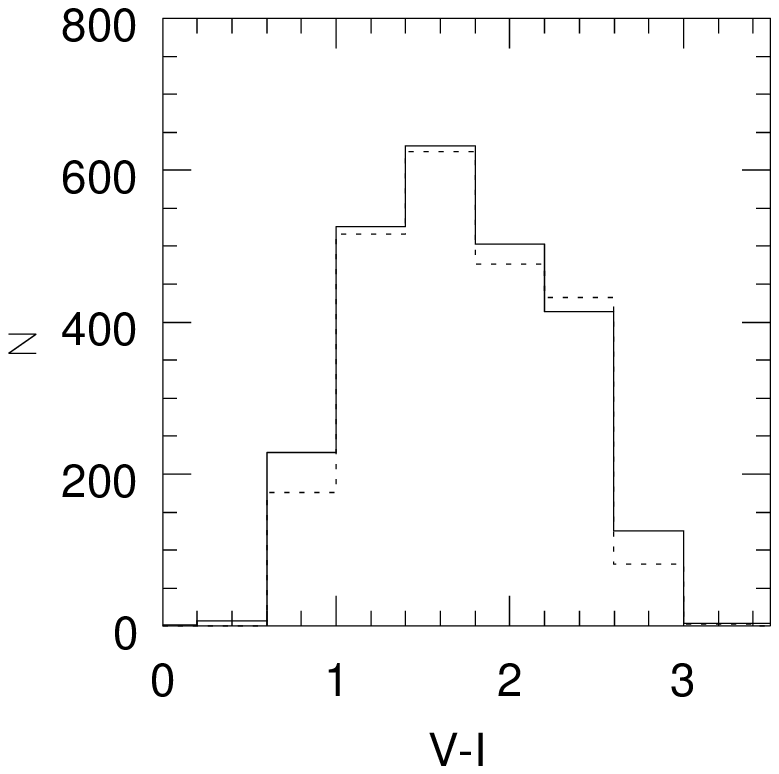}}
 \vspace{-1.5cm}
 \caption{Left. Color distribution of the parent photometric catalog (solid
 line) and the spectroscopic catalog corrected for completeness (dashed line)
 for the 9 clusters with B photometry. 
 Right. As left panel, for 
 clusters with V photometry.\label{sanitycheck}}
 \end{figure*}

11) There are numerous implications stemming from the evolutionary
scenario proposed. The parallelism between the observed evolution of
the star-forming fraction and the morphological evolution suggests that
the same physical interpretation of a link with the mass of the
environment at $z>2.5$ and with the accretion history on a cluster can be
applied to the formation of ellipticals and at least some of the S0s,
respectively. In fact, the scenario we propose can be used as an interpretative
tool for the variations of galaxy properties with environment in
general. Our results highlight the fact that not all trends with
environment are necessarily linked with environmental processes
truncating star formation in recently accreted galaxies, and that
primordial conditions (=the enviroment at very high redshift) are an
important factor in determining the trends observed with
environment. Related to this, we have discussed the consequences of
our results for the origin and evolution of the morphology-density and
star formation-density relations.

\acknowledgments

We warmly thank Jarle Brinchmann for helping us use the SDSS
products and throughout this project, Risa Wechsler, Frank van den
Bosch and Ray Carlberg for their valuable theoretical advice, Ian
Smail for his comments that helped us strenghten an
early version of this paper, and the referee, David Wilman, for his careful
and detailed report that led to several improvements.  
We acknowledge the usefulness of the WEB
pages maintained by James Bullock and Pieter van Dokkum at
http://www.physics.uci.edu/\~{}bullock/CVIR and
http://www.astro.yale.edu/dokkum/evocalc (astro-ph 0501236, van Dokkum
\& Franx 2001), respectively.  B.~M.~P. acknowledges financial support
from the ``Fondo per gli Investimenti nella Ricerca di Base'' of the
Italian Ministery of Education, University and Research (code
RBAU018Y7E).  G.~D.~L. thanks the Alexander von Humboldt Foundation,
the Federal Ministry of Education and Research, and the Programme for
Investment in the Future (ZIP) of the German Government for financial
support.






\appendix

\section{Spectroscopic completeness}

Here we describe our method for correcting 
the EDisCs and the Sloan spectroscopic samples for incompleteness.

During the EDisCS spectroscopic runs, slits were assigned to galaxies
giving preference, whenever possible, to the brightest targets.  The
Sloan spectroscopic sample was aimed to be complete down to $r=17.7$,
but about 6\% of the galaxies were missed due to fiber collision
(Strauss et al. 2002), an effect that is likely to be more relevant in
crowded regions such as cluster cores.  Therefore it is necessary to
quantify how the completeness of the spectroscopic samples varies as a
function of galaxy apparent magnitude and of distance from the cluster
center.

As a function of galaxy magnitude, this was done for each field
comparing the number of objects in the spectroscopic catalog with the
number in the parent photometric catalog in bins of I (for EDisCS) or
$g$ (for Sloan) magnitude. The parent catalog included all entries in
the EDisCS photometric catalog that were retained as targets for
spectroscopy (see Halliday et al. 2004).  The ratio of these two
numbers yielded a weight as a function of galaxy apparent magnitude
($W_{mag}$).  We preferred to compute these weights field by field
instead of binning all fields together, because depending on cluster
richness and number of masks observed, the behaviour of the
completeness functions can change significantly. As an example, in
Fig.~\ref{completeness} we show two clusters for which the
completeness decreases towards fainter magnitudes (Cl\,1040 and
Cl\,1216), a cluster where the variations are negligible (Cl\,1232)
and one of the only two cases in which the completeness increases
towards fainter magnitudes (Cl\,1411).

We also quantified the presence of eventual geometrical effects due to
possible variations in the sampling as a function of the
clustercentric radius. Geometrical effects can in principle affect a
spectroscopic sample of a cluster due to the fact that cluster
galaxies are indeed more ``clustered'' towards the cluster center
while observational constraints on the minimum distance between slits
or fibres could result in a lower sampling of these central
regions. This effect is expected to be small when several masks of the
same cluster, always centered on the cluster center, were taken, as it
was the case for EDisCS.  Neverthless, a geometrical completeness
$W_{geo}$ was computed, after applying the magnitude completeness
correction, comparing the number of galaxies in the spectroscopic and
in the parent photometric catalogs in 4 annuli with $R<1.4$,
$1/4<R<1/2$, $1/2<R<1$ and $R>1$ in units of $R_{200}$. 
The geometrical corrections hence include corrections for the areas
not covered by our FORS2 spectroscopy
in the few clusters with incomplete radial sampling
out to $R_{200}$. 

The magnitude and geometrical completeness functions, computed cluster
by cluster, were applied to the EDisCs and Sloan spectroscopic samples
to weight each galaxy accordingly before calculating the \oii
fractions and the EW(\oii) distributions described above.  
We found
that these weights do not alter significantly the \oii fractions, 
as evident from Tables~1 and ~3 listing the fractions computed with and
without completeness corrections.

\section{Color sanity check}

As a final check that the EDisCS spectroscopic sample is not biased
in any way that depends on the star formation properties of
the galaxies, we have compared the color distributions ($B-I$ or $V-I$
for the mid-z and high-z samples, respectively) of
the spectroscopic sample (with and without completeness
corrections) with the color distribution of the parent 
photometric sample. This has been done field by field,
and the resulting KS test probabilities are given in Table~6.
A small probability ($<5$\%) indicates that the two distributions are
significantly different, hence that a color bias might be present.
The table illustrates a number of results:

1) for all fields except Cl\,1202
probabilities are high even when
the spectroscopy has not been corrected for completeness (column 3).
Evidently, even the uncorrected samples do not show a strong
color bias. In principle, this was not guaranteed a priori because,
although targets for spectroscopy were drawn from the parent 
targeting catalog with no color criterion, slits were preferentially
assigned to the brightest galaxies first, and this could have 
introduced a difference in the color distribution of the two samples.

2) For 8 fields the probabilities 
increase (therefore the agreement
between the two color distributions improve) once the completeness
corrections are taken into account. This indicates that
indeed the corrections yield a sample even more closely
representative of the whole galaxy distributions. For another 8
clusters, the probabilities remain unchanged with and without
completeness corrections.

3) Two fields behave differently from the above. 
When including completeness corrections, the probability for Cl\,1037
slightly decreases, though it remains high (0.85) ruling out a color
bias. Cl\,1202 is the cluster with the lowest 
probabilities (0.058 and 0.018 without and with corrections, respectively).
The color distributions for Cl\,1202 are shown in Fig.~\ref{cl1202}. 
The plot shows that
there is no strong bias towards either red or blue galaxies, therefore
we do not expect the \oii fraction derived for this cluster to be strongly biased.

Figure~\ref{sanitycheck}  compares the color distributions of the 
spectroscopic catalogs
and the parent photometric catalogs for all clusters, grouping
together the mid-z clusters
(with B photometry, left panel) and 
the high-z clusters (V photometry, right
panel). As the numbers in Table~6, also the figure illustrates 
that no significant color bias is present in our
spectroscopic sample. We conclude that the \oii fractions derived in this
paper are representative within the errors of the ``true'' fractions
in the EDisCs clusters to the adopted magnitude limits.

\section{Velocity dispersion of Sloan clusters}

  As for the EDisCS sample, we rely on the biweight estimator of
  Beers et al. (1990) for determining the cluster redshift $z_{\rm C}$ and
  velocity dispersion  $\sigma_{\rm v}$ of Sloan clusters:
\begin{enumerate}
\item To begin with, 
  we chose galaxies within 2.2 Abell radii of the BCG and within
  $\pm 0.015$ from the cluster redshift as given by Struble \& Rood (1990).
  From these galaxies, first estimates of the cluster
  redshift $z_{\rm C}$ and the velocity dispersion $\sigma_{\rm z}$
  were calculated as the median and the median absolute deviation. If
  $\sigma_{\rm z}$ was larger than 0.0017 (corresponding to about
  $500 \, \rm km \, s^{-1} $ at $z=0$), we set it to this value.  This step was necessary
  to avoid too much contamination from surrounding structures.
\item From those galaxies within $\pm 3\sigma_{\rm z}$ from $z_{\rm C}$ and
  1.2~$R_{200}$ (calculated from $z$ and $\sigma_{\rm z}$ using
  Eq.~1), 
  we recalculated first $z$, then $\sigma_{\rm z}$
  via the biweight estimators given in Beers et al. (1990). 
\item This process was iterated until it reached convergence. At each round,
  every galaxy in the initial sample can re-enter the cluster sample if it
  meets the constraints on  redshift and velocity.
\item The error on the final $\sigma_{\rm z}$ is calculated from a bootstrap
  analysis of the galaxies that make up the final cluster sample.
\end{enumerate}




{}

\end{document}